# Memory-resolved prediction of ligand residence times on the minutes timescale

Hossein Batebi[1], Lucas Tepper[1] and Roland R. Netz[1]

[1] Department of Physics, Freie Universität Berlin, Arnimallee 14, 14195 Berlin, Germany

Corresponding authors, hbat@zedat.fu-berlin.de; rnetz@physik.fu-berlin.de

**Abstract**

AI-enabled computational drug discovery is advancing rapidly in predicting molecular structures, binding poses and affinities, but ligand dissociation kinetics remain harder to access. Bound-state residence time, important for biological response and kinetic selectivity, is not fixed by equilibrium affinity but depends on the dissociation free-energy landscape and dissipative friction effects. Here, we show that restrained umbrella-sampling trajectories can be used to calculate residence times on the minutes timescale. Time-resolved coordinate fluctuations yield friction profiles that incorporate non-Markovian effects on dissociation kinetics. Combining them with the free-energy profile, we predict a residence time of $\tau_{res}$ = 275 s for the adenosine $A_2A$ receptor with ZM241385 (comparable to the experimental value of 125 s) and reproduce the experimental ranking of wild type and two receptor mutants. For HSP90 with compound 18 and RNase S with the S15 peptide, we predict $\tau_{res}$ = 2,169 s and 2,259 s (experimental values are 833 s and 2,439 s). Neglecting memory effects on residence kinetics overestimates $\tau_{res}$ by factors of 2.6 to 29.

The drug–target residence time, the mean time a ligand remains bound before it dissociates, is an important determinant of drug action but remains difficult to predict computationally.[1,2] Affinity is an equilibrium quantity, whereas dissociation is a dynamical process. Affinity is determined by the ratio of the mean bound and unbound residence times at a given ligand concentration, or equivalently by the mean occupation fraction of the binding site, and does not involve friction: friction changes how fast equilibrium is reached, not the equilibrium itself. Ligands with comparable affinities can display markedly different residence times, with consequences for duration of action and kinetic selectivity.[1,2] Direct prediction of ligand dissociation with unbiased atomistic simulation is difficult because the relevant events commonly occur on seconds-to-minutes timescales, far beyond routine trajectory lengths. Dissociation can also involve several sequential steps, including disruption of interactions in the binding pocket, passage through intermediate vestibular regions, rearrangement of surrounding protein and solvent, and final release into bulk solution. For G-protein-coupled receptors (GPCRs), this kinetic problem can become more complex because the receptor itself samples multiple functional states during signalling. Receptor activation and G-protein coupling proceed through distinct receptor and receptor–transducer intermediates, including nucleotide-bound and nucleotide-free complexes, with substantial conformational rearrangements in both partners.[3,4] This creates distinct conformational backgrounds from which ligand dissociation may occur, adding another source of heterogeneity to the escape process.

Because direct trajectories rarely reach these timescales, dissociation kinetics are usually estimated with dedicated rare-event approaches. Long unbiased trajectories and Markov state models estimate rates from sampled transitions, whereas metadynamics, accelerated dissociation schemes, milestoning and weighted-ensemble approaches obtain kinetics from enhanced or partitioned sampling.[5–15] Rates and position-dependent diffusion can also be recovered from biased trajectories,[16,17] and umbrella sampling has been coupled to milestoning for ligand dissociation.[18] The direct output of umbrella sampling, however, is a potential of mean force (PMF), which does not determine $k_{off}$ by itself. The binding free energy fixes the ratio $k_{off}/k_{on}$ rather than $k_{off}$, and ligands with similar affinities can differ in residence time by orders of magnitude.[1,2] Barrier-crossing kinetics therefore contain two distinct contributions: the free-energy landscape and a prefactor set by friction along the dissociation coordinate. Friction here denotes the dissipative coupling of this coordinate to protein, solvent and membrane degrees of freedom; we use the terms friction and dissipation interchangeably. Two ligand–receptor complexes can have similar dissociation barriers yet different residence times if motion along the escape coordinate experiences different friction. A PMF captures the energetic contribution but not, by itself, the frictional one.

First-passage analysis connects the energetic and frictional contributions. Given the PMF $U(x)$ along the dissociation coordinate $x$ and the local friction, it returns an escape time without requiring the dissociation event to occur in any individual trajectory.[16–20] The additional quantity needed beyond the PMF is therefore the friction profile.

When a many-body trajectory is projected onto one dissociation coordinate, protein, solvent and membrane motions give rise to friction and non-Markovian memory effects in the projected dynamics.[21–23] For ligand binding, solvent fluctuations in hydrophobic pockets have been shown to produce such non-Markovian kinetics.[24,25] In protein folding, accounting for such memory substantially altered predicted barrier-crossing times, with friction changes that could rival or exceed changes in the free-energy barrier.[21,22] We therefore ask whether the same separation of energetic and frictional contributions can be obtained from restrained ligand trajectories.

Here we extract the friction profile $\gamma(x)$ from the same restrained trajectories that provide $U(x)$, using a non-Markovian construction that maps multiexponential coordinate relaxation onto barrier-crossing times and gives an effective friction $\gamma_{eff}(x)$.[20–23,26–30] We compare it with two alternative friction profiles computed from the same trajectories: a position-autocorrelation friction $\gamma_{pos}(x)$[19] and a solvent-force friction $\gamma_{sol}(x)$ obtained from the fluctuating force on the restrained coordinate.[31] All three friction profiles account for the non-Markovian memory effects on the friction profile, but only $\gamma_{eff}$ accounts for the effect of memory on the barrier-crossing time and hence on the residence time. How strongly memory effects change mean first-passage times also depends on where along the coordinate they act, in the bound well or at the barrier.[29,32] Our approach thus converts umbrella trajectories into a residence time and provides information on whether non-Markovian effects are relevant.

To evaluate the approach across systems rather than a single molecular example, we selected systems according to four criteria: quantitative experimental dissociation data; residence times on seconds-to-minutes scales beyond routine atomistic simulation; distinct molecular environments that challenge transferability; and, where applicable, direct relevance to drug discovery. The adenosine $A_2A$ receptor provides a class A GPCR with both an experimental wild-type (WT) residence time and a mutation series measured under identical conditions. GPCRs are among the protein families most frequently targeted by approved drugs; a comprehensive analysis found that approximately 34% of drugs approved by the US Food and Drug Administration (FDA) acted at GPCR targets,[33] and the $A_2A$ antagonist istradefylline is FDA-approved as adjunctive therapy for Parkinson's disease with OFF episodes.[34] The N-terminal domain of HSP90 (N-HSP90) provides a structurally unrelated soluble small-molecule target from a molecular-chaperone system extensively investigated in cancer therapeutics, where HSP90 supports numerous oncogenic client proteins.[35] RNase S with the S15 peptide (RNase S–S15) extends the test beyond conventional small-molecule drug targets to a protein–peptide complex with directly measured exchange kinetics. This set therefore tests the same calculation across membrane and soluble proteins, small-molecule and peptide ligands, and different experimental kinetic assays.

## Results

We implemented the analysis as a single workflow from restrained umbrella trajectories to predicted residence times (**Fig. 1**). The trajectories are processed in parallel to obtain the free-energy profile $U(x)$ and the friction profiles, which are then combined in the residence-time calculation (equation (3)). Both $\gamma_{eff}$ and $\gamma_{pos}$ are obtained from the position autocorrelation $C_{xx}(t)$ of each umbrella window (equation (6)), $\gamma_{pos}$ from its time integral (equation (11)) and $\gamma_{eff}$ from the time integral of its square (equation (7)). The solvent-force friction $\gamma_{sol}$ is obtained from the autocorrelation of the fluctuating force reconstructed from the same coordinate trajectory (equations (12) and (13)). Because the three friction profiles enter the same calculation with the same $U(x)$, the effect of memory on barrier crossing can be assessed directly. For $A_2A$–ZM241385, N-HSP90–compound 18 and RNase S–S15, the calculation predicted $\tau_{res}$ = 275, 2,169 and 2,259 s, compared with experimental values of 125, 833 and 2,439 s. Below we examine how the energetic and frictional contributions combine to produce these values, first in $A_2A$–ZM241385 and then in the two other systems.

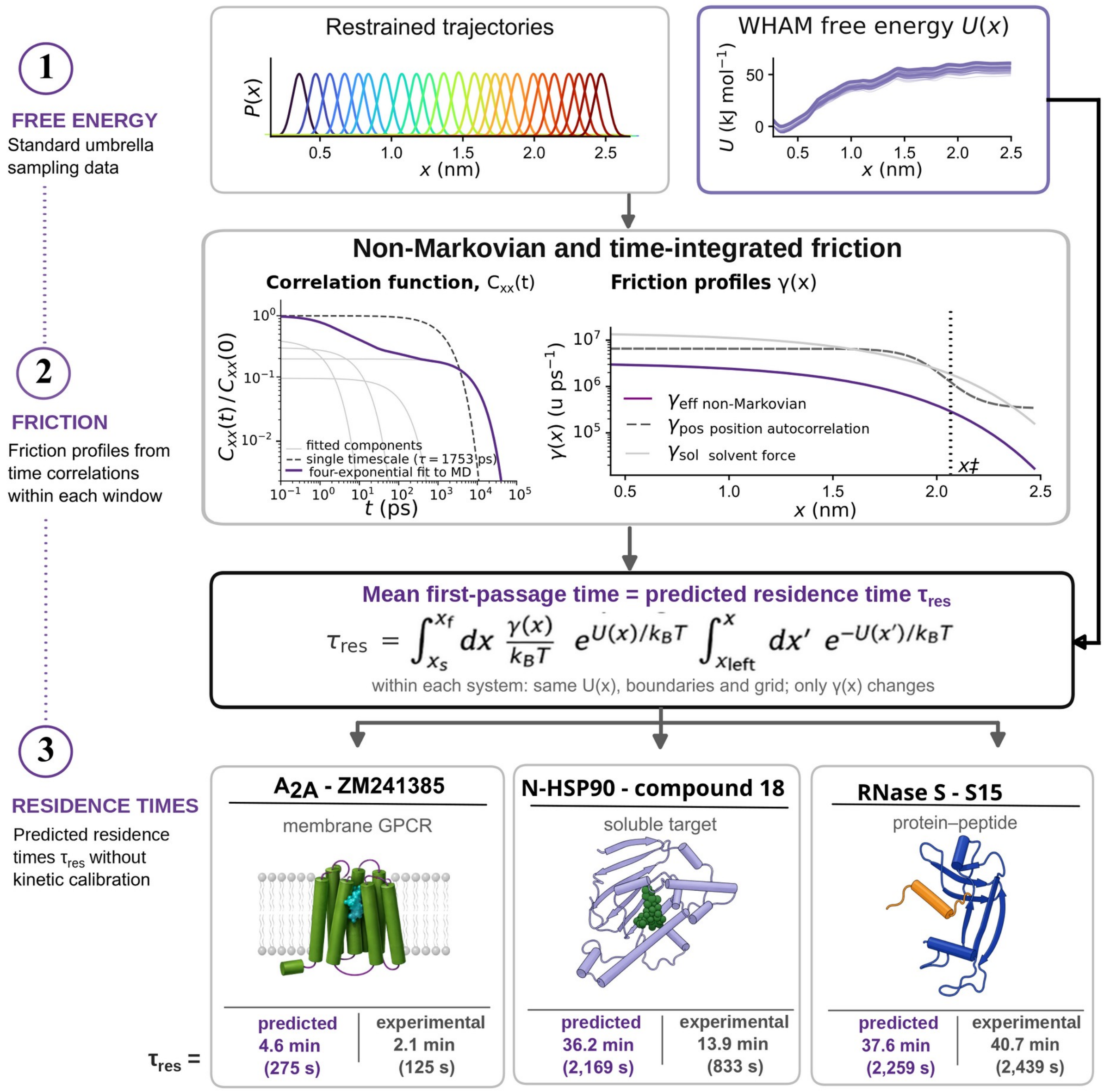

**Fig. 1 | Residence-time prediction from restrained trajectories.** Umbrella trajectories provide the free-energy profile U(x) and three friction profiles: non-Markovian $\gamma_{eff}(x)$ (equation (7)), position-autocorrelation $\gamma_{pos}(x)$ (equation (11)) and solvent-force $\gamma_{sol}(x)$ (equation (12)). Within each system, the same U(x), boundaries and numerical grid are used, and only γ(x) changes. Bottom, predicted and experimental residence times $\tau_{res}$ for the $A_2A$ receptor with ZM241385, the N-terminal domain of HSP90 with compound 18 and RNase S with the S15 peptide.

Of the three systems, $A_2A$ with the selective antagonist ZM241385 allows the most detailed test, because WT and receptor variants were analysed along the same dissociation coordinate and experimental data are available for the WT residence time and for the relative effects of T256A and E169Q. We therefore use $A_2A$ to separate the structural, energetic and frictional contributions before turning to the other two systems. ZM241385 occupies the orthosteric cavity and competes with the endogenous agonist adenosine. Along the sampled dissociation coordinate, ZM241385 exits through an extracellular vestibule containing E169, H264 and T256. This escape route provides a structurally interpretable coordinate along which both the free-energy profile and the friction profile can be evaluated.

The dissociation coordinate x is the separation along the membrane normal between the centre of mass of ZM241385 and the centre of mass of a binding-pocket reference group of ten Cα atoms (**Fig. 2** and Methods). Representative configurations at x = 0.35, 1.28 and 2.18 nm define the bound, vestibular and bulk-solvent regions (**Fig. 2**).

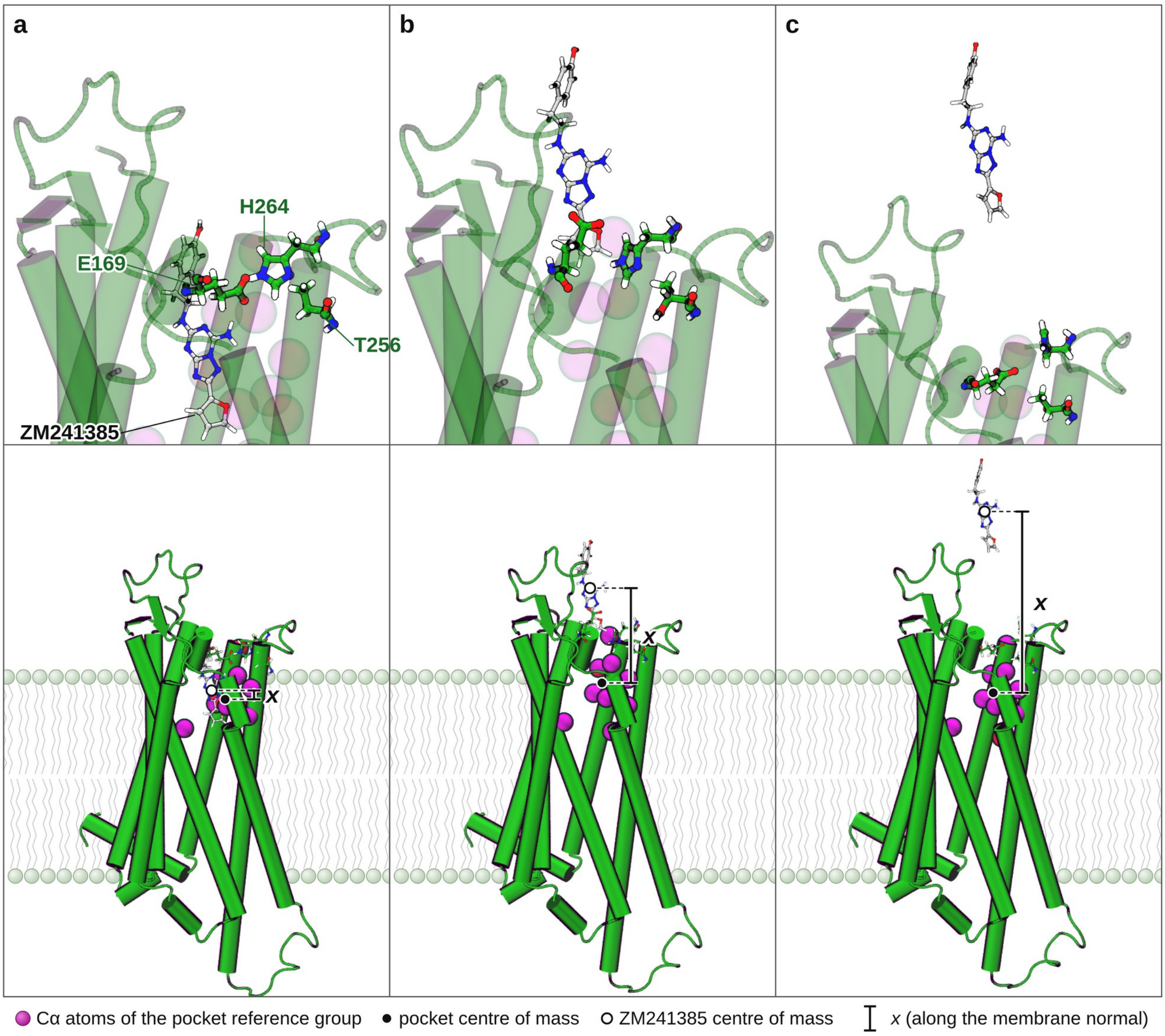

**Fig. 2 | ZM241385 dissociation coordinate in $A_2A$. a–c**, Bound (x = 0.35 nm), vestibular (x = 1.28 nm) and absorbing-boundary (x = 2.18 nm) configurations. Upper panels show the binding region with E169, H264 and T256; lower panels show the receptor in a schematic membrane. Magenta spheres (transparent in the upper panels) mark the Cα atoms of the pocket reference group. In the lower panels, the black and white circles mark the projected centres of mass of this group and of ZM241385 calculated from the coordinates, and the vertical bar indicates x, their separation along the membrane normal (Methods).

### The free-energy landscape sets the mutation ordering

We first examined how the mutations change the free-energy profile U(x) and then how they change the friction profile $\gamma_{eff}(x)$ (equation (7)). WT and mutants share the same coordinate definition and umbrella force constant (Methods), so their profiles can be compared directly along x.

The principal barrier $\Delta U^{\ddagger} = U(x^{\ddagger}) - U(x_s)$ decreased from 61.35 kJ mol$^{-1}$ for WT to 56.13 kJ mol$^{-1}$ for T256A, 49.31 kJ mol$^{-1}$ for E169Q and 45.58 kJ mol$^{-1}$ for E169A (**Fig. 3a,b**, **Supplementary Figs. 1 and 2** and **Supplementary Table 1**). Among constructs with experimental residence times, $\tau_{res}$ follows the same order: WT > T256A > E169Q.

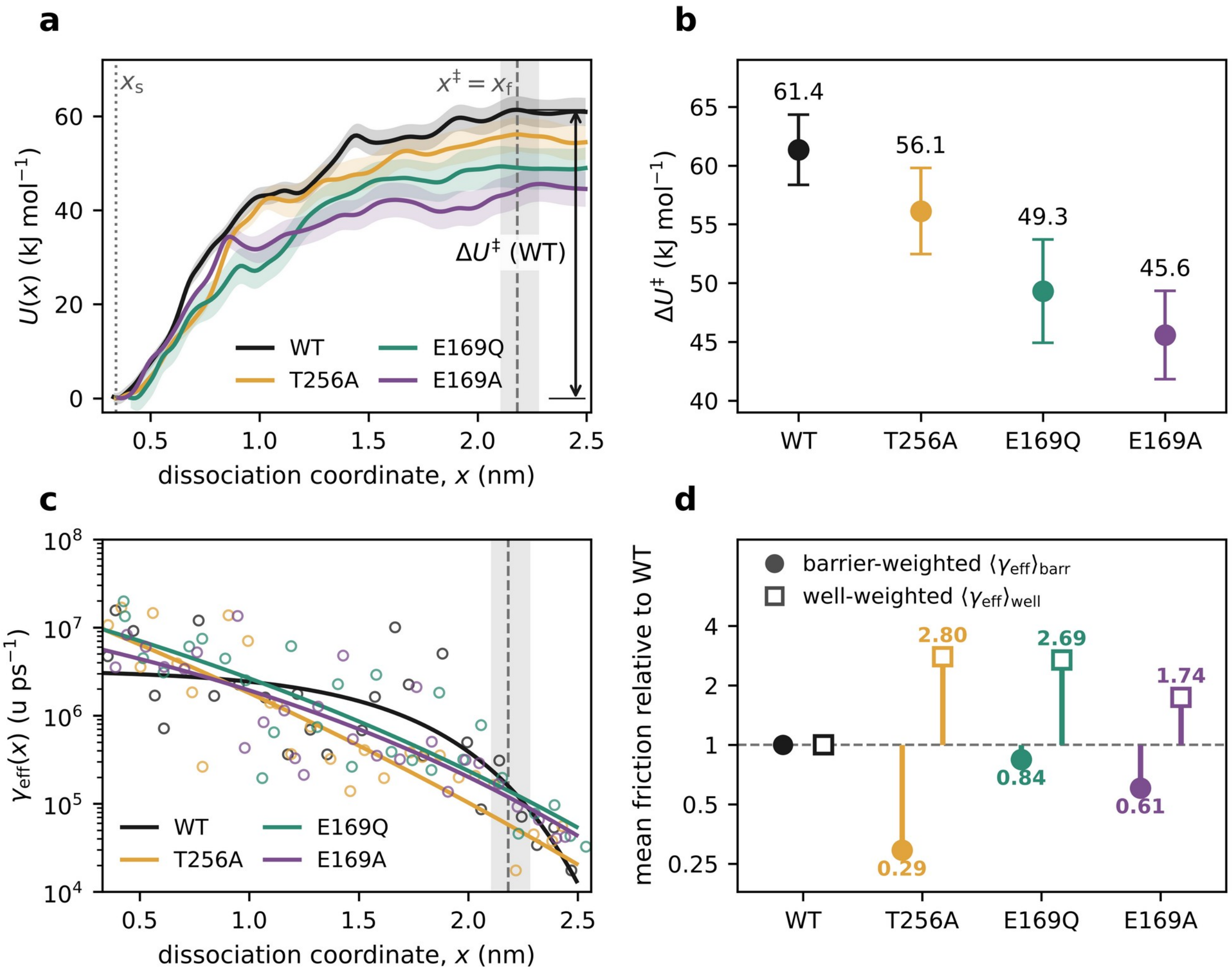

**Fig. 3 | $A_2A$ energetic and frictional response to mutation.** **a**, Free-energy profiles U(x) for WT, T256A, E169Q and E169A. Shading shows the bootstrap mean ± s.d. from 1,000 iterations; $x_s$ marks the bound-state minimum and $x‡ = x_f$ the principal barrier and absorbing boundary; the double arrow marks ΔU‡ for WT. **b**, Barrier heights ΔU‡ = U(x‡) − U($x_s$). Points show ΔU‡ and error bars the maximum pointwise bootstrap s.d. (1,000 iterations); each construct is represented by one umbrella array. **c**, Non-Markovian friction profiles $\gamma_{eff}(x)$ (equation (7)). Symbols show per-window estimates and lines fitted profiles (equation (10)). **d**, Barrier-weighted mean friction $\langle\gamma_{eff}\rangle_{barr}$ (equation (4); filled circles) and well-weighted mean friction $\langle\gamma_{eff}\rangle_{well}$ (equation (5); open squares) relative to WT on a logarithmic scale; dashed line, WT value. Predicted residence times with $\gamma_{pos}$ and $\gamma_{sol}$ are compared in **Fig. 4**.

In all constructs, $\gamma_{eff}(x)$ (equation (7)) was largest near the binding pocket and decreased toward the solvent (**Fig. 3c** and **Supplementary Figs. 3–8**). The mutations change the friction in opposite directions in the barrier region and in the bound well (**Supplementary Table 11**). In the barrier region, quantified by the barrier-weighted mean $\langle\gamma_{eff}\rangle_{barr}$ (equation (4)), the friction of T256A falls to 0.29 of WT while its barrier decreases by only 5.2 kJ mol$^{-1}$, and E169Q retains 0.84 of WT friction despite a 12.0 kJ mol$^{-1}$ lower barrier (**Fig. 3b,d**). In the bound well, quantified by the well-weighted mean $\langle\gamma_{eff}\rangle_{well}$ (equation (5)), all three substitutions increase the friction, by factors of 2.80 (T256A), 2.69 (E169Q) and 1.74 (E169A) (**Fig. 3d**). Because equation (3) weights γ(x) most strongly in the barrier region, the predicted $\tau_{res}$ depends on $\langle\gamma_{eff}\rangle_{barr}$ (Methods). $\langle\gamma_{eff}\rangle_{well}$ does not enter equation (3); it instead summarizes friction in the bound well, the region expected to control the MFPT in the long-memory limit.[32] Within this description, the two mutations accelerate dissociation through different combinations of barrier lowering and reduced barrier friction.

### Non-Markovian friction sets the residence-time scale

For single-exponential relaxation of the position autocorrelation $C_{xx}(t)$ (equation (6)), $\gamma_{eff}$ and $\gamma_{pos}$ coincide. The $A_2A$ dissociation coordinate departs strongly from this limit: near the WT barrier, $C_{xx}(t)$ requires four relaxation components spanning 1.6 ps to 8.8 ns (**Fig. 4a** and **Supplementary Figs. 3 and 4**). With the normalized correlation $c(t) = C_{xx}(t)/C_{xx}(0)$, equation (11) gives $\gamma_{pos} = K\int c(t)dt$ and equation (7) gives $\gamma_{eff} = 2K\int c(t)^2 dt$ (Methods). The two expressions are identical for single-exponential relaxation but differ when relaxation is multiscale. All three friction profiles account for the non-Markovian memory effects on the friction profile, but only $\gamma_{eff}$ accounts for the effect of memory on the barrier-crossing time; $\gamma_{pos}$ (equation (11)) and $\gamma_{sol}$ (equation (12)) are time-integrated frictions, which neglect this effect (Methods). For WT, $\gamma_{eff}/\gamma_{pos}$ is 0.46 for the well-weighted means (equation (5)) and 0.19 for the barrier-weighted means (equation (4); **Fig. 4a**). At the WT barrier, $\gamma_{eff}$ (equation (7)) is $1.59 \times 10^5$ u ps$^{-1}$, compared with $5.80 \times 10^5$ u ps$^{-1}$ for $\gamma_{pos}$ (equation (11)) and $1.10 \times 10^6$ u ps$^{-1}$ for $\gamma_{sol}$ (equation (12)) (**Fig. 1** and **Supplementary Fig. 8**).

Because the three calculations share the same U(x) and $\tau_{res}$ is proportional to $\langle\gamma\rangle_{barr}$ (equation (4)) at fixed U(x) (Methods), the ratios of the predicted residence times equal the ratios of the barrier-weighted friction: for WT, $\tau_{res}$ is 274.7 s with $\gamma_{eff}$ (equation (7)), 1,476.8 s with $\gamma_{pos}$ (equation (11)) and 1,565.7 s with $\gamma_{sol}$ (equation (12)) (**Supplementary Fig. 9** and **Supplementary Table 2**). The experimental $\tau_{res}$ at 25 °C is 125 ± 8 s.[36] The non-Markovian prediction is therefore 2.2 times longer than the experimental value, compared with 11.8 and 12.5 times for $\gamma_{pos}$ and $\gamma_{sol}$.

### Predicted and experimental effects of receptor mutations

The experimental WT residence time and the mutation series test different aspects of the calculation. The 25 °C WT value tests the absolute residence-time scale, whereas the 4 °C series, in which WT and mutants were measured under identical conditions, tests whether the calculation reproduces the direction and magnitude of the mutation effects. T256A dissociates 16.8 times faster than WT and E169Q approximately 60 times faster.[37] The corresponding predicted accelerations are 19.5 and 99.8 times (**Fig. 4b** and **Supplementary Tables 2 and 3**). With $\gamma_{pos}$ (equation (11)) or $\gamma_{sol}$ (equation (12)) instead of $\gamma_{eff}$ (equation (7)), the predicted residence times of all four constructs are approximately five to seven times longer (**Supplementary Table 2**). E169A was not included in the experimental mutation series because negligible radioligand binding was reported for this substitution. We therefore retain E169A as a computational perturbation only. Its predicted $\tau_{res}$ is 0.34 s, about 800 times shorter than WT.

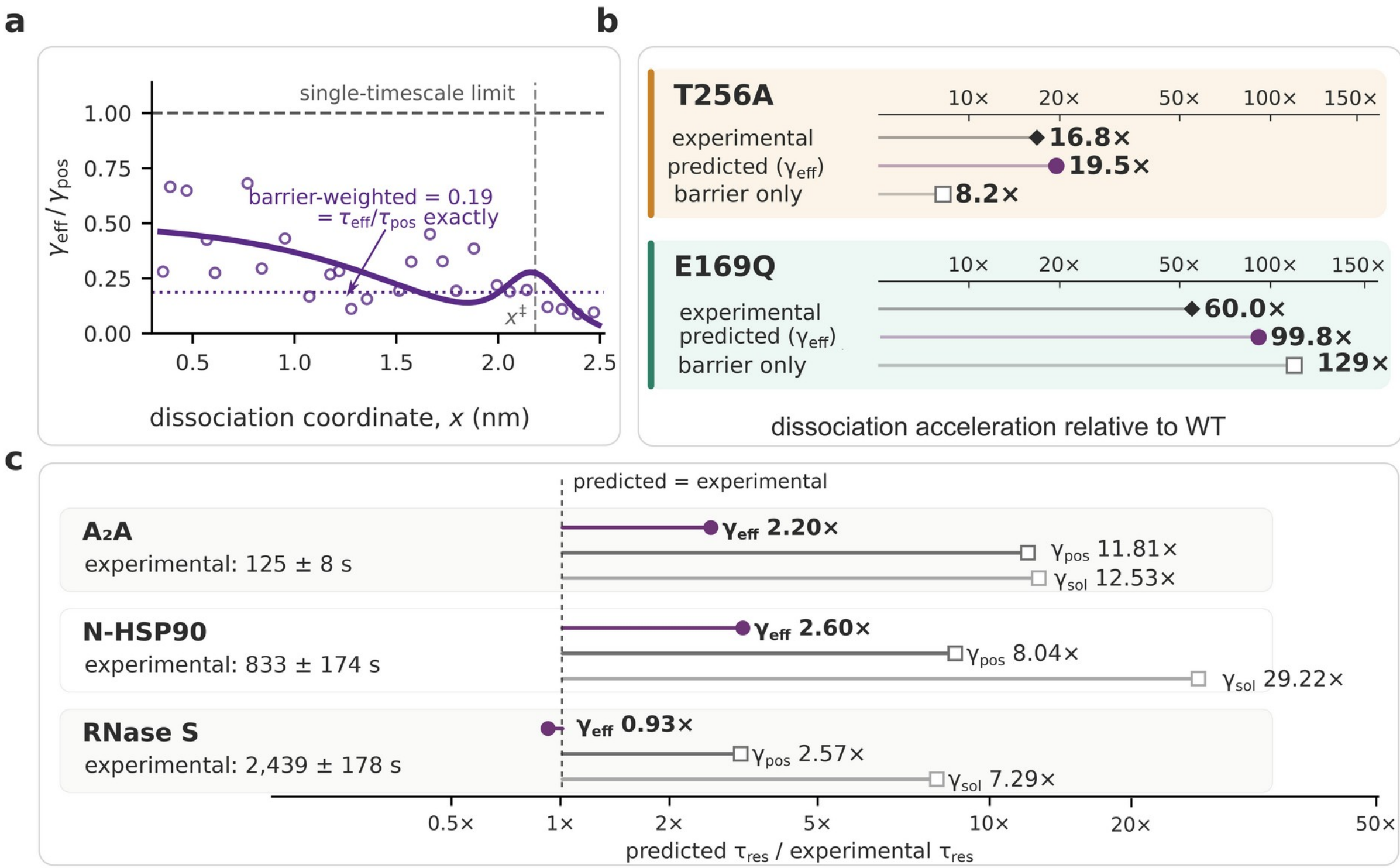

**Fig. 4 | Predicted residence times for the three friction profiles.** **a**, WT $\gamma_{eff}(x)/\gamma_{pos}(x)$ along the dissociation coordinate. Circles show per-window ratios and the solid line a guide to the eye; the dashed horizontal line marks the single-exponential limit and the dotted line the barrier-weighted ratio $\langle\gamma_{eff}\rangle_{barr}/\langle\gamma_{pos}\rangle_{barr} = 0.19$ (equation (4)). **b**, T256A and E169Q dissociation acceleration relative to WT: experimental (black diamonds), full non-Markovian prediction (purple circles) and barrier-only estimate from the free-energy barrier difference with the WT friction (open squares). **c**, Ratio of predicted to experimental $\tau_{res}$ for $A_2A$ WT, N-HSP90–compound 18 and RNase S–S15 with $\gamma_{eff}$, $\gamma_{pos}$ and $\gamma_{sol}$; the dashed line marks equality. Horizontal connectors are visual guides, not error bars.

To isolate the contribution of barrier lowering, we also calculated a barrier-only acceleration. This estimates how much faster dissociation would become if the mutation changed the free-energy barrier but left the friction unchanged from WT. The friction then cancels in the relative acceleration, leaving the Boltzmann factor $\exp[(\Delta U^{\ddagger}WT - \Delta U^{\ddagger}mut)/k_BT]$. This gives an acceleration of 8.2 times for T256A, compared with 19.5 times from the full calculation, and 129 times for E169Q, compared with 99.8 times (**Fig. 4b**). The predicted residence times were stable once the projected relaxation was represented by three or more exponential components (**Supplementary Figs. 10–12** and **Supplementary Table 5**).

## Residence-time predictions transfer to different systems

For N-HSP90–compound 18 (Protein Data Bank entry 5J86), the non-Markovian calculation with $\gamma_{eff}$ (equation (7)) predicts $\tau_{res}$ = 2,169 s (36.2 min), compared with 6,696 s with $\gamma_{pos}$ (equation (11)) and 24,337 s with $\gamma_{sol}$ (equation (12)). The experimental $\tau_{res}$ from 25 °C surface plasmon resonance is 833 ± 174 s.[38]

For the protein–peptide complex RNase S–S15 (Protein Data Bank entry 1RNU),[39] the non-Markovian calculation with $\gamma_{eff}$ (equation (7)) predicts $\tau_{res}$ = 2,259 s (37.6 min), compared with 6,278 s with $\gamma_{pos}$ (equation (11)) and 17,782 s with $\gamma_{sol}$ (equation (12)). The experimental $\tau_{res}$ from direct peptide exchange at 4 °C is 2,439 ± 178 s.[40]

Across the three systems, whose experimental residence times span 125 to 2,439 s, the non-Markovian prediction lies within a factor of 2.6 of the experimental value and is 7% below it for RNase S. The predictions with $\gamma_{pos}$ (equation (11)) and $\gamma_{sol}$ (equation (12)) are 2.6 to 29 times longer than the experimental values (**Fig. 4c** and **Supplementary Fig. 23**).

## Discussion

For $A_2A$ WT, the same free-energy profile gives $\tau_{res}$ = 275 s with $\gamma_{eff}$ (equation (7)), compared with 1,477 s with $\gamma_{pos}$ (equation (11)) and 1,566 s with $\gamma_{sol}$ (equation (12)) (**Fig. 4c** and **Supplementary Table 2**). All three friction profiles are extracted from the same trajectories and account for non-Markovian memory effects; the difference arises because only $\gamma_{eff}$ accounts for the effect of memory on the barrier-crossing time. Because the free-energy profile, boundaries and numerical treatment are unchanged, this comparison isolates the effect of the friction description. A systematic error in the free-energy profile could shift all three predicted times, but it cannot generate the difference produced by changing the friction profile alone.

The mutation series shows that energetic and frictional changes do not track one another. T256A lowers the free-energy barrier by 5.2 kJ mol$^{-1}$ and reduces the barrier-weighted friction $\langle\gamma_{eff}\rangle_{barr}$ (equation (4)) to 0.29 of WT, whereas E169Q lowers the barrier by 12.0 kJ mol$^{-1}$ while retaining 0.84 of WT $\langle\gamma_{eff}\rangle_{barr}$ (**Fig. 3b,d**). In equation (3), these different combinations reproduce the experimental direction and approximate magnitude of the T256A and E169Q accelerations (**Fig. 4b**). In the bound well, by contrast, all substitutions increase the friction, with $\langle\gamma_{eff}\rangle_{well}$ (equation (5)) 1.7 to 2.8 times that of WT (**Fig. 3d** and **Supplementary Table 11**). Whether the friction in the bound well also contributes to the crossing time depends on how the memory times compare with the time spent in the barrier region,[32] which we did not determine for each system; if it does, the higher well friction of the mutants would partly offset the effect of barrier lowering. Structural perturbations can therefore alter dissociation through different combinations of energetic destabilization and friction, a molecular realization of the distinction between an activation barrier and a friction-dependent prefactor. Protein, solvent and membrane degrees of freedom that are not resolved by the dissociation coordinate reappear as friction and memory after projection onto it.[21–23,30]

Across $A_2A$, N-HSP90 and RNase S, the same non-Markovian calculation remains within a factor of 2.6 of the experimental residence time, whereas the predictions with $\gamma_{pos}$ (equation (11)) and $\gamma_{sol}$ (equation (12)) are systematically longer (**Fig. 4c**). This supports transfer of the calculation across distinct molecular environments without system-specific kinetic calibration. It should not be read as a general error estimate. With only three experimental residence times, compensation among force-field error, reaction-coordinate incompleteness, free-energy error and the friction description cannot be excluded, particularly because the experimental assays differ across systems (**Supplementary Notes 4 and 5**).

One systematic contribution to the remaining deviation could arise from the umbrella restraint. Harmonic confinement has been demonstrated to increase the friction of molecular solutes in water by a factor that depends on the solute and on the confinement strength.[31,41] Based on these findings, we expect a maximal increase by a factor of about 2 for the restrained coordinates studied here.

Such an increase would lengthen the predicted residence times and could account for much of the deviation for $A_2A$ and HSP90 (factors of 2.2 and 2.6), but not for RNase S, where our prediction is 7% below the experimental value. Varying the umbrella force constant would quantify this effect; no correction is applied to the reported predictions. Because all $A_2A$ constructs were sampled with the same coordinate definition and force constant, a similar friction increase in each construct is expected to cancel in the relative mutation accelerations.

Each molecular system is represented by one umbrella array, so bootstrap and model-order analyses assess internal stability rather than independent reproducibility. The experimental $A_2A$ WT residence time and mutation series were measured at different temperatures, the HSP90 free-energy profile is less tightly constrained than the $A_2A$ and RNase S profiles, and the three experimental values were obtained with different assays. A temperature-matched kinetic series with independent simulation replicates would provide a substantially stronger test (**Supplementary Note 6** and **Supplementary Table 10**).

All residence-time calculations in this study use the same restrained trajectories that provide the PMF (**Fig. 1**). When those trajectories retain sufficient temporal resolution, the kinetic analysis can therefore be added to an existing umbrella sampling workflow. The added computation is primarily friction estimation and first-passage integration, and the main additional requirement is storage of high-frequency coordinate information. The residence time then becomes an additional output of restrained sampling rather than requiring a separate rare-event simulation. Dedicated kinetic methods remain preferable when transition pathways themselves or direct kinetic ensembles are the objective[5–18] (Supplementary Note 1).

A prospective congeneric series measured under a common protocol would provide the next test of whether the separation into energetic and frictional contributions resolves kinetic changes during chemical optimization. For computational drug discovery, the non-Markovian first-passage analysis turns an established thermodynamic simulation workflow into a route to residence times on the minutes timescale, which are ordinarily inaccessible to routine atomistic simulation, using temporal information already contained in the same restrained trajectories and without system-specific kinetic calibration.

## Methods

All systems were analysed with the same sequence: free-energy reconstruction with the weighted histogram analysis method (WHAM), friction estimation and calculation of the residence time as a mean first-passage time (MFPT), using fixed system-specific boundaries. Additional implementation and numerical details are provided in the Supplementary Methods.

### System preparation and molecular dynamics

The starting structure was the inactive human adenosine $A_2A$ receptor bound to ZM241385 (Protein Data Bank entry 3EML).[42] Missing side chains and short loops were rebuilt and the systems were prepared with CHARMM-GUI.[43] The receptor was embedded in a palmitoyl-oleoyl-phosphatidylcholine bilayer, solvated with TIP3P water and neutralized with 0.15 M NaCl. WT, E169Q, T256A and E169A were prepared independently. The CHARMM36m protein force field, CHARMM36 lipid parameters and CHARMM General Force Field ligand parameters were used.[44–46]

Simulations were performed with GROMACS 2026.1.[47] Bonds involving hydrogen were constrained with the linear constraint solver (LINCS) algorithm, permitting a 2 fs time step.[48] Long-range electrostatics were treated with particle-mesh Ewald and short-range electrostatic and van der Waals cutoffs were 1.2 nm. Temperature was maintained at 298 K with velocity rescaling and pressure at 1 bar with semi-isotropic Parrinello-Rahman coupling.[49,50] Each umbrella window was simulated for 800 ns after equilibration. The final arrays contained 26 windows for WT, 28 for T256A, 27 for E169Q and 27 for E169A, corresponding to 20.8–22.4 µs of production sampling per construct.

### Transfer systems and sampling

The soluble small-molecule transfer system was the human N-terminal HSP90 domain bound to compound 18 from Protein Data Bank entry 5J86.[38] The system was prepared with CHARMM-GUI using CHARMM36 protein parameters, CHARMM General Force Field (CGenFF)-compatible parameters for compound 18, TIP3P water and 0.15 M NaCl. We analysed 24 umbrella windows at 298 K. The analysed production segment spanned 300–1,100 ns in each window (800 ns per window), and the primary four-exponential position-autocorrelation fit used the first 400 ns of that segment. The umbrella force constant was 1,000 kJ $mol^{-1}$ $nm^{-2}$.

The protein–peptide transfer system was RNase S containing the S15 peptide (residues 1–15) from Protein Data Bank entry 1RNU, in which the resolved S-protein spans residues 24–124.[39] The system was prepared with CHARMM-GUI using CHARMM36 protein parameters for both S-protein and S15, TIP3P water and 0.10 M NaCl. An elongated solvent box was used along the dissociation direction to avoid interactions with periodic images near the absorbing boundary. Calculations were performed at 277.15 K ($k_BT$ = 2.304 kJ mol$^{-1}$) with 22 umbrella windows and a 1,000 kJ mol$^{-1}$ nm$^{-2}$ restraint. The friction analysis used approximately 250 ns per window, with the primary autocorrelation integral evaluated over the first approximately 125 ns. A cumulative WHAM convergence scan through 300 ns per window showed no material change in the barrier after 200 ns.

**Reaction coordinate, umbrella sampling and WHAM**

For $A_2A$, x was the separation along the membrane normal (z) between the centre of mass of all ZM241385 heavy atoms and the centre of mass of a binding-pocket reference group formed by the Cα atoms of residues 85, 177, 246, 249, 250, 253, 267, 270, 271 and 274. It was restrained with the GROMACS pull code (umbrella potential, distance geometry restricted to the z component) and a force constant of 1,000 kJ mol$^{-1}$ nm$^{-2}$. The same coordinate definition and force constant were used for WT and all mutants. For HSP90 and RNase S, x was defined in the same way, as the z component of the separation between the two centres of mass, with z along the long axis of the elongated simulation box. For HSP90, the ligand group comprised all heavy atoms of 6GW (compound 18) and the reference group the Cα atoms of residues 51, 54, 93, 96, 97, 102, 103, 162, 184 and 186. For RNase S, the peptide-ligand group comprised the Cα atoms of S15 residues 8–15 and the protein reference comprised Cα atoms of residues 25, 29, 33, 34, 35, 44–51, 54 and 117–120; peptide position restraints were likewise limited to Cα atoms to avoid constraining side-chain degrees of freedom. Umbrella centres were generated along the corresponding dissociation direction. WHAM[51] reconstructed the potential of mean force (PMF), U(x), and the same sampled coordinates used for the PMF were subsequently used for friction extraction and MFPT integration within each system.

**MFPT and spatial friction framework**

When the residence time exceeds the accessible trajectory length, dissociation is not observed in unbiased simulations. It can nevertheless be calculated from equilibrium properties along the dissociation coordinate, because the dissociation rate is determined by an activation free energy and a prefactor set by friction,

$$k_{\mathrm{off}} = A \exp\left(-\frac{\Delta G^{\ddagger}}{k_B T}\right) \tag{1}$$

where A is the prefactor and ΔG‡ the activation free energy. Restrained sampling supplies both quantities as functions of x: U(x) determines the energetic cost of crossing and γ(x) the local friction. In the reduced one-dimensional description used here, the probability density follows a Smoluchowski equation with position-dependent friction,[20,52]

$$\frac{\partial P(x,t)}{\partial t} = \frac{\partial}{\partial x}\left[\frac{k_B T}{\gamma(x)}\, e^{-U(x)/(k_B T)}\, \frac{\partial}{\partial x}\left(P(x,t)\, e^{U(x)/(k_B T)}\right)\right] \tag{2}$$

For a trajectory starting at the bound-state position $x_s$ and absorbed at $x_f$, the mean first-passage time is[20,52]

$$\tau_{\mathrm{MFPT}}(x_s, x_f) = \int_{x_s}^{x_f} dx\, \frac{\gamma(x)}{k_B T}\, e^{U(x)/(k_B T)} \int_{x_{\mathrm{left}}}^{x} dx'\, e^{-U(x')/(k_B T)} \tag{3}$$

with a reflecting boundary at $x_{left}$. This reflecting boundary was the lower edge of the sampled coordinate grid. We define the predicted residence time as $\tau_{res} = \tau_{MFPT}(x_s, x_f)$, and the experimental residence time as $\tau_{res} = 1/k_{off}$, where $k_{off}$ is the experimental first-order dissociation rate constant.

The bound-state minimum defined $x_s$. The absorbing boundary $x_f$ was placed at the outer edge of the final principal barrier, where the cumulative MFPT approached a plateau. For the $A_2A$ profiles, which rise into a terminal plateau, the reported principal-barrier position x‡ coincides with $x_f$. Numerical integration used the common free-energy and friction grid with trapezoidal quadrature. $k_BT$ was 2.478 kJ mol$^{-1}$ for the 298 K $A_2A$ and HSP90 calculations and 2.304 kJ mol$^{-1}$ for the 277.15 K RNase S calculation.

**Barrier-weighted and well-weighted mean friction**

To compare the friction of different ligand–receptor pairs, such as WT and the receptor mutants, we characterize each friction profile by a single number, using two different weighted averages over the friction profile. The barrier-weighted mean uses the inverse Boltzmann weight,

$$\langle\gamma\rangle_{\text{barr}} = \frac{\int_{x_{\text{left}}}^{x_f} dx\, \gamma(x)\, e^{U(x)/(k_B T)}}{\int_{x_{\text{left}}}^{x_f} dx\, e^{U(x)/(k_B T)}} \qquad (4)$$

and puts maximal weight on the barrier region. The well-weighted mean uses the Boltzmann weight,

$$\langle\gamma\rangle_{\text{well}} = \frac{\int_{x_{\text{left}}}^{x_f} dx\, \gamma(x)\, e^{-U(x)/(k_B T)}}{\int_{x_{\text{left}}}^{x_f} dx\, e^{-U(x)/(k_B T)}} \qquad (5)$$

and puts maximal weight on the well region. From the mathematical form of equation (3), where γ(x) enters multiplied by the first-passage weight $\exp[U(x)/k_B T]\int dx'\exp[-U(x')/k_B T]$, it transpires that the mean first-passage time is dominated by the friction at the barrier and thus should be proportional to $\langle\gamma\rangle_{\text{barr}}$. For non-Markovian dynamics with memory times longer than the time the system spends on the barrier, the memory in the well instead determines the MFPT,[32] which is also why Grote–Hynes theory, which considers only the barrier region,[28] does not correctly predict the MFPT in the long-memory limit.[32] Which limit applies depends on how the memory times compare with the time spent in the barrier region, which we did not determine for each system. We therefore report $\langle\gamma\rangle_{\text{well}}$ alongside $\langle\gamma\rangle_{\text{barr}}$ as a measure of how the value of the memory time can influence the impact of friction on the mean first-passage time (**Supplementary Table 11**).

**Non-Markovian effective friction**

For each umbrella window, the mean-subtracted position autocorrelation was

$$C_{xx}(t) = \langle [x(0) - \langle x\rangle][x(t) - \langle x\rangle]\rangle \qquad (6)$$

The effective friction developed for multi-timescale projected dynamics, which accounts for the effect of memory on barrier crossing, was calculated from the squared autocorrelation integral,[21,22,30]

$$\gamma_{\text{eff}}(\langle x\rangle) = \frac{2k_B T}{[C_{xx}(0)]^3}\int_0^\infty dt\, [C_{xx}(t)]^2 \qquad (7)$$

The autocorrelation was represented by a positive multi-exponential model,

$$C_{\text{fit}}(t) = \sum_i \frac{g_i}{\tau_i}\, e^{-t/\tau_i} \qquad (8)$$

which permits analytic evaluation of the squared integral,

$$\int_0^\infty dt\, [C_{\text{fit}}(t)]^2 = \sum_i \frac{g_i^2}{2\tau_i} + 2\sum_{i<j}\frac{g_i g_j}{\tau_i + \tau_j} \qquad (9)$$

The primary analysis used four components and 10,000 logarithmically spaced lag times over the first half of each trajectory. Models with two to six components were evaluated as a sensitivity analysis (**Supplementary Figs. 10–12** and **Supplementary Table 5**). Continuous friction profiles were obtained from

$$\log_{10}\gamma_{\text{fit}}(x) = \frac{\sigma}{1 + e^{(x - x_0)/w}} + c \qquad (10)$$

where σ is the amplitude, $x_0$ is the midpoint, w is the transition width and c is the outer coordinate baseline of $\log_{10}$ γ. We also report the pointwise ratio $\gamma_{\text{eff}}(x)/\gamma_{\text{pos}}(x)$. Writing $c(t) = C_{xx}(t)/C_{xx}(0)$ and using $C_{xx}(0) = k_B T/K$, equations (7) and (11) give $\gamma_{\text{eff}} = 2K\int c(t)^2 dt$ and $\gamma_{\text{pos}} = K\int c(t)dt$, so the ratio is unity for a single exponential and below unity when relaxation contains multiple timescales.

**Position-autocorrelation and solvent-force friction**

The position-autocorrelation friction used the restrained-coordinate construction,[19]

$$\gamma_{\rm pos}(x) = \frac{K^2}{k_BT}\int_0^\infty dt\, C_{xx}(t), \qquad K = \frac{k_BT}{\langle [x-\langle x\rangle]^2\rangle} \tag{11}$$

where K is the variance-derived local stiffness. The solvent-force friction used a Green–Kubo force-correlation construction,[31]

$$\gamma_{\rm sol}(x) = \frac{1}{k_BT}\int_0^\infty dt\, \langle \delta F(0)\, \delta F(t)\rangle \tag{12}$$

with the instantaneous reconstructed force and effective mass defined as

$$F(t) = m_{\rm eff}\, a(t) + K[x(t) - \langle x\rangle], \qquad m_{\rm eff} = \frac{k_BT}{\langle v^2\rangle} \tag{13}$$

Velocity and acceleration were calculated from the unstrided coordinate trajectory with Savitzky–Golay derivatives.[53] The velocity was mean-centred before evaluating the variance used for the effective-mass estimate in equation (13), and the reconstructed force was mean-centred before its autocorrelation was calculated. Solvent-force autocorrelations were represented by positive three-exponential fits. Per-window autocorrelation functions, running integrals, effective masses and continuous profiles are provided in **Supplementary Figs. 3–8, 14–16 and 19–21**, with model-selection and MFPT-sensitivity diagnostics in **Supplementary Figs. 10–12, 17 and 22**. Both $\gamma_{pos}$ and $\gamma_{sol}$ estimate the time-integrated friction $\int\Gamma(t)dt$, where $\Gamma(t)$ is the memory kernel of the projected dynamics. They therefore contain the full memory contribution to the friction, but in equation (3) they correspond to the Markovian limit of barrier crossing and neglect the effect of memory on the crossing time; $\gamma_{eff}$ (equation (7)) instead accounts for the effect of finite memory times on barrier crossing.

### Uncertainty, robustness and experimental reference data

Autocorrelation models containing two to six exponential components were fitted independently. Akaike and Bayesian information criteria, relative fit errors, running-integral behaviour and MFPT sensitivity were evaluated. Umbrella windows are coordinate-resolved sampling units rather than independent replicates. No inferential test was applied across windows, and no independent umbrella realization was generated for a construct. WHAM bootstraps therefore quantify resampling sensitivity within one trajectory array. Primary software and analysis settings for all three systems are summarized in **Supplementary Table 4**.

The 25 °C [$^3$H]ZM241385 dissociation rate constant of 0.48 ± 0.03 min$^{-1}$ ($\tau_{res}$ = 125 ± 8 s) was used as the experimental $A_2A$ WT residence time.[36] Relative mutation effects used the 4 °C WT, T256A and E169Q residence times of 84 ± 2, 5.0 ± 0.2 and 1.4 ± 0.1 min, measured under identical conditions.[37] The absolute 4 °C $A_2A$ residence times were not compared with the 298 K predictions; only the relative mutation accelerations were compared. E169A remained a computational perturbation. For HSP90 with compound 18, the experimental value was the 25 °C surface plasmon resonance dissociation rate constant $k_{off}$ = (1.2 ± 0.25) × 10$^{-3}$ s$^{-1}$ measured for WT N-HSP90 ($\tau_{res}$ = 833 ± 174 s).[38] For RNase S, the experimental value was the direct $^{13}$C nuclear magnetic resonance peptide-exchange rate constant $k_{off}$ = (4.1 ± 0.3) × 10$^{-4}$ s$^{-1}$ measured at 4 °C for the synthetic RNase(1–15) peptide ($\tau_{res}$ = 2,439 ± 178 s).[40] The three assays report first-order dissociation kinetics but are different experimental modalities; the cross-system comparisons test transferability rather than define a pooled statistic.

### Software implementation and availability

MemEngine-MD was implemented in Python to read WHAM PMFs and umbrella trajectories, calculate $\gamma_{eff}$, $\gamma_{pos}$ and $\gamma_{sol}$, fit continuous spatial profiles, integrate MFPTs, propagate PMF sensitivity, perform model-selection and convergence analyses, and generate standardized numerical and graphical outputs. Primary analysis settings are listed in **Supplementary Table 4**. MemEngine-MD v0.1.1 is publicly released with source code, an installable wheel, tests, example inputs, reference outputs and reproducibility documentation, and is permanently archived on Zenodo (DOI: 10.5281/zenodo.22015045).[54]

### Use of large language models

Claude (Anthropic) was used during manuscript revision for language editing and to assist in drafting analysis scripts. All AI-assisted outputs were checked by the authors, who take full responsibility for the manuscript and analyses.

**Data availability**

The PMF profiles, bootstrap uncertainty profiles, per-window friction estimates, continuous friction profiles, MFPT profiles and numerical summary tables underlying the $A_2A$, HSP90 and RNase S analyses are provided in the Supplementary Information and versioned MemEngine-MD release.[54] The public software archive contains compact reference outputs and the analysis settings needed to audit the reported calculations. Full-resolution production trajectory arrays are maintained separately because the solvent-force estimator requires the original high-frequency coordinate sampling; these files are available from the authors on request and will be deposited as a companion dataset upon journal publication. Starting structures are available from the Protein Data Bank under accessions 3EML ($A_2A$–ZM241385), 5J86 (N-HSP90–compound 18) and 1RNU (RNase S–S15).

**Code availability**

MemEngine-MD v0.1.1, used to calculate the three friction profiles, integrate mean first-passage times, propagate PMF sensitivity and generate the analysis outputs, is openly available from GitHub and permanently archived at Zenodo (https://doi.org/10.5281/zenodo.22015045).[54] The archived release contains source code, tests, example inputs, reference outputs and reproducibility documentation.

**Acknowledgements**

The authors gratefully acknowledge the scientific support and HPC resources provided by the Erlangen National High Performance Computing Center (NHR@FAU) of the Friedrich-Alexander-Universität Erlangen-Nürnberg (FAU) under the NHR project c108cb. NHR funding is provided by federal and Bavarian state authorities. NHR@FAU hardware is partially funded by the German Research Foundation (DFG), 440719683.

**Author contributions**

H.B. conceived the study, prepared and simulated the systems, performed the analyses, contributed to methodology and software, interpreted the results, prepared the figures and wrote the manuscript. L.T. contributed to the friction methodology and software and revised the manuscript. R.R.N. led the methodological development, supervised the study, interpreted the results and acquired funding. All authors approved the submitted version.

**Competing interests**

The authors declare no competing interests.

**Additional information**

Correspondence and requests for materials should be addressed to H.B. or R.R.N.

**Supplementary Information**

# Memory-resolved prediction of ligand residence times on the minutes timescale

Hossein Batebi[1], Lucas Tepper[1] and Roland R. Netz[1]

[1] Department of Physics, Freie Universität Berlin, Arnimallee 14, 14195 Berlin, Germany

Corresponding authors, hbat@zedat.fu-berlin.de; rnetz@physik.fu-berlin.de

## Contents

## Supplementary Methods

The Supplementary Information provides the numerical and methodological detail for the $A_2A$, N-HSP90–compound 18 and RNase S–S15 analyses. $A_2A$ provides the mutation test, whereas HSP90 and RNase S extend the calculation to different systems. The sections below report sampling, fitting, sensitivity and diagnostic outputs.

### Common umbrella and coordinate treatment

For all systems, x was the z component of the separation between the ligand and reference centres of mass (GROMACS pull code,[1] distance geometry with only the z dimension active). The analysed reaction coordinate was the sampled mean restrained separation in each umbrella window rather than the nominal umbrella centre. Continuous friction profiles were fitted against the sampled mean positions. The $A_2A$ receptor calculations used a nominal window spacing of 0.08 nm; the transfer systems used the corresponding spacing needed to cover the selected dissociation coordinate. Umbrella windows are coordinate-resolved sampling units and are not independent simulation replicates.

WHAM reconstructed the PMF U(x) using 1,000 bootstrap iterations.[2] The retained WHAM output contains pointwise PMF means and standard deviations. We propagated $U(x) - \sigma U(x)$ and $U(x) + \sigma U(x)$ through the MFPT expression with the same friction profile, reflecting and absorbing boundaries and integration grid. These limits are deterministic sensitivity envelopes because covariance among PMF bins and independent bootstrap profile realizations were not retained; they are not confidence intervals.

### Autocorrelation and friction processing

Position autocorrelations were calculated by a mean-subtracted fast Fourier transform. The primary model used four positive exponential components and 10,000 logarithmically spaced lag times over the first half of the analysed trajectory segment. The non-Markovian effective friction was evaluated from the squared position-autocorrelation integral,[3,4] the position friction from the linear autocorrelation integral[5] and the solvent-force friction from a Green–Kubo force-correlation integral.[6] Effective and position friction coincide in the single-exponential limit; the ratio $\gamma_{eff}/\gamma_{pos}$ therefore reports departure from single-timescale projected relaxation. A ratio below one does not by itself distinguish intrinsic bath memory from slow coordinates omitted by the one-dimensional coordinate.

The solvent-force calculation used the unstrided coordinate trajectory to reconstruct velocity and acceleration with Savitzky–Golay derivatives (window 11, polynomial order 3). Solvent-force autocorrelations were represented by positive three-exponential fits. Effective masses were estimated from equipartition. The corresponding per-window autocorrelation functions, running integrals, effective masses and continuous spatial profiles are shown in Supplementary Figs. 3-8 for $A_2A$, Supplementary Figs. 14-16 for HSP90, and Supplementary Figs. 19-21 for RNase S.

**Model-order sensitivity and exclusions**

Autocorrelation models containing two, three, four, five and six positive exponential components were fitted independently. Akaike and Bayesian information criteria, relative fit errors, running-integral behaviour and final MFPT sensitivity were evaluated. The primary manuscript uses the four-exponential model because it provides stable MFPTs while avoiding unnecessary model-order dependence. No $A_2A$ umbrella window was excluded from the final arrays, and the transfer-system archives retain all windows reported in Supplementary Tables 6-9.

**$A_2A$ receptor system construction and equilibration**

WT, E169Q, T256A and E169A were derived from the same $A_2A$ receptor-ZM241385 starting structure (PDB 3EML) and prepared independently in a POPC bilayer with CHARMM-GUI.[7] The same protonation scheme, force-field family, ionic strength and reaction-coordinate definition were retained across constructs to isolate receptor substitutions. The receptor used CHARMM36m, POPC used CHARMM36 lipid parameters, ZM241385 used CGenFF-compatible parameters, and the systems were solvated with TIP3P water and 0.15 M NaCl. Each included umbrella window contributed 800 ns of production sampling at 298 K.

**N-HSP90–compound 18 transfer system**

The soluble small-molecule transfer system used the human N-terminal HSP90 domain bound to compound 18 from PDB 5J86. The experimental study assigned compound 18 to the helical binding-site conformation and measured binding kinetics by SPR at 25 °C.[8] We analysed 24 umbrella windows at 298 K. The analysed production segment spanned 300–1,100 ns in each window (800 ns per window), and the primary four-exponential position-autocorrelation fit used the first 400 ns of that segment. The umbrella force constant was 1,000 kJ $mol^{-1}$ $nm^{-2}$.

The HSP90 system was prepared with CHARMM-GUI using CHARMM36 protein parameters, CGenFF-compatible parameters for compound 18, TIP3P water and 0.15 M NaCl. The ligand CV group comprised 6GW heavy atoms, and the protein reference was the Cα group of residues 51, 54, 93, 96, 97, 102, 103, 162, 184 and 186. Exact coordinates, topology files and run inputs are preserved in the archived release.

**RNase S–S15 transfer system**

The protein–peptide transfer system used RNase S from PDB 1RNU. The structure contains the truncated S15 peptide (residues 1–15) and resolved S-protein residues 24–124 at 1.60 Å resolution.[9] The experimental reference used the same synthetic RNase(1-15) peptide and was measured directly by 13C-NMR peptide exchange at 4 °C.[10] We analysed 22 umbrella windows at 277.15 K ($k_BT$ = 2.304 kJ $mol^{-1}$) with a 1,000 kJ $mol^{-1}$ $nm^{-2}$ restraint. The friction analysis used approximately 250 ns per window, with the primary autocorrelation integral evaluated over the first approximately 125 ns. A separate cumulative WHAM scan was extended through 300 ns per window and showed no material barrier change after 200 ns.

For the RNase S reaction coordinate, the peptide-ligand group comprised the Cα atoms of S15 residues 8-15. The protein reference comprised Cα atoms of residues 25, 29, 33, 34, 35, 44-51, 54 and 117-120. Peptide position restraints were limited to Cα atoms rather than all heavy atoms, allowing side-chain and backbone heteroatom relaxation. Index groups retained global PDB atom numbering, whereas position-restraint ITP files used molecule-local indices for the separate peptide and S-protein molecule types. An elongated solvent box was used along the dissociation direction to prevent contact with periodic images near the absorbing boundary.

**Experimental kinetic data**

The published 25 °C [3H]ZM241385 dissociation rate of 0.48 ± 0.03 $min^{-1}$ was used as the experimental $A_2A$ WT residence time.[11] Relative mutation effects used the matched 4 °C WT, T256A and E169Q residence times.[12] E169A is retained as a computational perturbation because the matched study reported negligible radioligand binding for this substitution. HSP90–compound 18 was measured by kinetic-titration SPR at 25 °C on immobilized recombinant N-HSP90. RNase S used direct 13C-NMR peptide exchange for the synthetic 1–15 peptide at 4 °C. The assays all report dissociation kinetics but use different experimental modalities, so the cross-system comparisons are not pooled into a single statistic.

## Supplementary Notes

The following notes delimit the scope of the kinetic interpretation, clarify the comparison with established methods and state the principal limitations of the current validation.

### Supplementary Note 1 | Relation to established kinetic approaches

First-passage analysis, kinetic reconstruction from biased simulations and position-dependent diffusion are established approaches.[13,14] The present work does not claim the first extraction of kinetics from biased or restrained simulations. Its specific use case is a restrained dissociation dataset that already exists for thermodynamic analysis: the non-Markovian friction is extracted from the same trajectories and compared with two friction profiles that represent the time-integrated friction under one common free-energy landscape and first-passage calculation. For pre-existing restrained datasets, kinetic analysis becomes an additional use of the sampling investment rather than a replacement for Markov state models, milestoning, weighted ensembles or other rare-event methods. The implementation used for all reported analyses is MemEngine-MD v0.1.1, permanently archived on Zenodo (DOI: 10.5281/zenodo.22015045).

### Supplementary Note 2 | Interpretation of projected memory

Finite projected memory changes the friction relevant to barrier crossing without changing the PMF. In the intermediate memory-acceleration regime, the crossing-relevant non-Markovian friction is lower than the zero-frequency or Markovian friction. Multi-exponential relaxation can also arise from hidden coordinates, so the present analysis uses the term projected memory and does not claim that every slow component is an intrinsic solvent or protein memory mode.

### Supplementary Note 3 | Scope of the transfer systems

HSP90 and RNase S were selected to test transfer to different systems rather than to maximize the number of targets. HSP90 adds an unrelated soluble drug target with a resolved inhibitor complex and direct temperature-matched SPR kinetics. RNase S adds a flexible peptide ligand and direct peptide-exchange kinetics. Together with the membrane-embedded $A_2A$ receptor and its mutation series, these systems span distinct environments without reframing the study as a systematic survey.

### Supplementary Note 4 | Interpretation of agreement with experimental residence times

The non-Markovian estimate is closest to the experimental residence time in all three systems, but the numerical agreement should not be interpreted as a universal error model. $A_2A$ uses radioligand dissociation in membranes, HSP90 uses SPR on immobilized recombinant protein and RNase S uses NMR peptide exchange in solution. Differences in assay environment, force-field accuracy, reaction-coordinate completeness and PMF uncertainty remain system specific.

### Supplementary Note 5 | Systematic free-energy error and limits of the cross-system comparison

Closer agreement with the experimental residence times obtained with effective friction could in principle reflect partial compensation between systematic errors. If the force-field and reaction-coordinate description overestimated dissociation barriers while non-Markovian friction systematically shortened the calculated times, the two effects could partially cancel. With three experimental residence times this possibility cannot be excluded, and the present work does not claim to exclude it. Two observations nevertheless constrain that interpretation. First, for each construct all three residence times are calculated with the same $U(x)$, reflecting and absorbing boundaries and integration grid; only $\gamma(x)$ differs. An additive constant in $U(x)$ cancels identically from the MFPT expression and cannot alter the predictions. A systematic PMF-shape or barrier error could shift all three calculated times together, but it cannot by itself generate their within-system separation. Second, the friction correction is not a construct-independent scalar. The barrier-weighted non-Markovian friction $\langle\gamma_{eff}\rangle_{barr}$ relative to WT is 0.29 for T256A, 0.84 for E169Q and 0.61 for E169A, whereas the well-weighted friction $\langle\gamma_{eff}\rangle_{well}$ is 2.80, 2.69 and 1.74 times that of WT (Supplementary Table 11). Here, the barrier-only Boltzmann acceleration is the mutant to WT acceleration predicted from the PMF barrier difference alone, with the friction taken to be unchanged from WT. The ratio of the full non-Markovian acceleration to this barrier-only acceleration is 2.37 for T256A, 0.77 for E169Q and 1.40 for E169A, spanning unity and differing in direction as well as magnitude across the series.

**Supplementary Note 6 | Diagnostic consistency across systems**

The transfer systems are reported with the same diagnostic depth as the $A_2A$ series so that transfer can be assessed from the underlying sampling and correlation-function fits as well as from the final residence times. For HSP90 and RNase S, the Supplementary Figures include umbrella distributions, free-energy profiles, position-autocorrelation fits, running integrals, solvent-force correlations, effective masses, friction profiles, cumulative residence-time profiles and model-order analyses. Supplementary Table 10 summarizes the corresponding diagnostic coverage and exclusions.

**Supplementary Tables**

**Supplementary Table 1 |** $A_2A$ system and PMF summary.

| **System** | **Windows** | **Bound state, $x_s$ (nm)** | **Absorbing boundary, $x_f$ (nm)** | Barrier (kJ $mol^{-1}$) | **Barrier ($k_BT$)** |
|---|---|---|---|---|---|
| WT | 26 | 0.343 | 2.182 | 61.35 | 24.76 |
| T256A | 28 | 0.354 | 2.182 | 56.13 | 22.65 |
| E169Q | 27 | 0.421 | 2.104 | 49.31 | 19.90 |
| E169A | 27 | 0.365 | 2.283 | 45.58 | 18.39 |

**Supplementary Table 2 |** $A_2A$ MFPTs and deterministic PMF sensitivity envelopes.

| **System** | **Friction** | **MFPT (s)** | **Lower (s)** | **Upper (s)** | **Ratio to effective** |
|---|---|---|---|---|---|
| WT | Effective | 274.7 | 143.3 | 537.5 | 1.00 |
| WT | Position | 1477 | 767.8 | 2896 | 5.38 |
| WT | Solvent force | 1566 | 812.4 | 3079 | 5.70 |
| T256A | Effective | 14.09 | 3.628 | 55.24 | 1.00 |
| T256A | Position | 87.03 | 22.19 | 344 | 6.18 |
| T256A | Solvent force | 100.7 | 25.63 | 398.5 | 7.15 |
| E169Q | Effective | 2.752 | 1.427 | 5.985 | 1.00 |
| E169Q | Position | 14.11 | 7.222 | 31.11 | 5.13 |
| E169Q | Solvent force | 14.87 | 7.656 | 32.55 | 5.40 |
| E169A | Effective | 0.3378 | 0.088 | 1.324 | 1.00 |
| E169A | Position | 1.639 | 0.4121 | 6.608 | 4.85 |
| E169A | Solvent force | 1.824 | 0.4684 | 7.229 | 5.40 |

**Supplementary Table 3 |** Experimental kinetic data used in the manuscript.

| **System** | **Temperature** | **$k_{off}$** | **Residence time** | **Assay / use** | **Reference** |
|---|---|---|---|---|---|
| $A_2A$ WT | 25 °C | 0.48 ± 0.03 $min^{-1}$ | 125 ± 8 s | Radioligand; absolute | Guo et al., 2012 |
| $A_2A$ WT | 4 °C | 0.0119 ± 0.0006 $min^{-1}$ | 84 ± 2 min | Matched mutation reference | Guo et al., 2016 |
| $A_2A$ T256A | 4 °C | 0.1994 ± 0.0120 $min^{-1}$ | 5.0 ± 0.2 min | Relative acceleration | Guo et al., 2016 |
| $A_2A$ E169Q | 4 °C | 0.7329 ± 0.0616 $min^{-1}$ | 1.4 ± 0.1 min | Relative acceleration | Guo et al., 2016 |
| N-HSP90 / 18 | 25 °C | (1.2 ± 0.25) x 10-3 s-1 | ~833 ± 174 s | SPR; absolute transfer | Amaral et al., 2017 |
| RNase S / S15 | 4 °C | (4.1 ± 0.3) x 10-4 s-1 | ~2439 ± 178 s | 13C NMR exchange; absolute transfer | Niu et al., 1980 |

**Supplementary Table 4 |** Software and primary analysis settings.

| **Item** | $A_2A$ | **HSP90 / 18** | **RNase S / S15** |
|---|---|---|---|
| MemEngine-MD | v0.1.1 | v0.1.1 | v0.1.1 |

| Item | $A_2A$ | **HSP90 / 18** | **RNase S / S15** |
|---|---|---|---|
| GROMACS | 2026.1 | 2026.1 | 2026.1 |
| Temperature | 298 K | 298 K | 277.15 K |
| Windows | 26-28 | 24 | 22 |
| Analysed production/window | 800 ns | 800 ns (300-1100 ns) | ~250 ns |
| Umbrella force constant | 1000 kJ $mol^{-1}$ $nm^{-2}$ | 1000 kJ $mol^{-1}$ $nm^{-2}$ | 1000 kJ $mol^{-1}$ $nm^{-2}$ |
| WHAM bootstrap | 1000 | 1000 | 1000 |
| Position ACF primary model | 4 exponentials | 4 exponentials | 4 exponentials |
| Position ACF fit interval | first 400 ns | first 400 ns | first ~125 ns |
| Solvent-force fit | 3 exponentials | 3 exponentials | 3 exponentials |
| Savitzky–Golay | window 11; order 3 | window 11; order 3 | window 11; order 3 |
| Preparation | CHARMM-GUI | CHARMM-GUI | CHARMM-GUI |
| Protein force field | CHARMM36m | CHARMM36 | CHARMM36 |
| Ligand / peptide parameters | CGenFF-compatible ZM241385 | CGenFF-compatible compound 18 | CHARMM36 protein parameters (S15) |
| Water model | TIP3P | TIP3P | TIP3P |
| NaCl | 0.15 M | 0.15 M | 0.10 M |

**Supplementary Table 5 |** $A_2A$ effective-friction MFPT sensitivity to exponential components.

| System | 2 exp (s) | 3 exp (s) | 4 exp (s) | 5 exp (s) | 6 exp (s) |
|---|---|---|---|---|---|
| WT | 317.2 | 283.9 | 275.8 | 274.9 | 274.8 |
| T256A | 15.78 | 14.2 | 14.06 | 14.09 | 14.09 |
| E169Q | 3.107 | 2.782 | 2.755 | 2.753 | 2.752 |
| E169A | 0.3825 | 0.3435 | 0.3375 | 0.3226 | 0.3375 |

**Supplementary Table 6 |** Transfer-system structural and sampling summary.

| Item | N-HSP90 / compound 18 | RNase S / S15 |
|---|---|---|
| Starting structure | PDB 5J86; ligand 6GW; 1.87 Å | PDB 1RNU; S15 residues 1–15; 1.60 Å |
| Temperature | 298 K | 277.15 K |
| Windows | 24 | 22 |
| Bound state $x_s$ | 0.3924 nm | 0.4042 nm |
| Absorbing boundary $x_f$ | 1.8950 nm | 2.0349 nm |
| Barrier | 66.98 kJ mol−1 (27.03 $k_BT$) | 64.02 kJ mol−1 (27.79 $k_BT$) |
| Max pointwise PMF s.d. | 5.66 kJ mol−1 | 2.34 kJ mol−1 |
| Ligand CV group | 6GW heavy atoms | S15 Cα, residues 8-15 |
| Protein reference | Cα 51,54,93,96,97,102,103,162,184,186 | Cα 25,29,33,34,35,44-51,54,117-120 |

**Supplementary Table 7 |** N-HSP90–compound 18 MFPTs and deterministic PMF sensitivity envelopes.

| Friction | MFPT (s) | Lower (s) | Upper (s) | Prediction / experiment |
|---|---|---|---|---|
| Effective | 2169.5 | 236.6 | 20013.8 | 2.60 |
| Position | 6696.4 | 736.4 | 61281.9 | 8.04 |
| Solvent force | 24336.8 | 2650.9 | 224762.4 | 29.20 |

**Supplementary Table 8 |** RNase S–S15 MFPTs and deterministic PMF sensitivity envelopes.

| Friction | MFPT (s) | Lower (s) | Upper (s) | Prediction / experiment |
|---|---|---|---|---|
| Effective | 2258.5 | 987.4 | 5212.4 | 0.93 |

| Friction | MFPT (s) | Lower (s) | Upper (s) | Prediction / experiment |
|---|---|---|---|---|
| Position | 6278.1 | 2758.8 | 14412.1 | 2.57 |
| Solvent force | 17782.4 | 7885.5 | 40456.0 | 7.29 |

**Supplementary Table 9 |** Transfer-system effective-friction MFPT sensitivity to exponential components.

| System | 2 exp (s) | 3 exp (s) | 4 exp (s) | 5 exp (s) | 6 exp (s) |
|---|---|---|---|---|---|
| N-HSP90 / 18 | 1983.3 | 2170.5 | 2168.2 | 2169.5 | 2167.5 |
| RNase S / S15 | 1619.1 | 2222.4 | 2248.8 | 2250.8 | 2400.6 |

**Supplementary Table 10 |** Diagnostic coverage and reproducibility summary.

| System | Windows retained | Primary ACF interval | Excluded windows | Diagnostics reported | MFPT stability summary |
|---|---|---|---|---|---|
| $A_2A$ WT | 26 | first 400 ns | 0 | Figs. 1-12 | 4-6 exp: 274.8-275.8 s |
| $A_2A$ T256A | 28 | first 400 ns | 0 | Figs. 1-12 | 4-6 exp: 14.06-14.09 s |
| $A_2A$ E169Q | 27 | first 400 ns | 0 | Figs. 1-12 | 4-6 exp: 2.752-2.755 s |
| $A_2A$ E169A | 27 | first 400 ns | 0 | Figs. 1-12 | 4-6 exp: 0.3226-0.3375 s |
| N-HSP90 / 18 | 24 | first 400 ns | 0 | Figs. 13-17 | 3–6 exp: 2167.5-2170.5 s |
| RNase S / S15 | 22 | first ~125 ns | 0 | Figs. 18-22 | 3-5 exp: 2222.4-2250.8 s; 6 exp 2400.6 s |

**Supplementary Table 11 |** Barrier-weighted and well-weighted mean friction. $\langle\gamma\rangle_{barr}$ and $\langle\gamma\rangle_{well}$ were evaluated with equations (4) and (5) of the main text over $[x_{left}, x_f]$ from the fitted continuous friction profiles and the WHAM free-energy profile of each system.

| System | Mean | $\gamma_{eff}$ (u ps$^{-1}$) | $\gamma_{pos}$ (u ps$^{-1}$) | $\gamma_{sol}$ (u ps$^{-1}$) | $\gamma_{eff}$ relative to WT |
|---|---|---|---|---|---|
| WT | $\langle\gamma\rangle_{barr}$ | $5.01 \times 10^5$ | $2.69 \times 10^6$ | $2.85 \times 10^6$ | 1.00 |
| WT | $\langle\gamma\rangle_{well}$ | $3.02 \times 10^6$ | $6.58 \times 10^6$ | $1.35 \times 10^7$ | 1.00 |
| T256A | $\langle\gamma\rangle_{barr}$ | $1.47 \times 10^5$ | $9.06 \times 10^5$ | $1.05 \times 10^6$ | 0.29 |
| T256A | $\langle\gamma\rangle_{well}$ | $8.45 \times 10^6$ | $1.69 \times 10^7$ | $2.94 \times 10^7$ | 2.80 |
| E169Q | $\langle\gamma\rangle_{barr}$ | $4.23 \times 10^5$ | $2.17 \times 10^6$ | $2.29 \times 10^6$ | 0.84 |
| E169Q | $\langle\gamma\rangle_{well}$ | $8.14 \times 10^6$ | $1.25 \times 10^7$ | $3.30 \times 10^7$ | 2.69 |
| E169A | $\langle\gamma\rangle_{barr}$ | $3.03 \times 10^5$ | $1.47 \times 10^6$ | $1.64 \times 10^6$ | 0.61 |
| E169A | $\langle\gamma\rangle_{well}$ | $5.26 \times 10^6$ | $9.88 \times 10^6$ | $2.10 \times 10^7$ | 1.74 |
| N-HSP90 / 18 | $\langle\gamma\rangle_{barr}$ | $6.21 \times 10^4$ | $1.92 \times 10^5$ | $6.97 \times 10^5$ | – |
| N-HSP90 / 18 | $\langle\gamma\rangle_{well}$ | $2.97 \times 10^6$ | $6.66 \times 10^6$ | $1.09 \times 10^7$ | – |
| RNase S / S15 | $\langle\gamma\rangle_{barr}$ | $3.88 \times 10^4$ | $1.08 \times 10^5$ | $3.05 \times 10^5$ | – |
| RNase S / S15 | $\langle\gamma\rangle_{well}$ | $8.03 \times 10^5$ | $1.63 \times 10^6$ | $6.41 \times 10^6$ | – |

**Supplementary Fig. 1 |** Umbrella-window position distributions and local harmonic reference distributions. WT, T256A, E169Q and E169A are arranged in separate groups. Numerical coordinate distributions and the corresponding harmonic reference distributions are plotted for each window.

**WT**

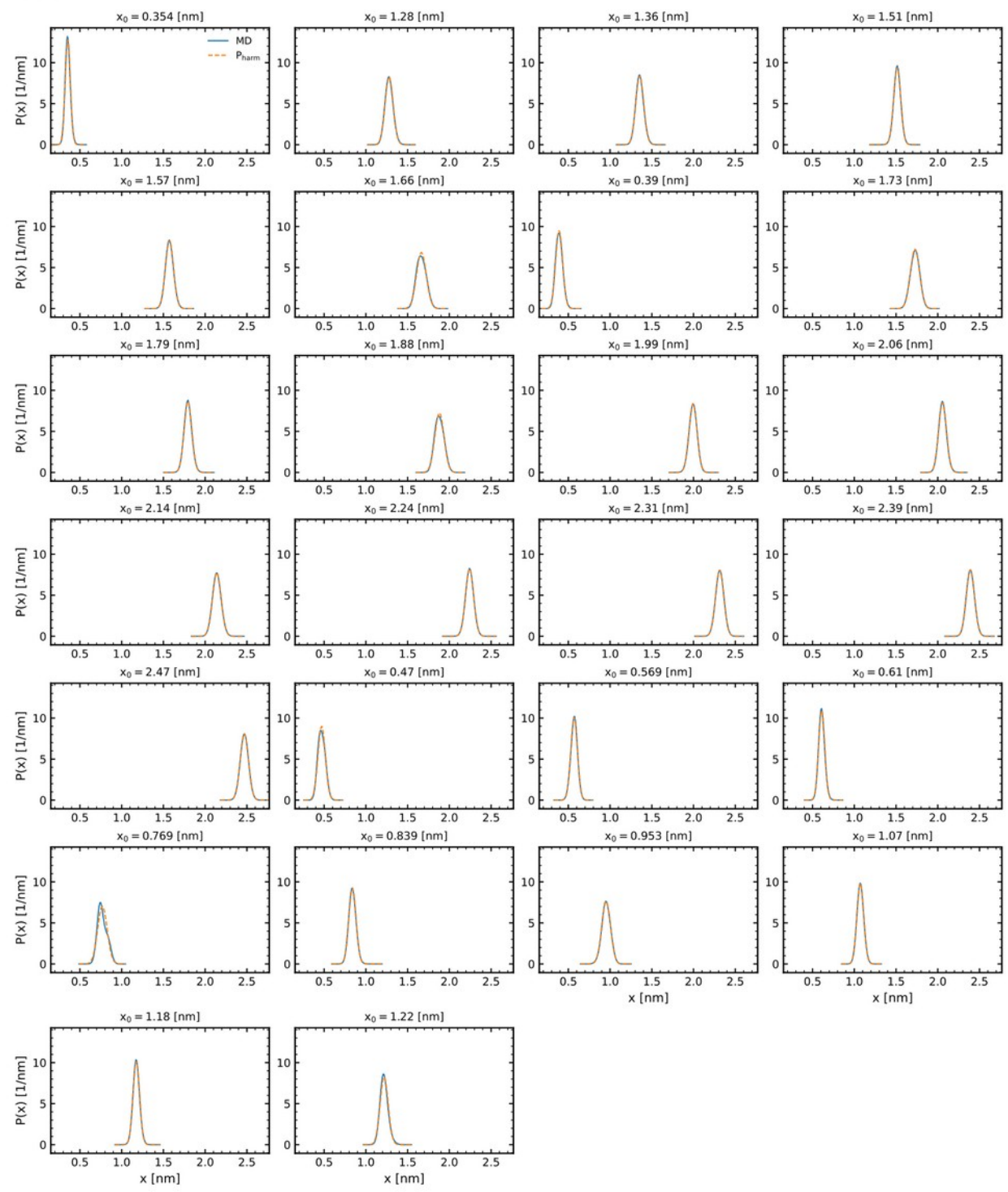

**T256A**

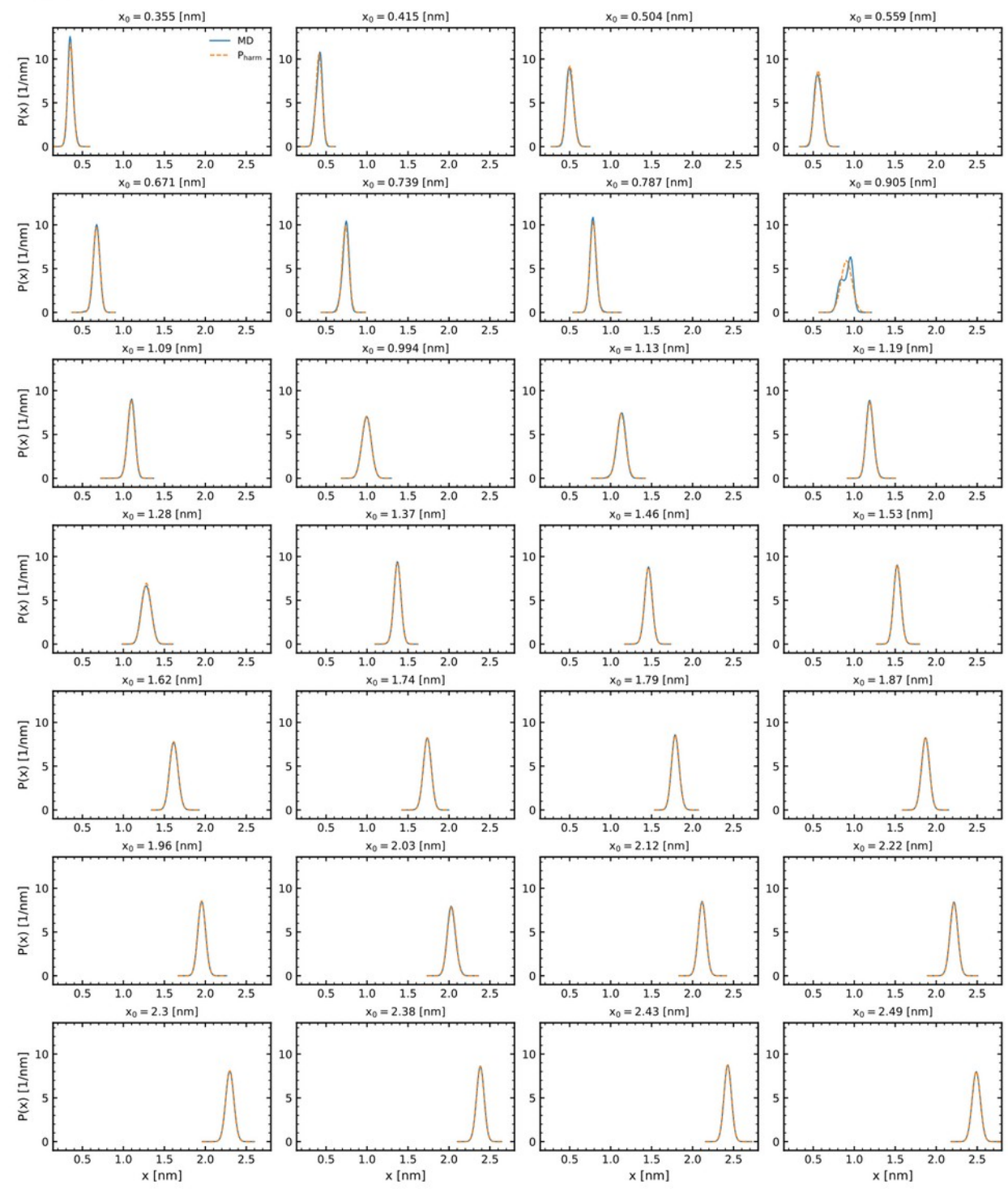

**E169Q**

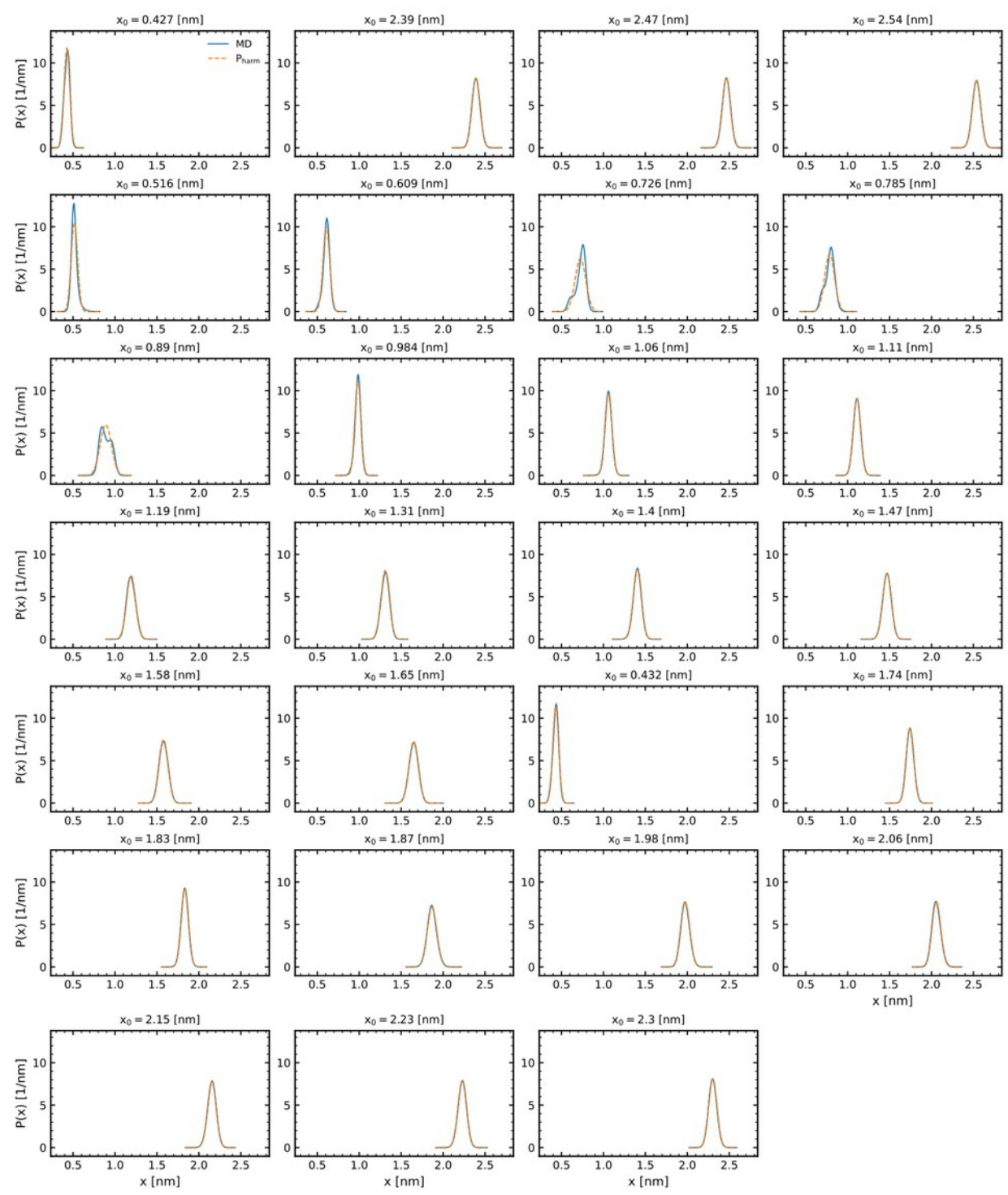

**E169A**

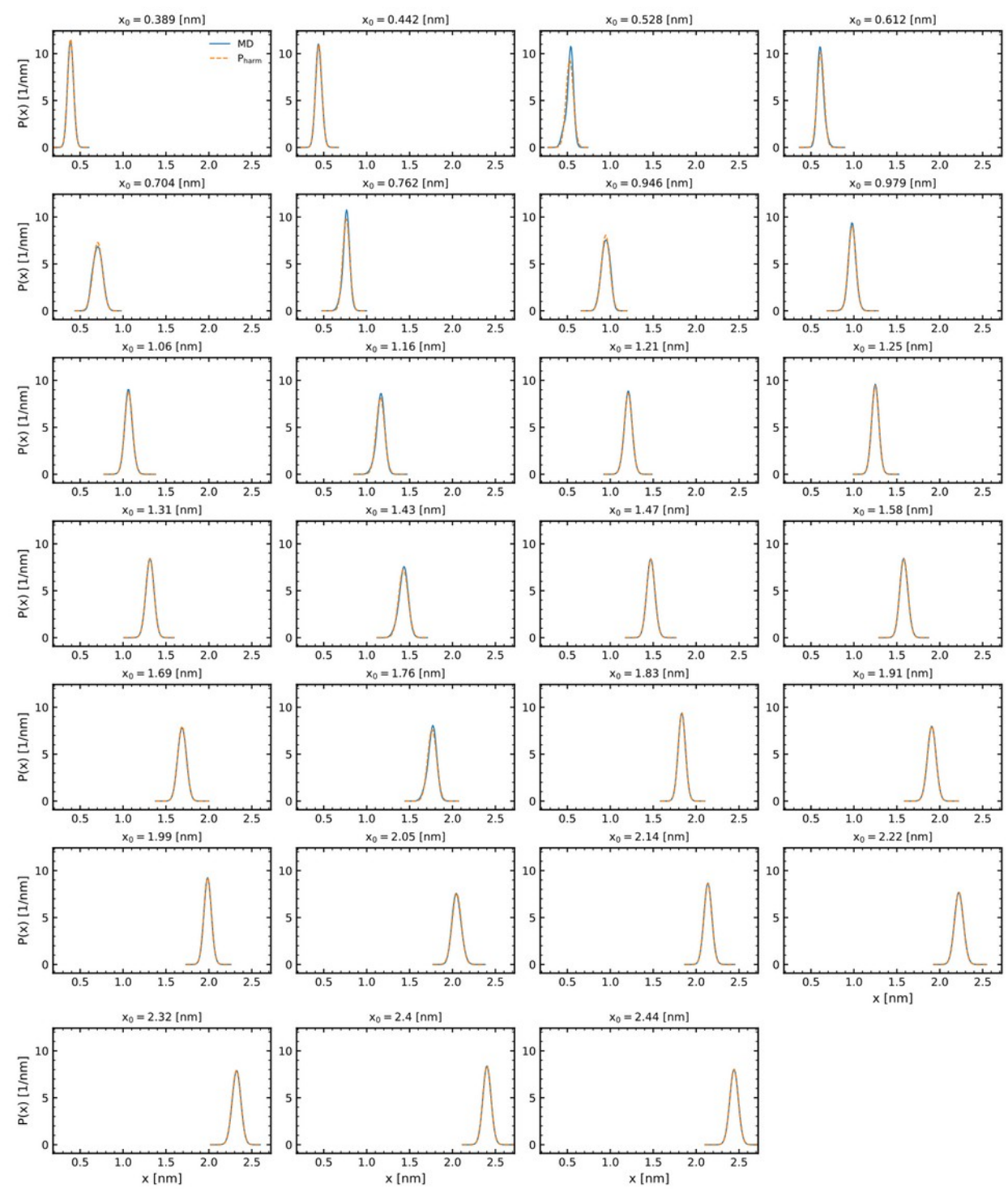

**Supplementary Fig. 2 |** WHAM PMFs and pointwise bootstrap uncertainty for WT, T256A, E169Q and E169A. Curves represent the pointwise PMF mean, and shaded regions represent mean ± one standard deviation from 1,000 WHAM bootstrap iterations.

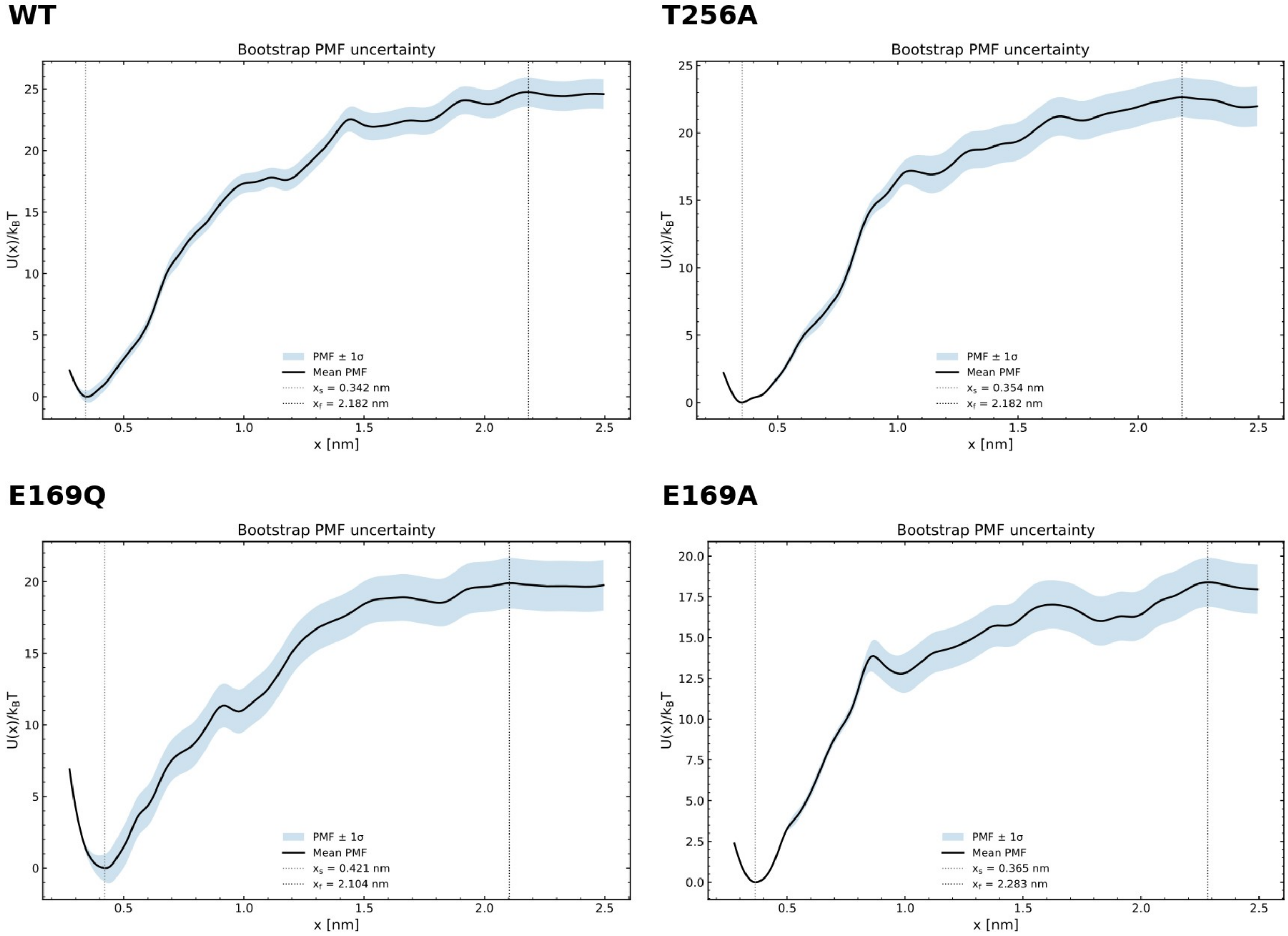

**Supplementary Fig. 3 |** Position-autocorrelation functions and multi-exponential fits. Each panel corresponds to one umbrella window. Curves contain the trajectory autocorrelation and the four-exponential fit used in the primary analysis over the first half of each 800 ns trajectory.

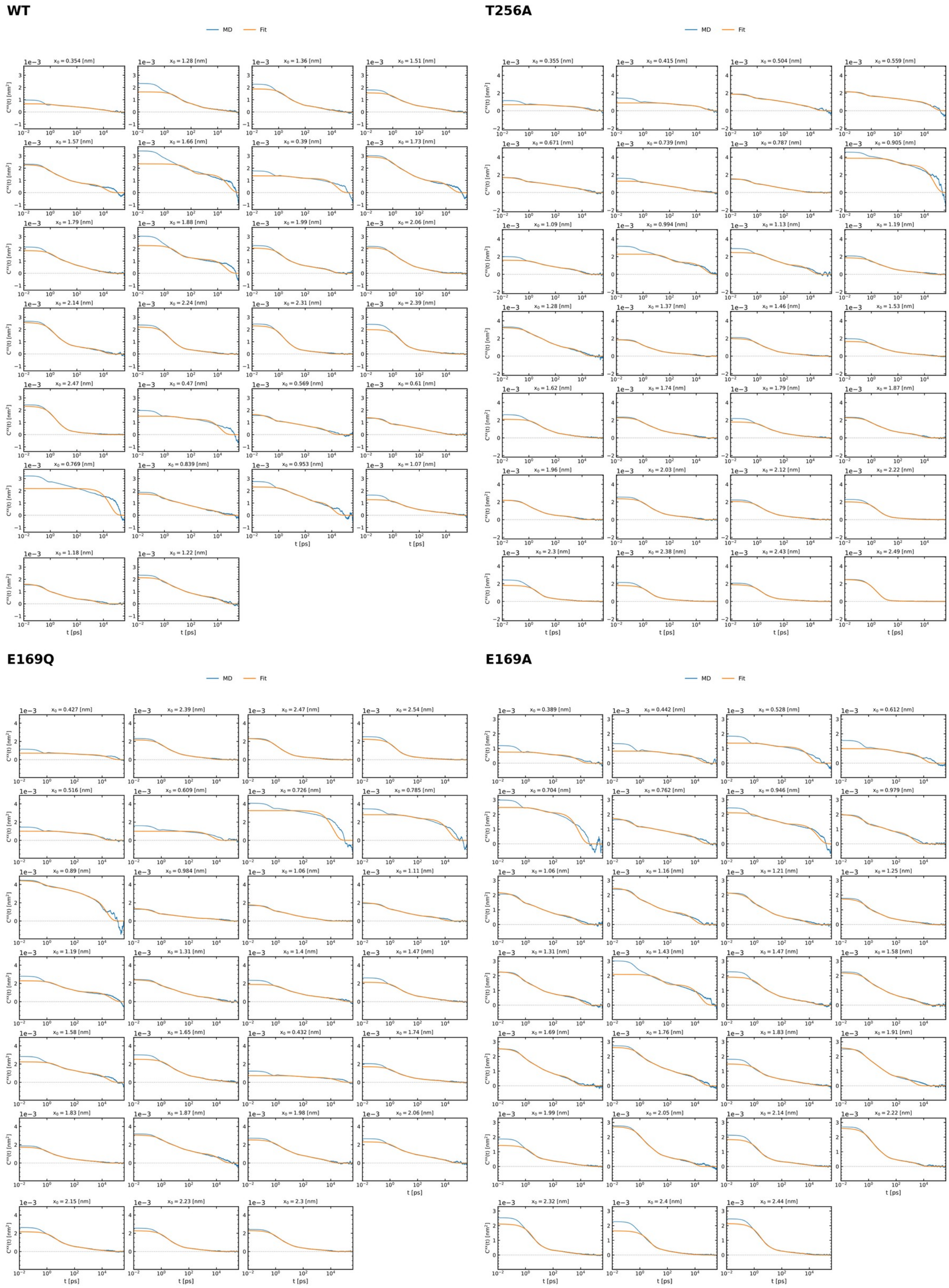

**Supplementary Fig. 4 |** Running integrals of the squared position autocorrelation for each umbrella window and receptor construct.

**WT**

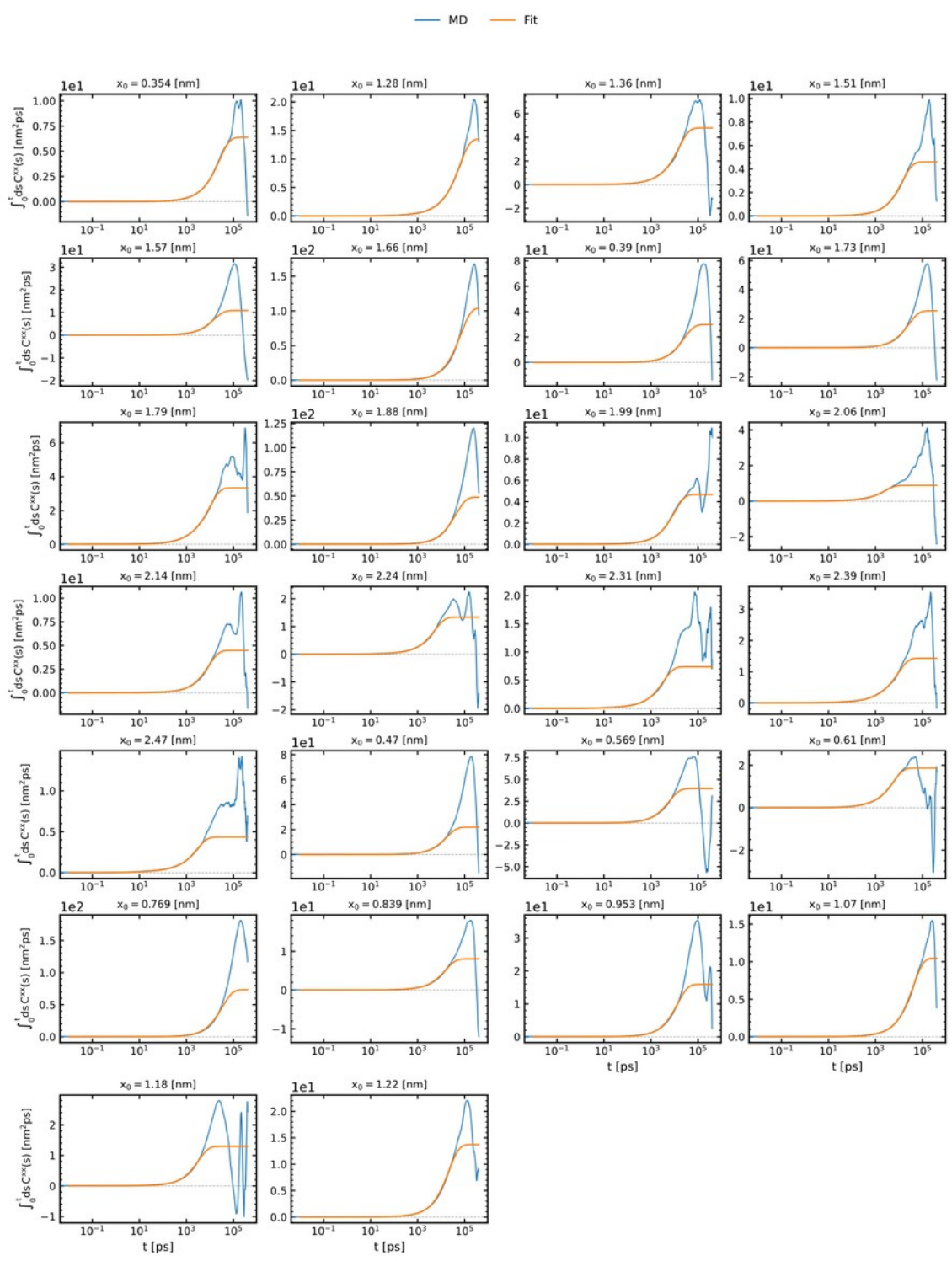

**T256A**

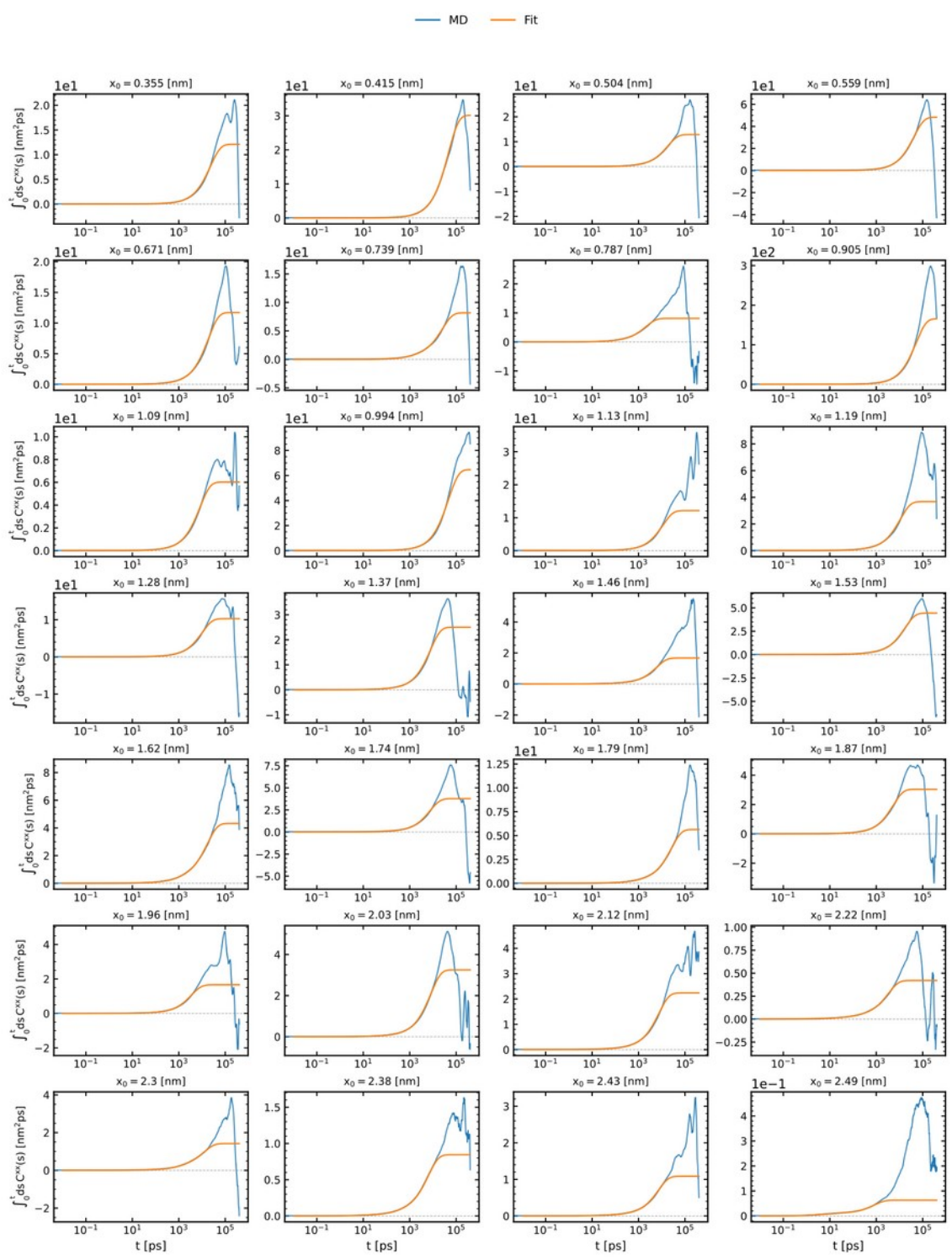

**E169Q**

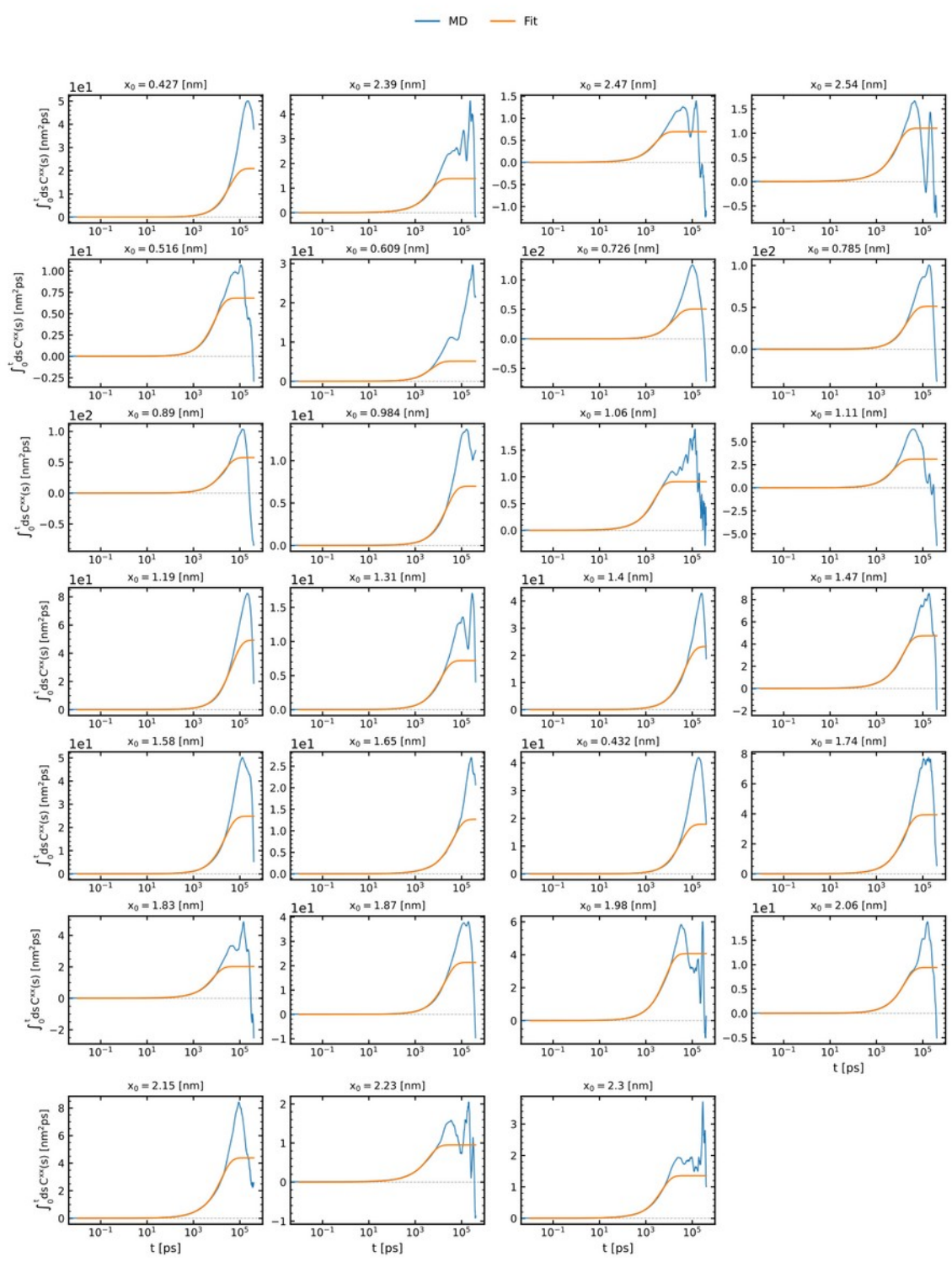

**E169A**

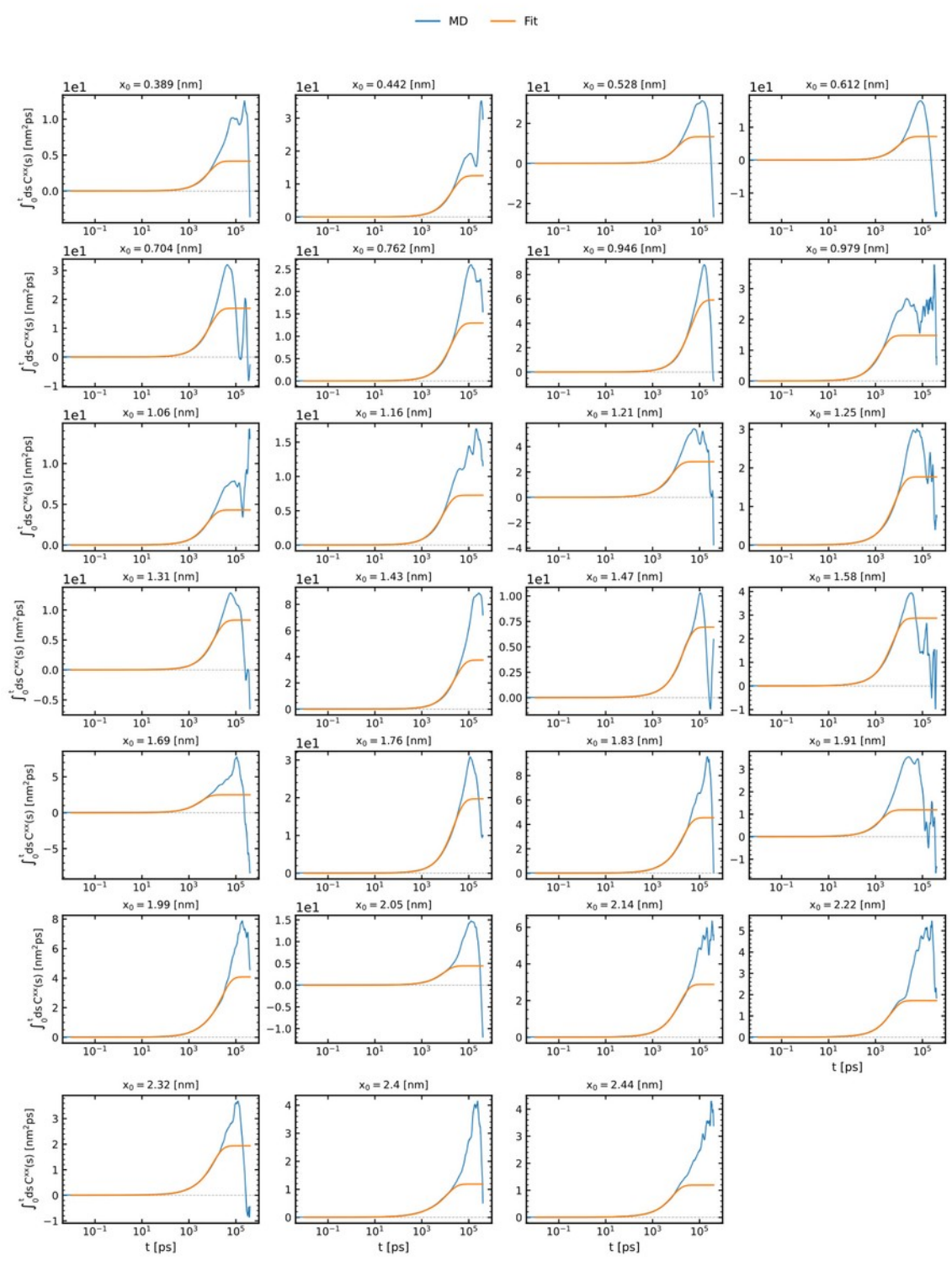

**Supplementary Fig. 5 |** Solvent-force autocorrelation functions and three-exponential fitted decay curves for each umbrella window. Force reconstruction used the full-resolution coordinate trajectory.

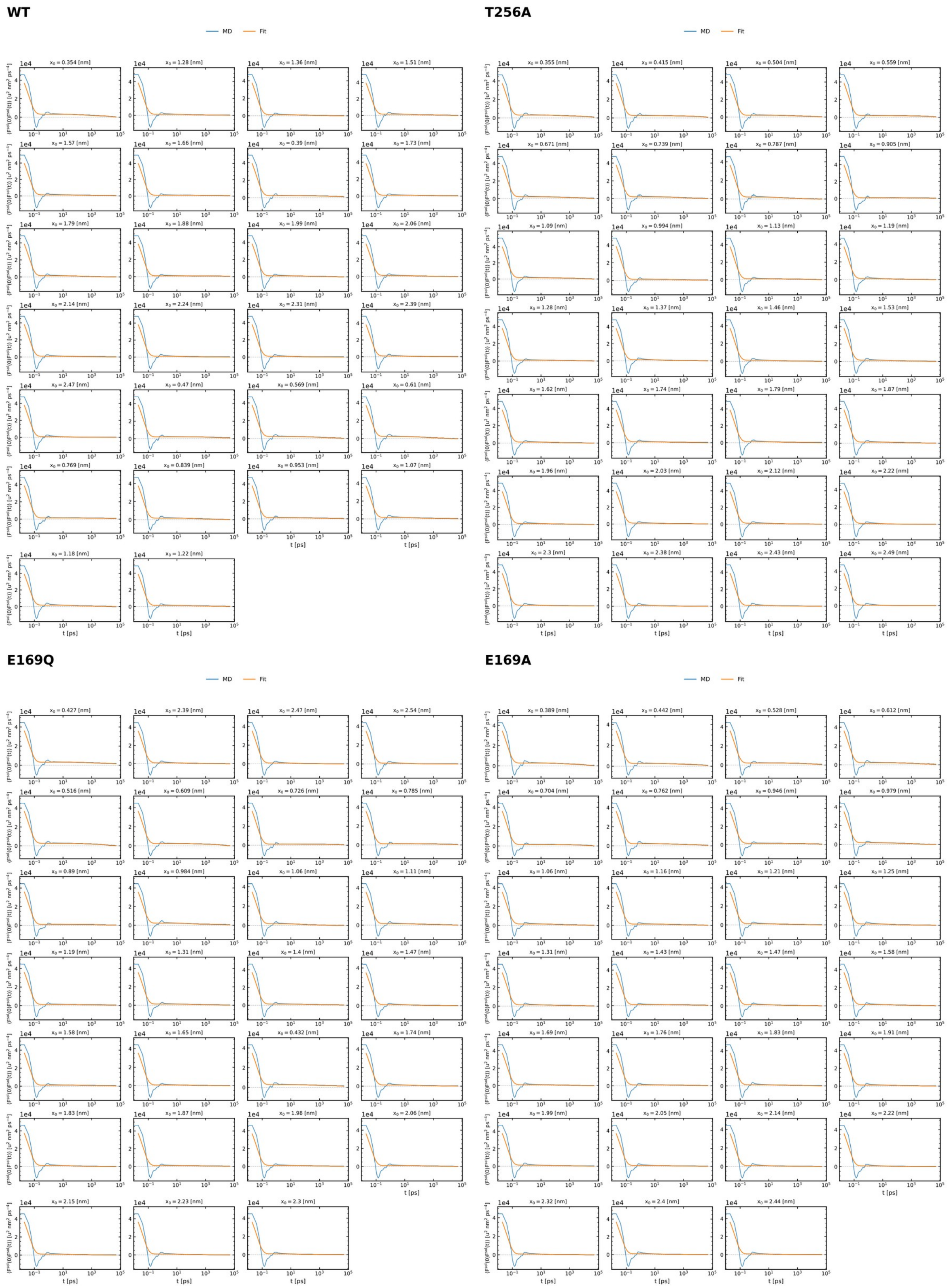

**Supplementary Fig. 6 | ** Running integrals of the solvent-force autocorrelation for each umbrella window. Reported solvent-force friction values were calculated from the analytic integrals of the fitted autocorrelations.

**WT**

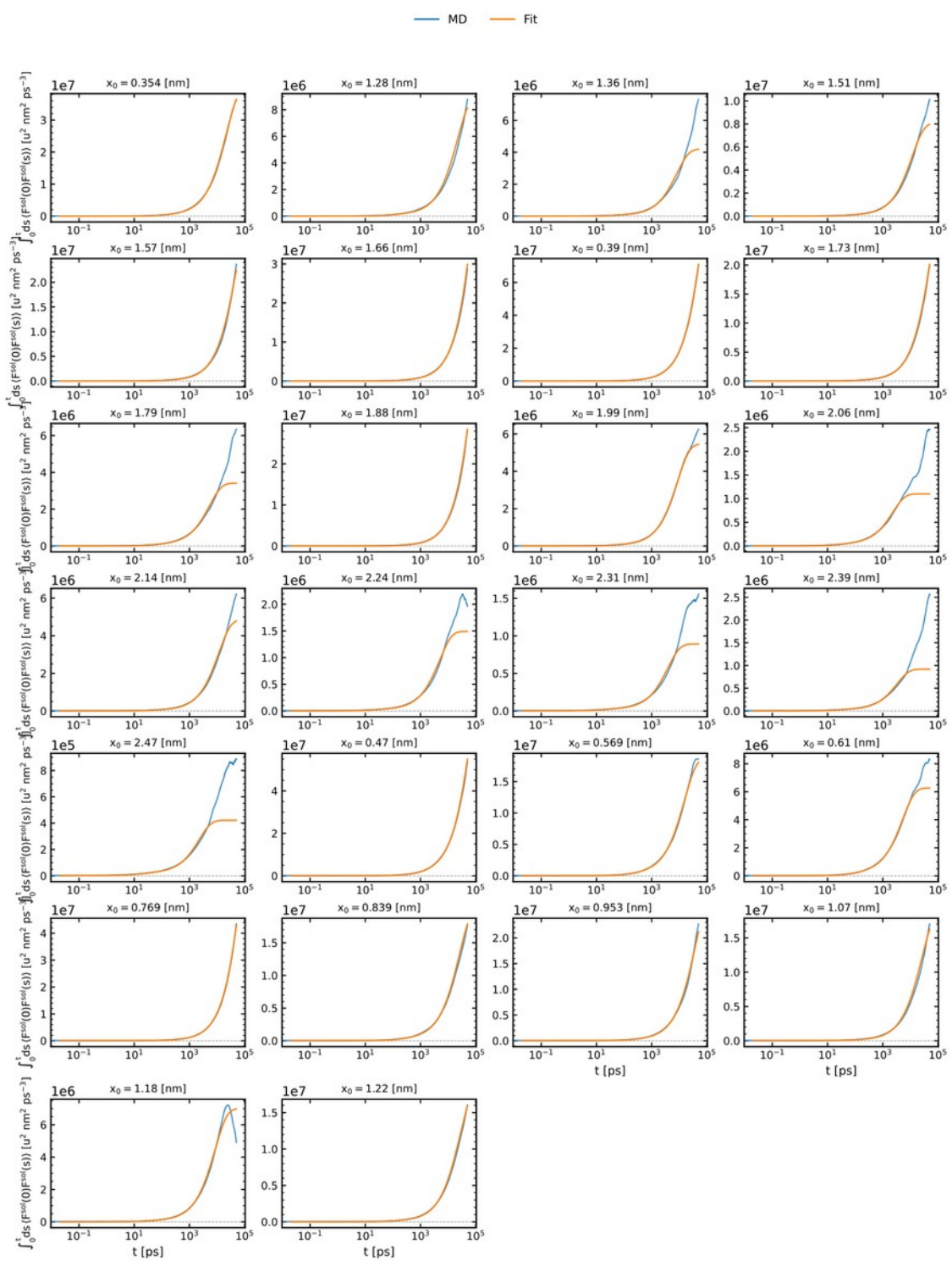

**T256A**

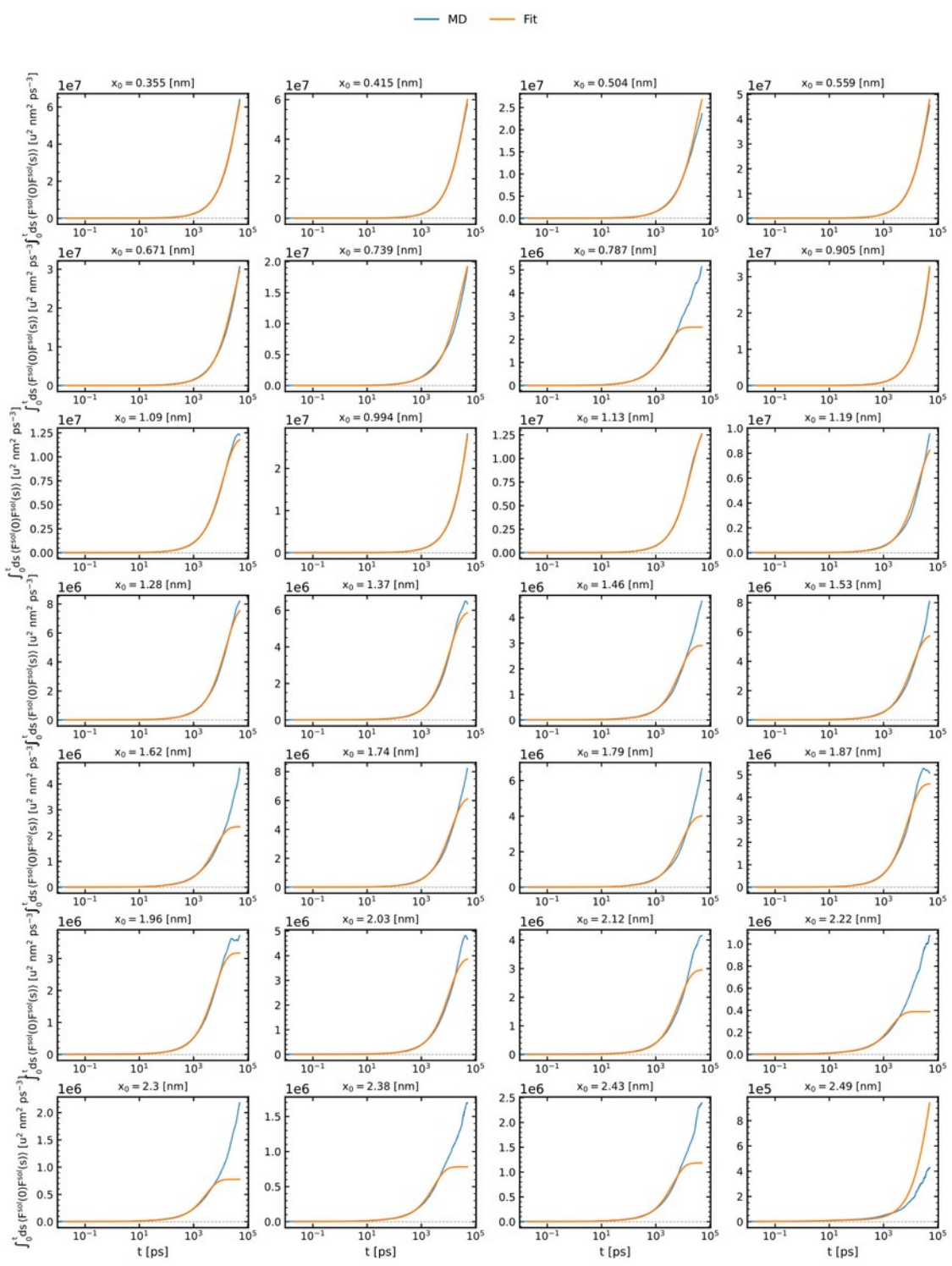

**E169Q**

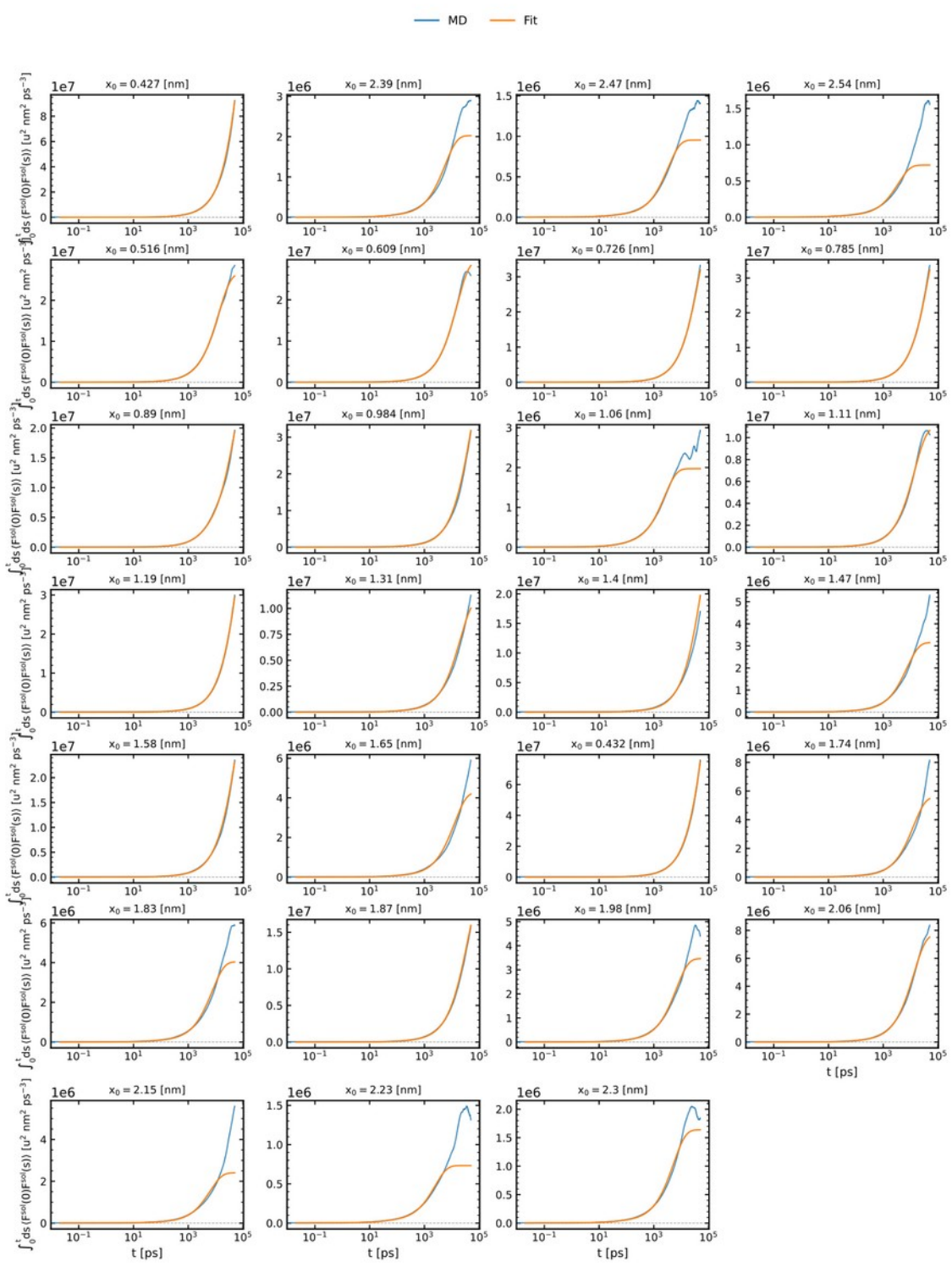

**E169A**

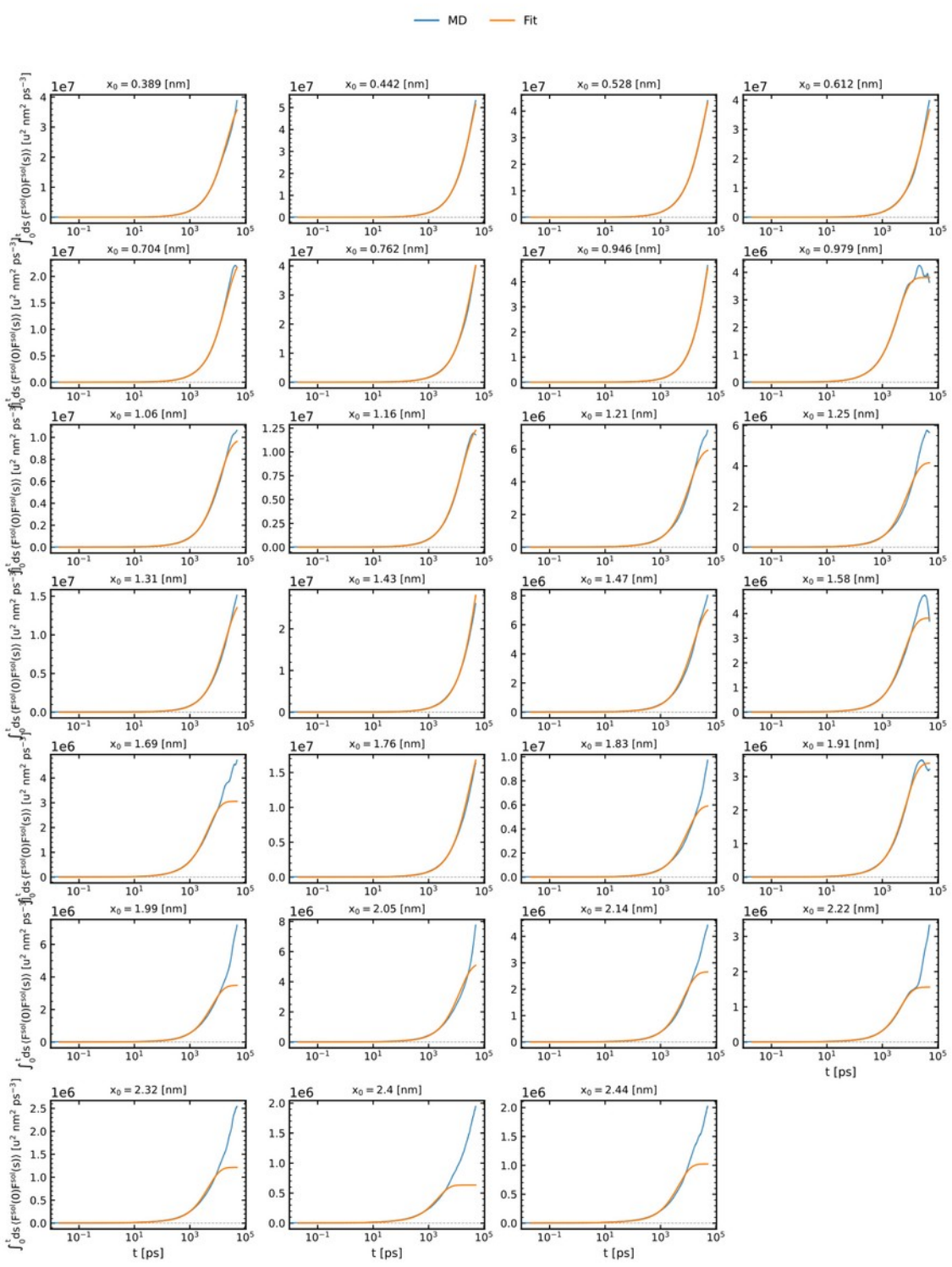

**Supplementary Fig. 7 |** Effective mass estimated by equipartition for each umbrella window and receptor construct.

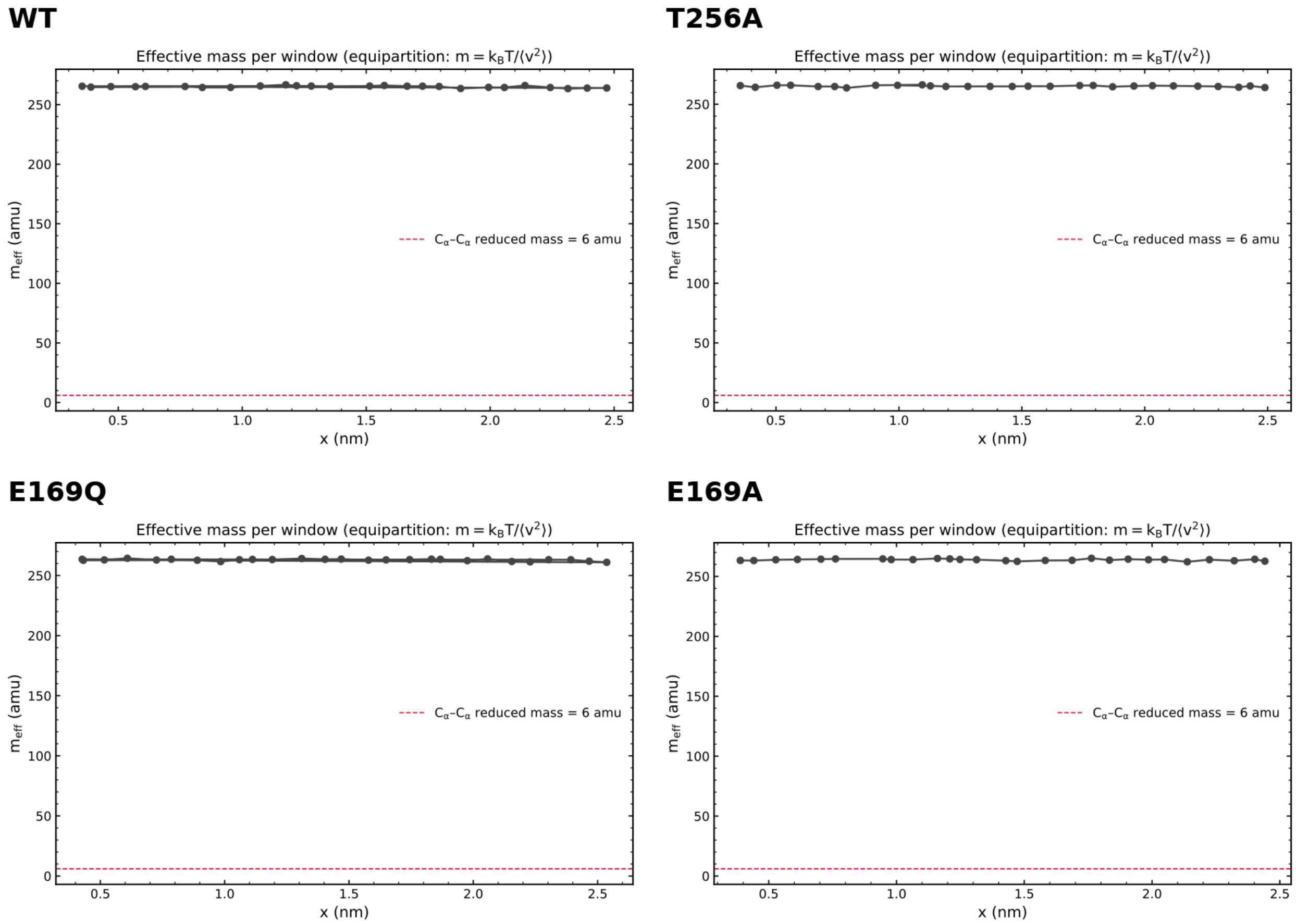

**Supplementary Fig. 8 |** Effective, position-autocorrelation and solvent-force friction profiles for WT, T256A, E169Q and E169A. Symbols represent per-window estimates and lines represent continuous fitted profiles.

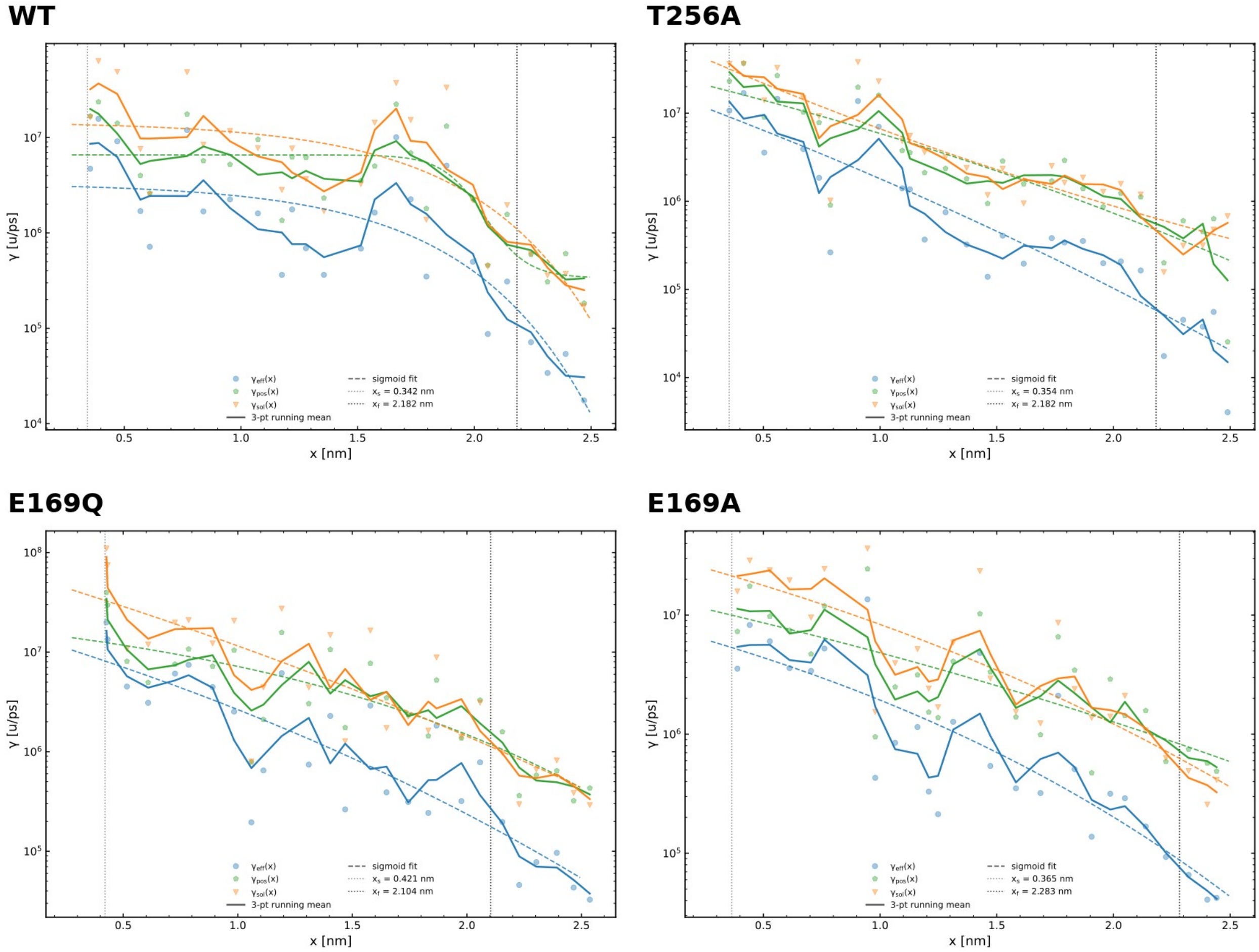

**Supplementary Fig. 9 |** Cumulative MFPT profiles for WT, T256A, E169Q and E169A calculated with the effective, position-autocorrelation and solvent-force friction profiles.

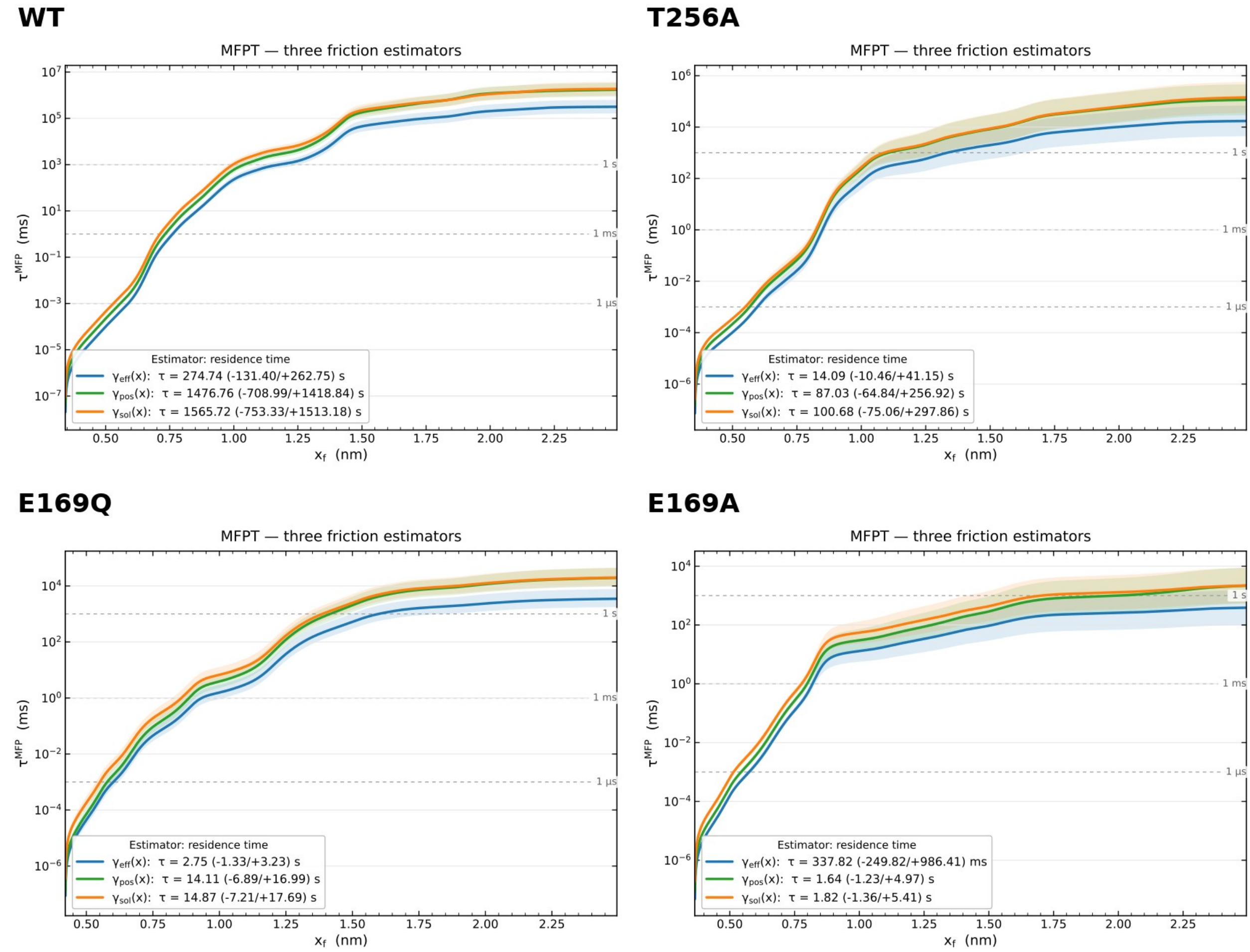

**Supplementary Fig. 10 |** Fit-quality and information-criterion outputs for position-autocorrelation models containing two to six exponential components.

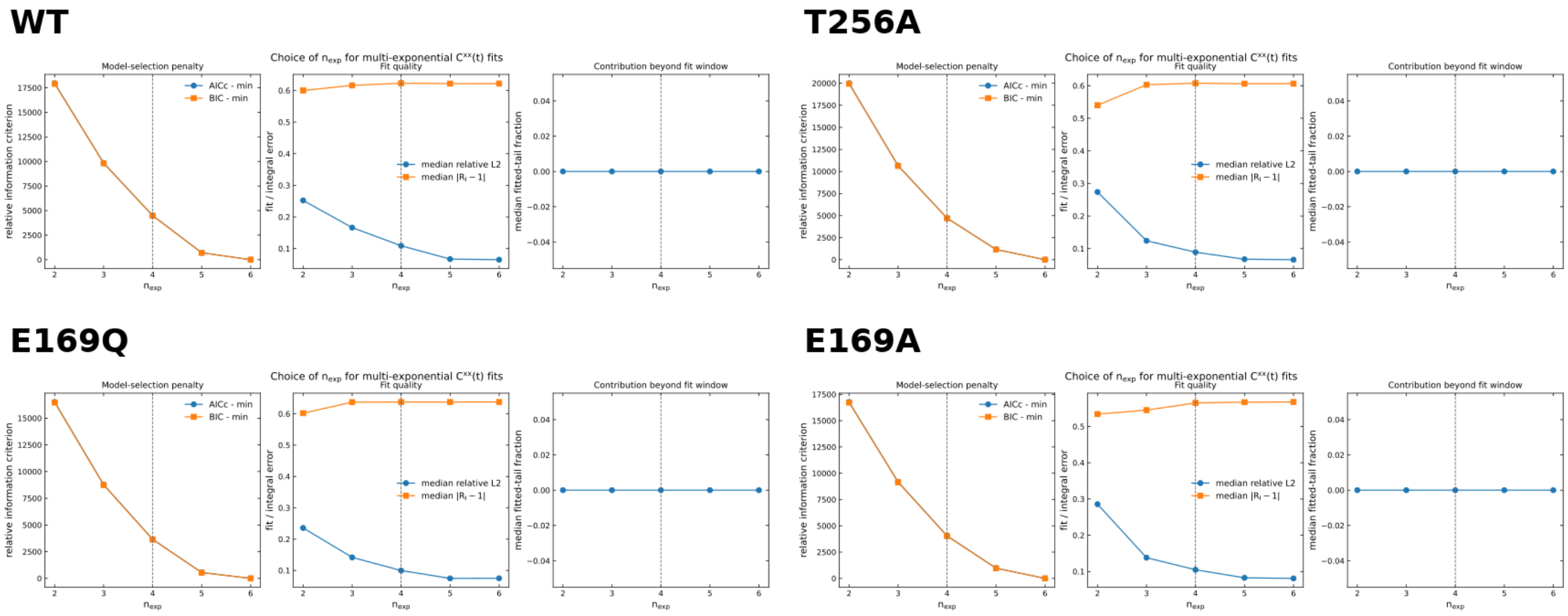

**Supplementary Fig. 11 | ** Effective-friction profiles and final MFPT values calculated with position-autocorrelation fits containing two to six exponential components.

**WT**

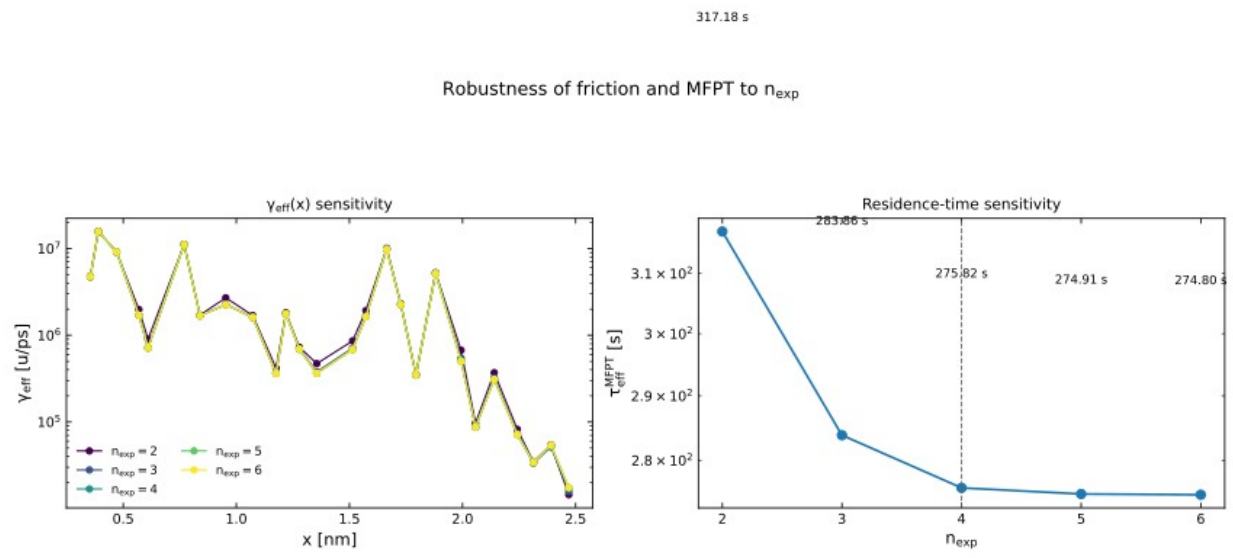

**T256A**

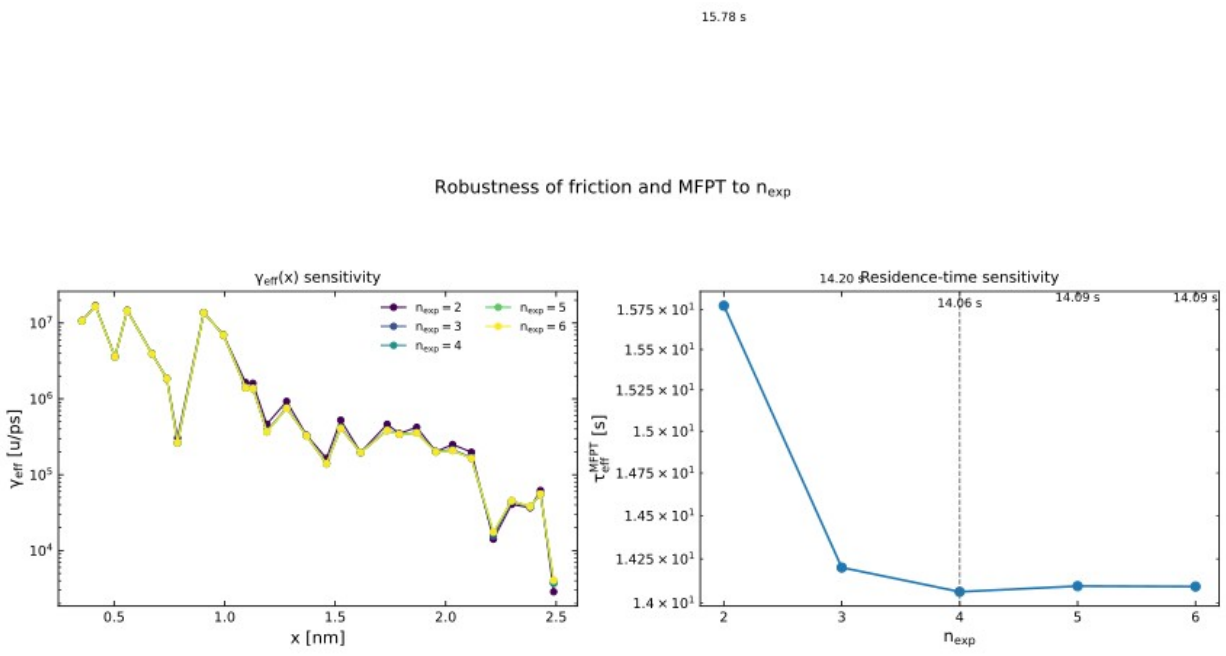

**E169Q**

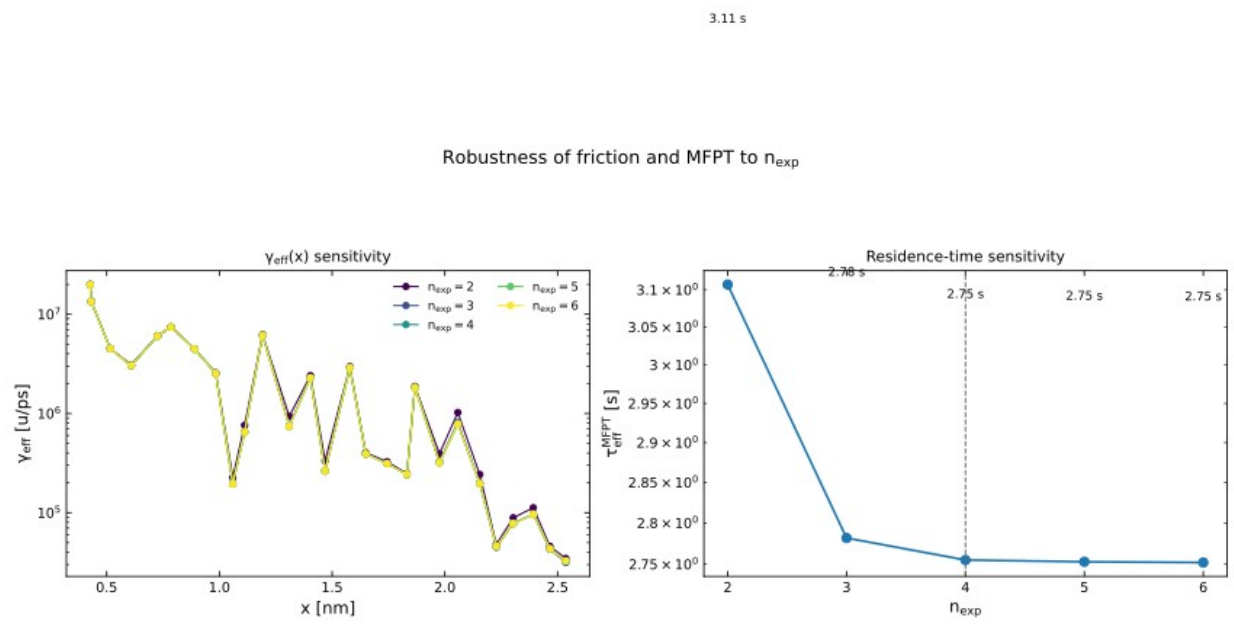

**E169A**

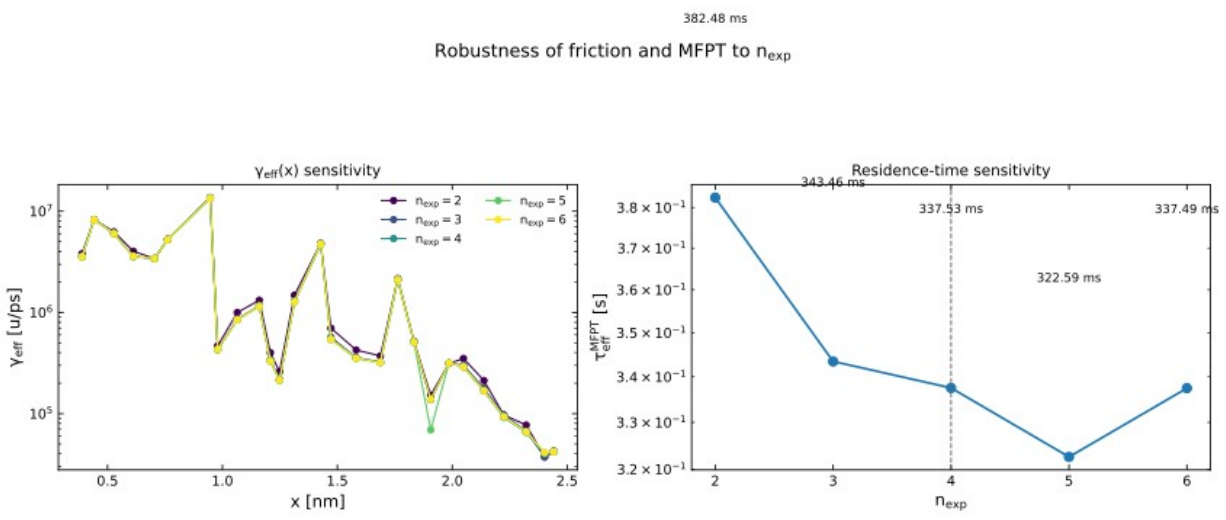

**Supplementary Fig. 12 |** Summary of friction-profile and MFPT sensitivity analyses for WT, T256A, E169Q and E169A.

## WT

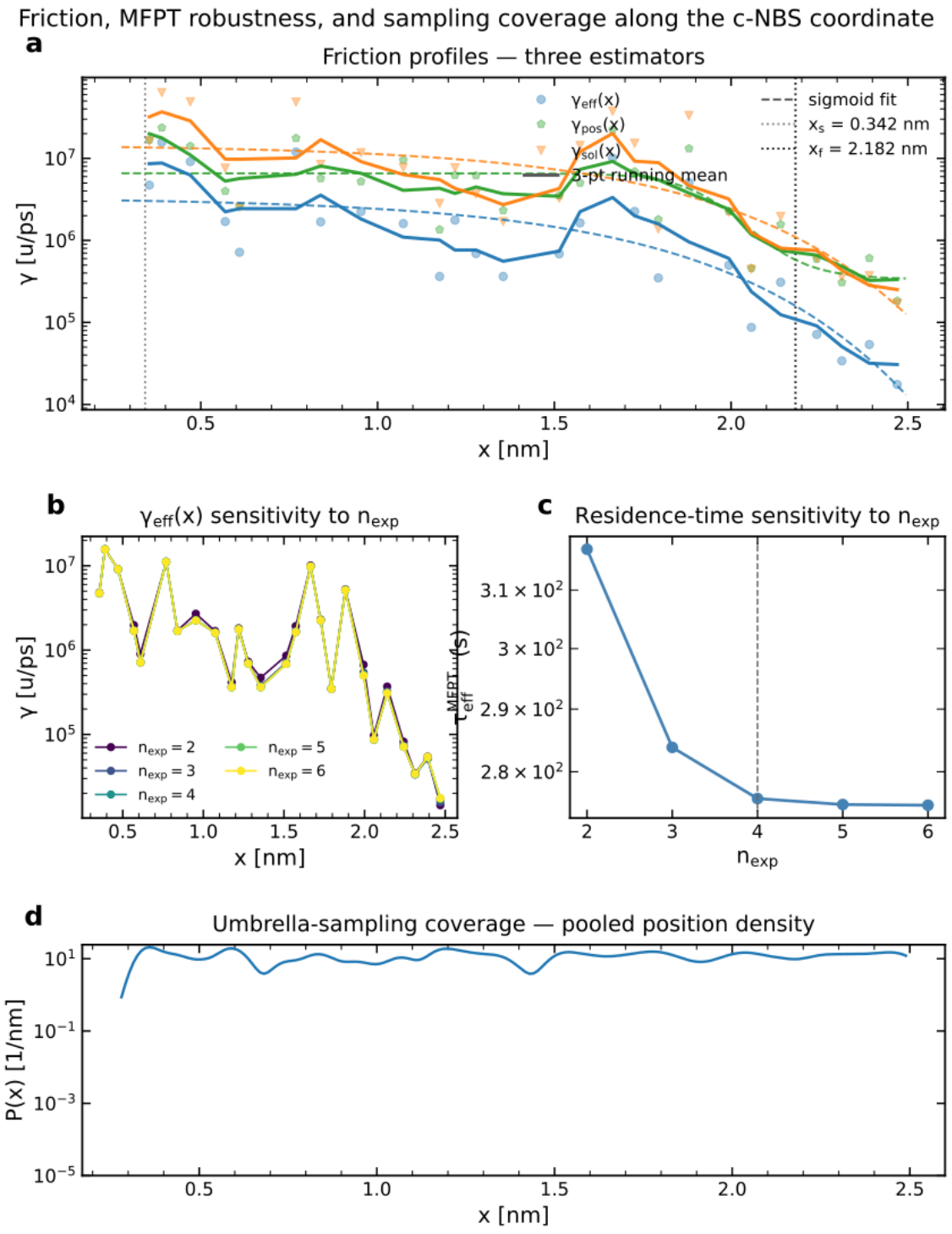

## T256A

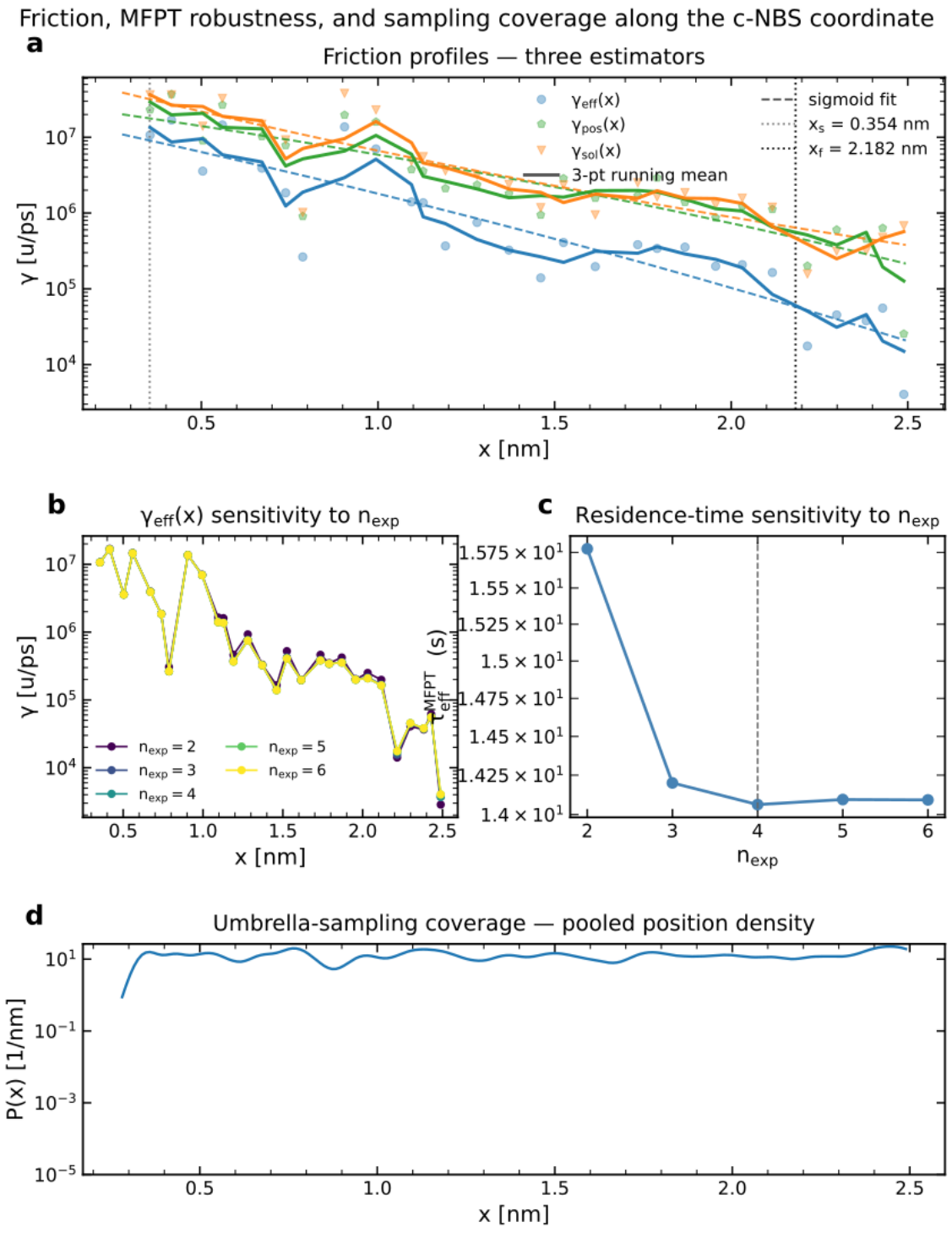

## E169Q

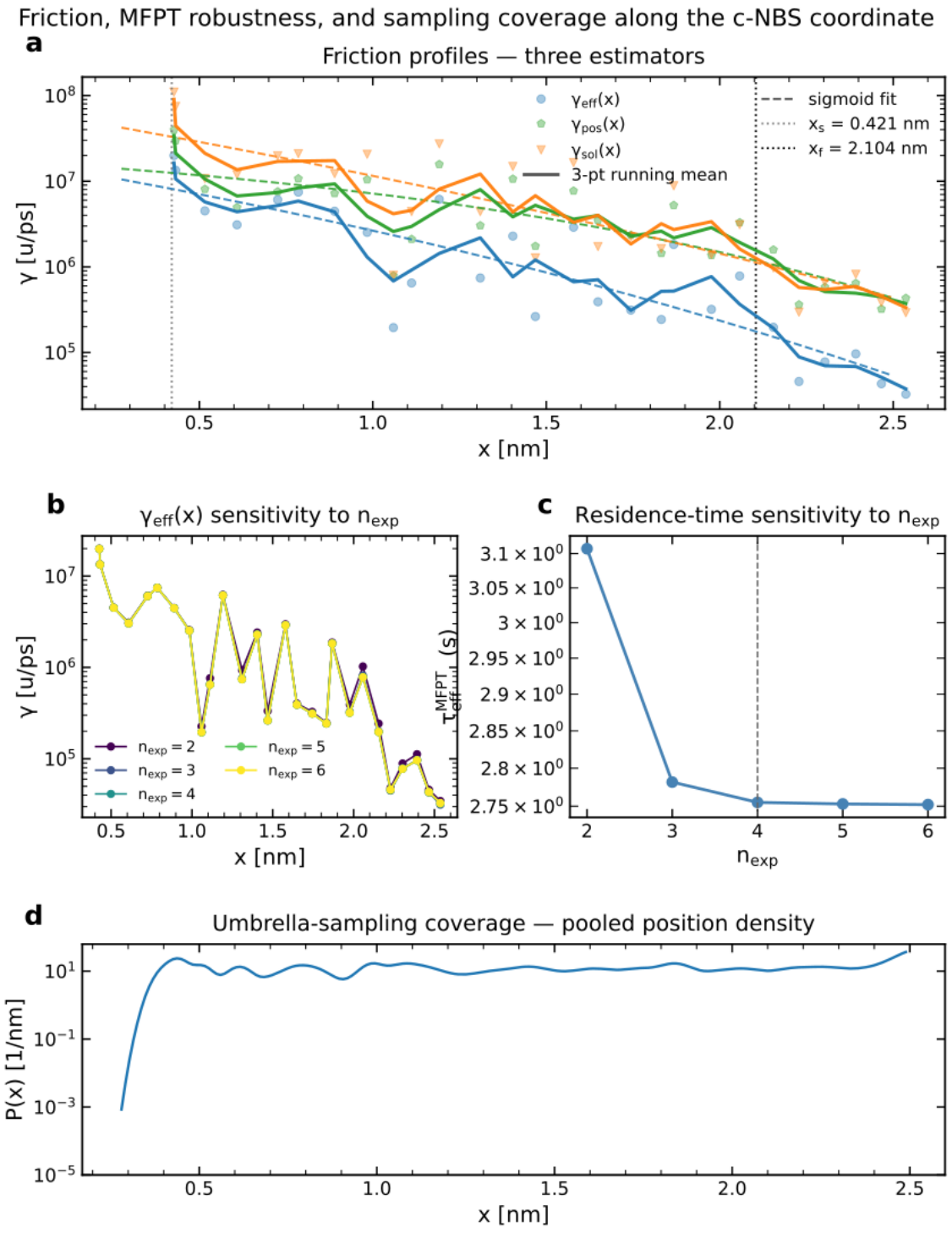

## E169A

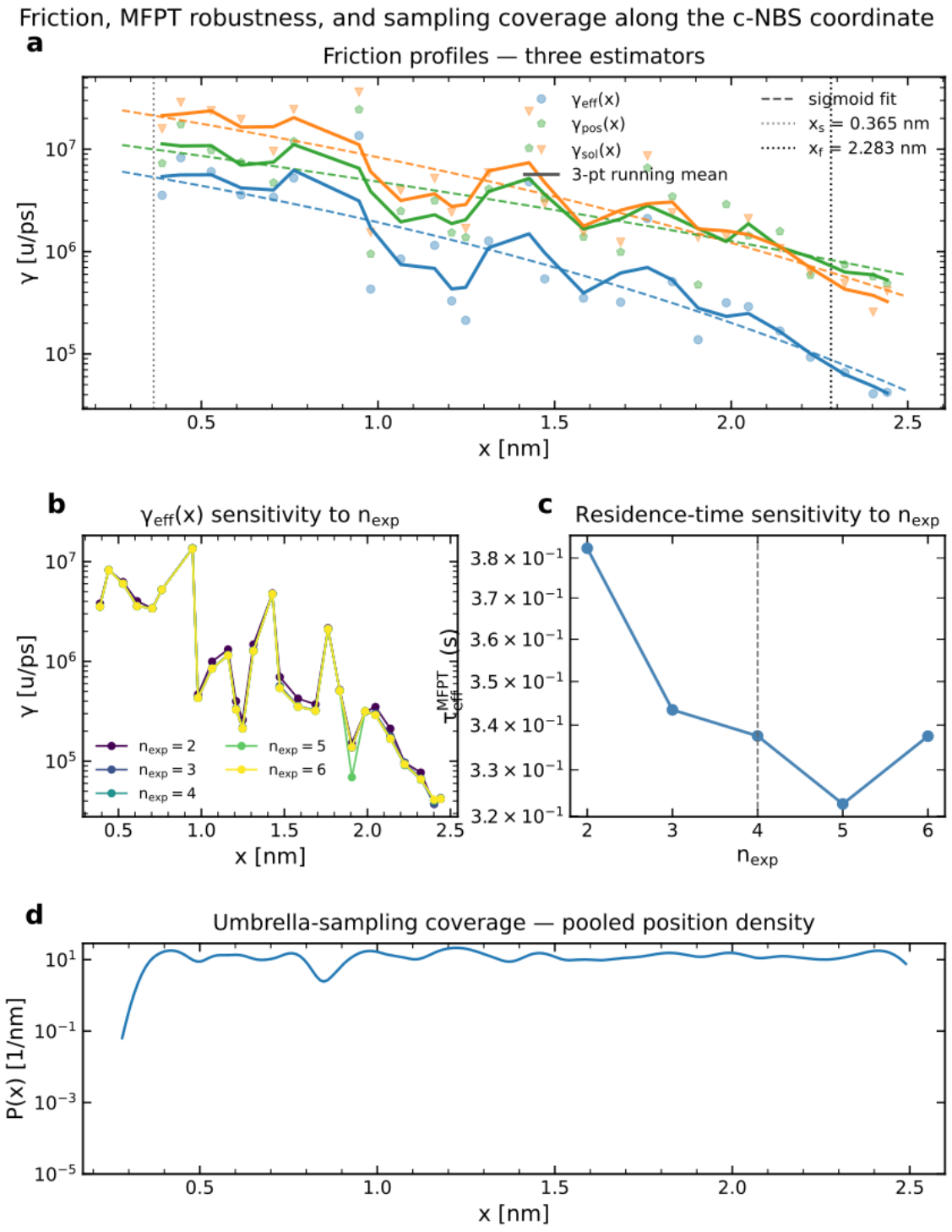

**Supplementary Fig. 13 | N-HSP90–compound 18 umbrella sampling and PMF diagnostics.** Panels show umbrella-window coordinate distributions with harmonic reference distributions, the WHAM PMF with pointwise bootstrap mean ± one standard deviation, and cumulative coordinate coverage.

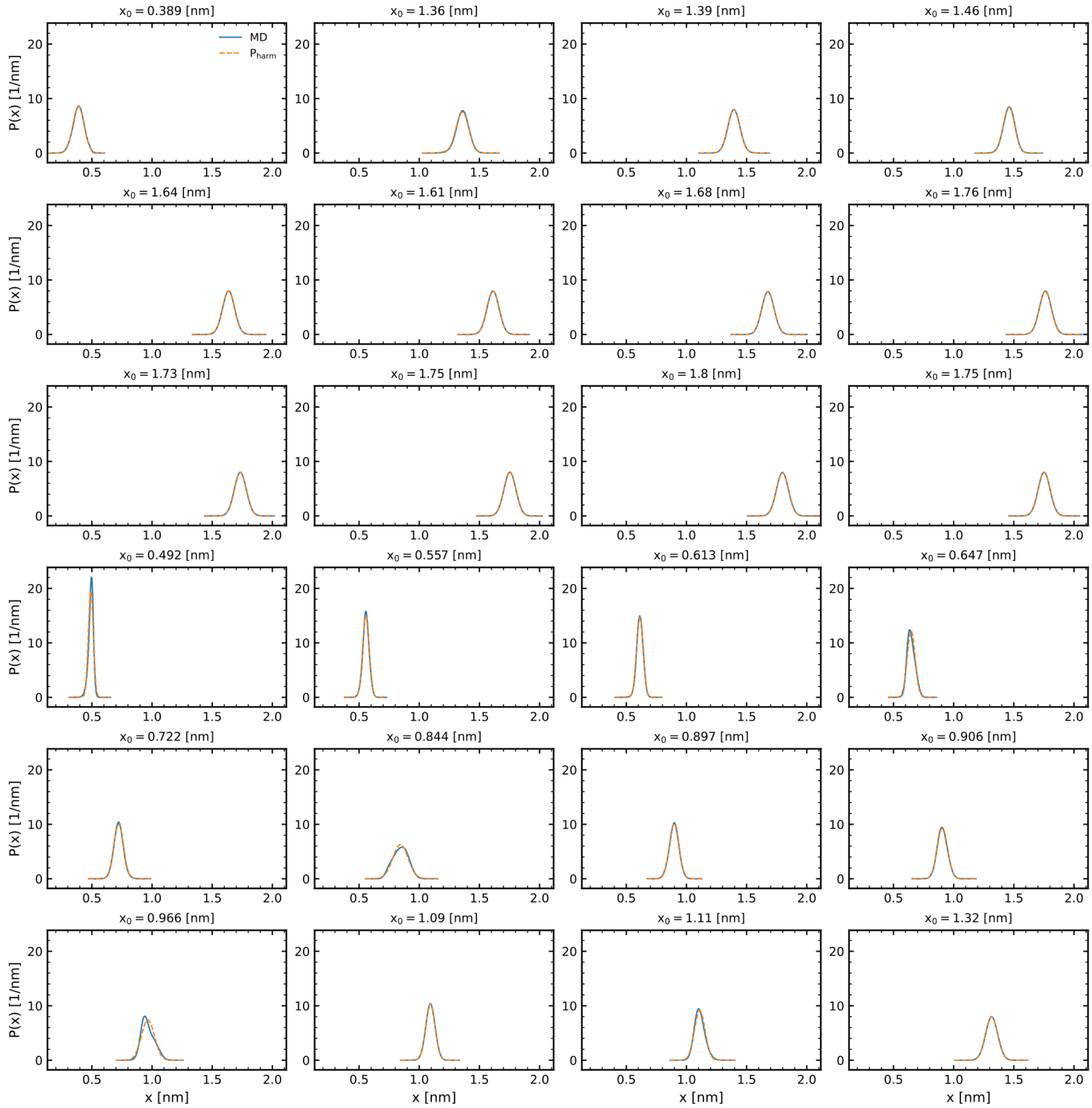

Supplementary Fig. 13 continued.

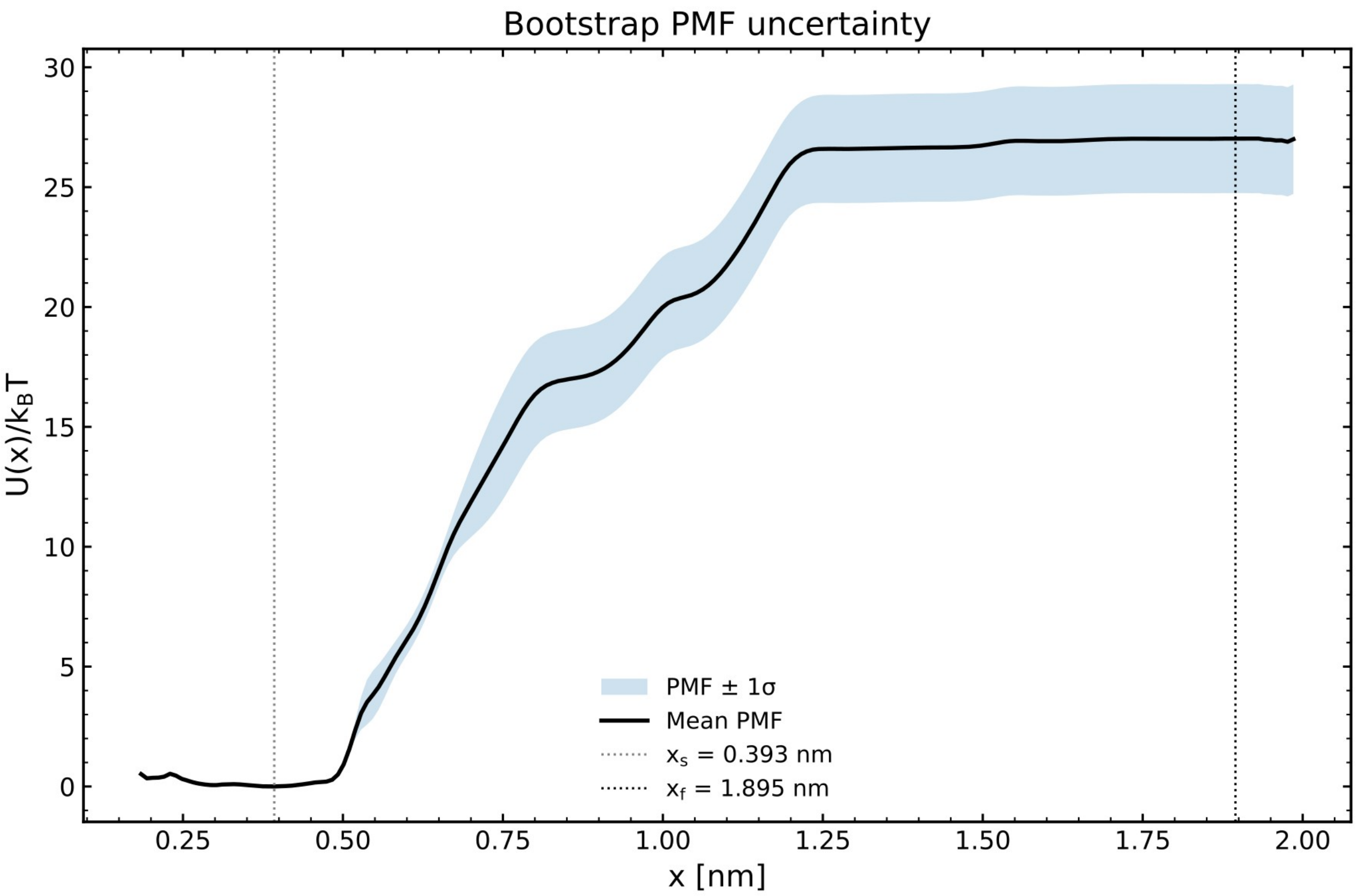

Supplementary Fig. 13 continued.

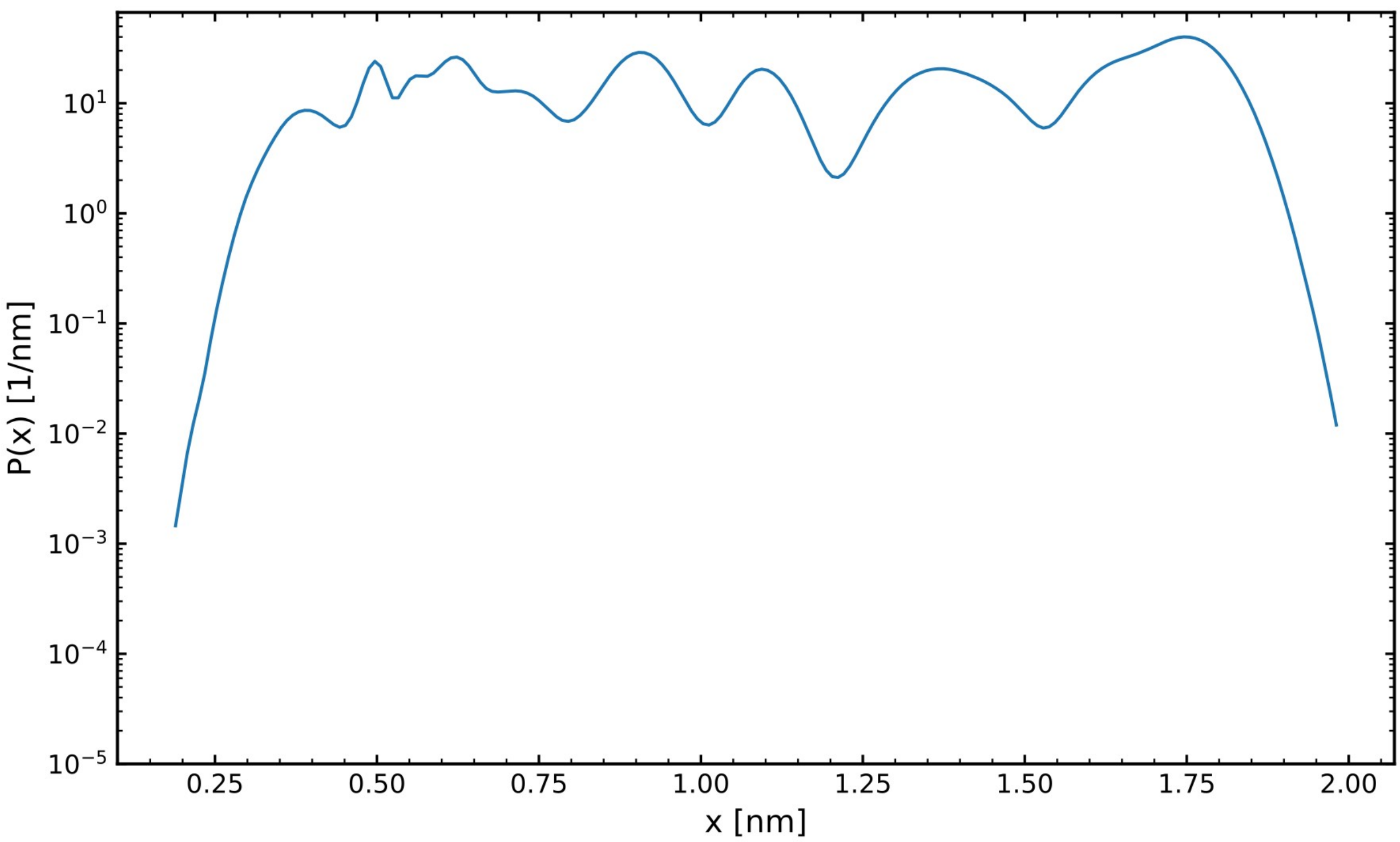

**Supplementary Fig. 14 | N-HSP90–compound 18 position-autocorrelation diagnostics.** Panels show per-window position-autocorrelation fits, running integrals and squared-autocorrelation running integrals used for the non-Markovian effective-friction calculation.

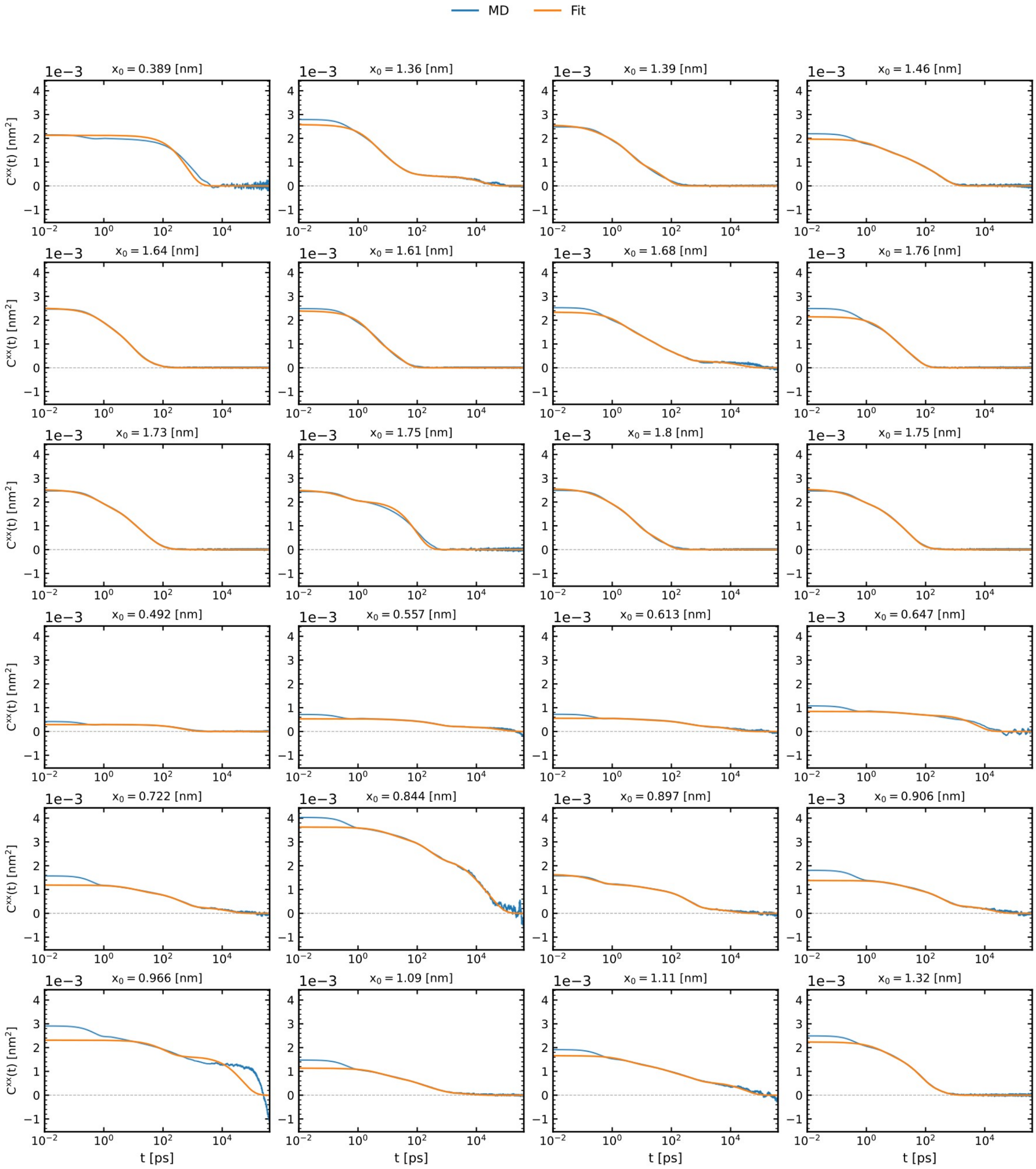

Supplementary Fig. 14 continued.

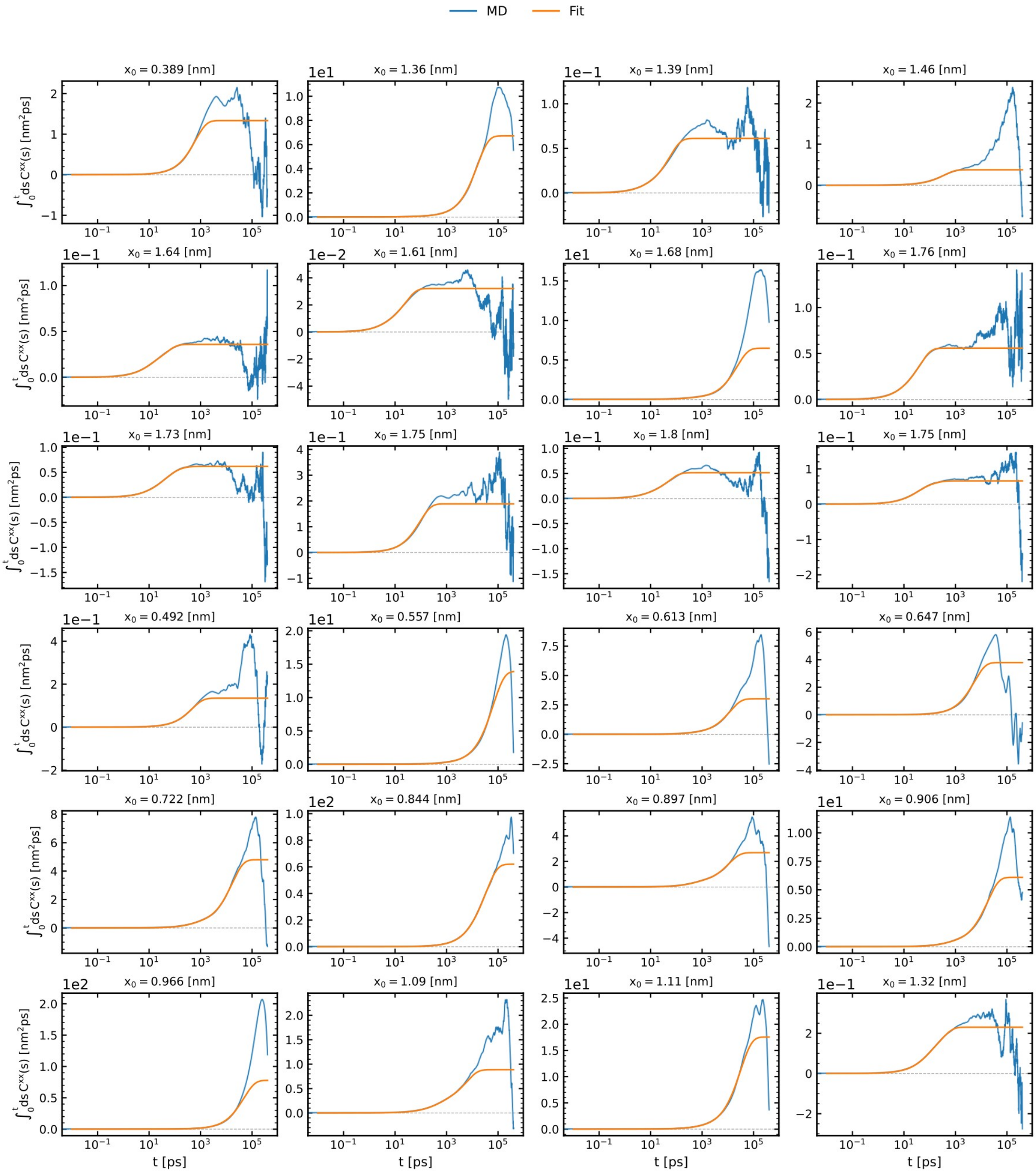

Supplementary Fig. 14 continued.

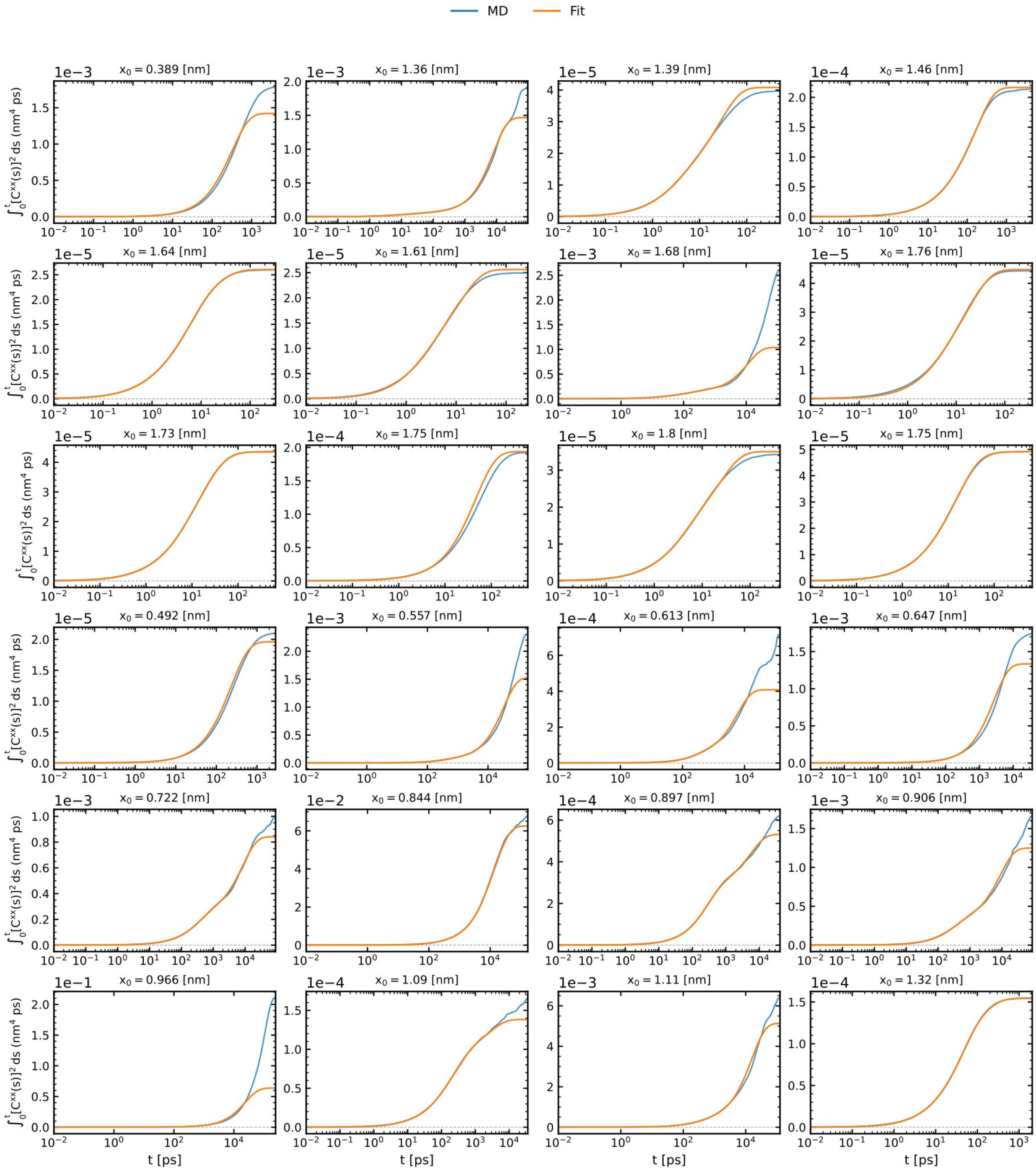

**Supplementary Fig. 15 | N-HSP90–compound 18 solvent-force and effective-mass diagnostics.** Panels show solvent-force autocorrelation fits, solvent-force running integrals and the effective mass estimated for each umbrella window.

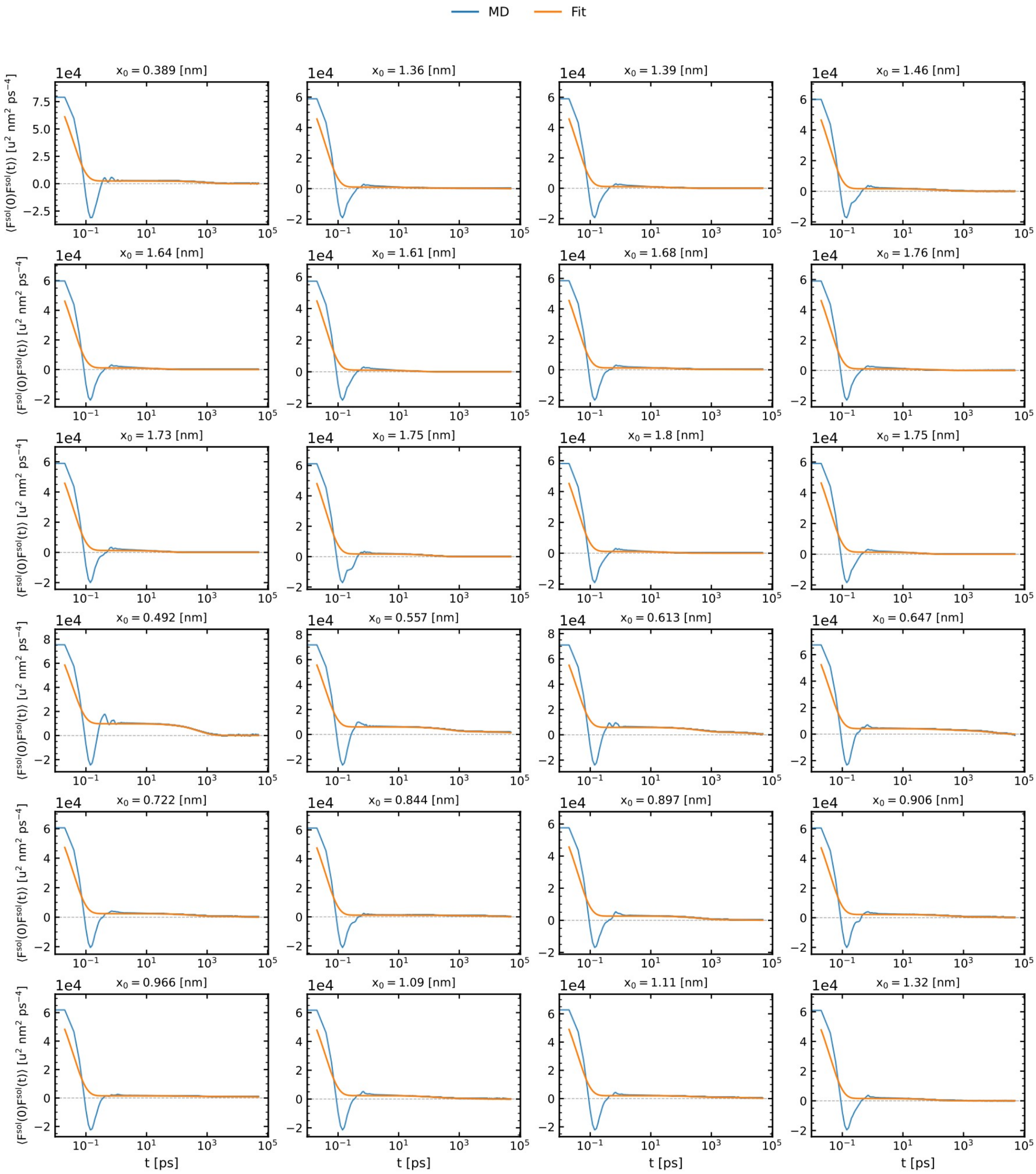

Supplementary Fig. 15 continued.

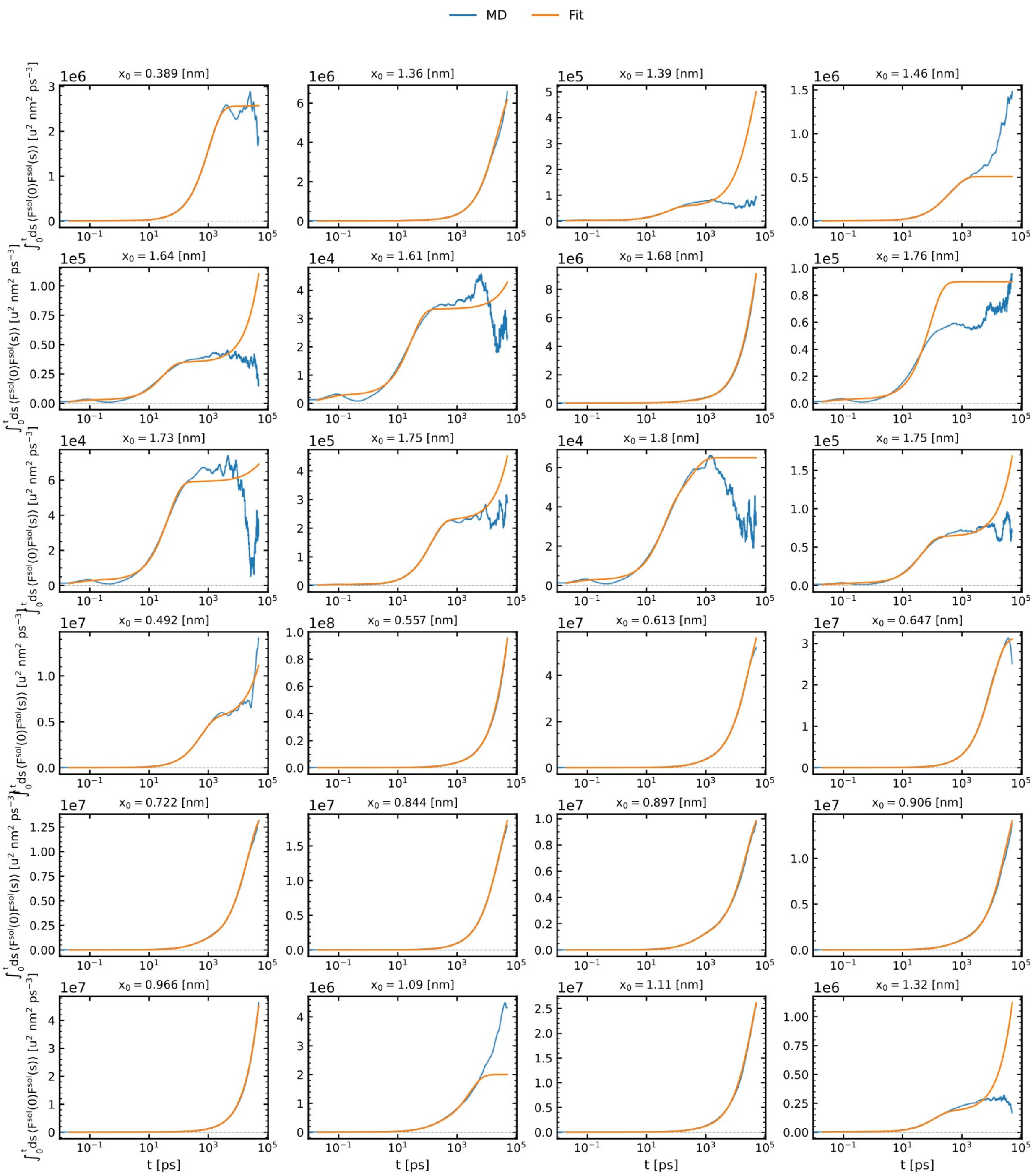

Supplementary Fig. 15 continued.

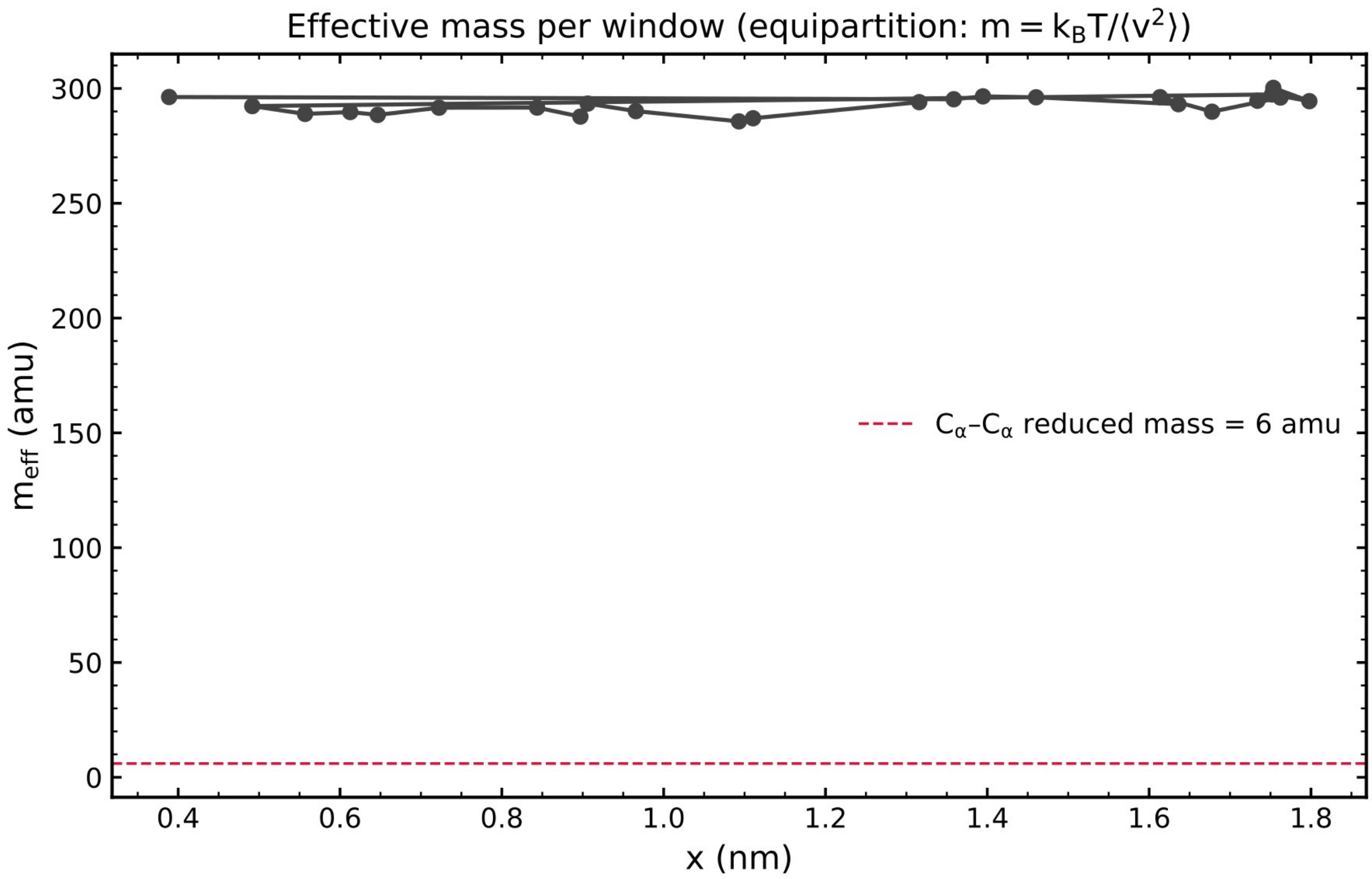

**Supplementary Fig. 16 | N-HSP90–compound 18 friction and MFPT diagnostics.** Panels show effective, position-autocorrelation and solvent-force friction profiles, cumulative MFPT accumulation and the final MFPT comparison among the three friction descriptions.

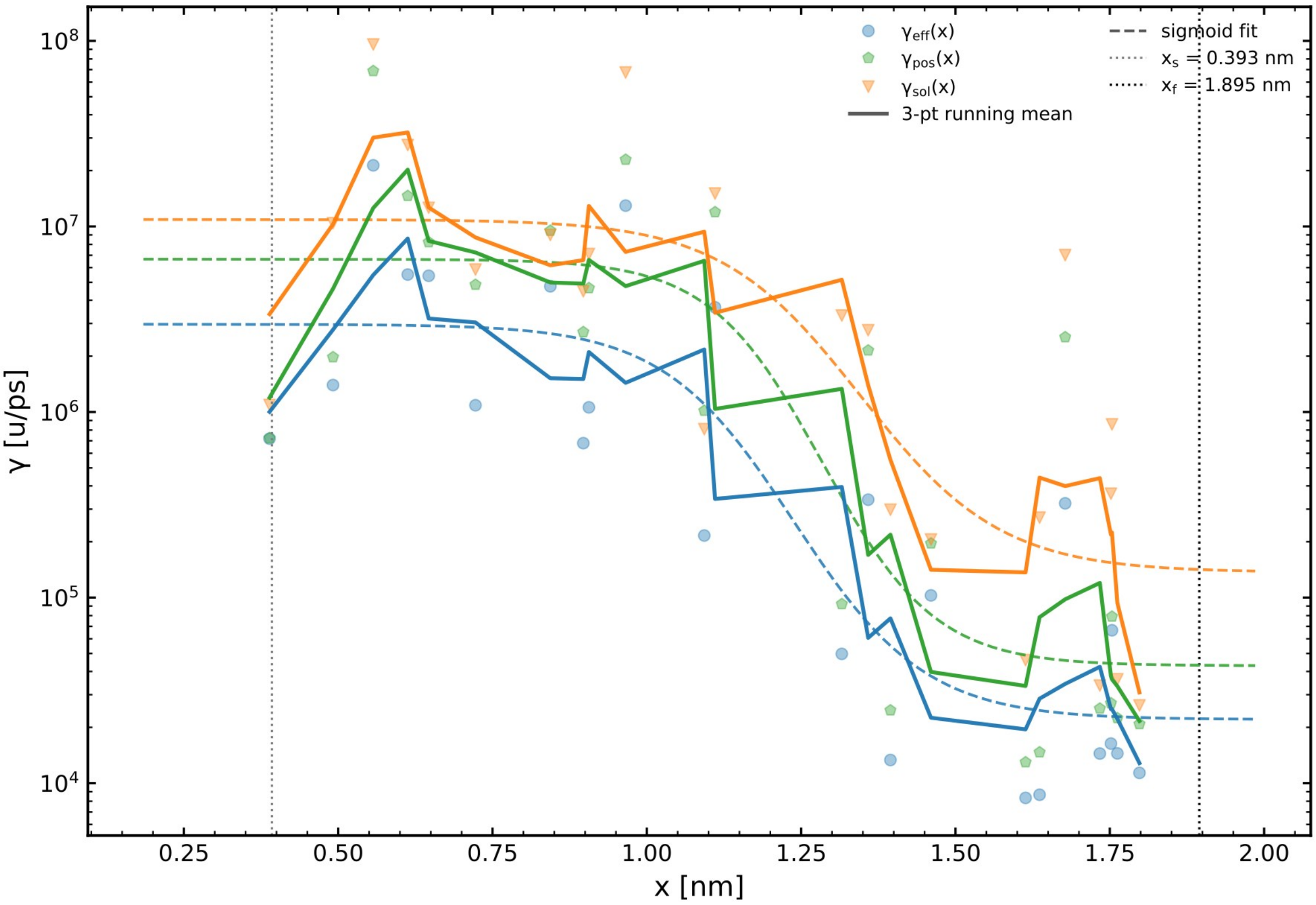

Supplementary Fig. 16 continued.

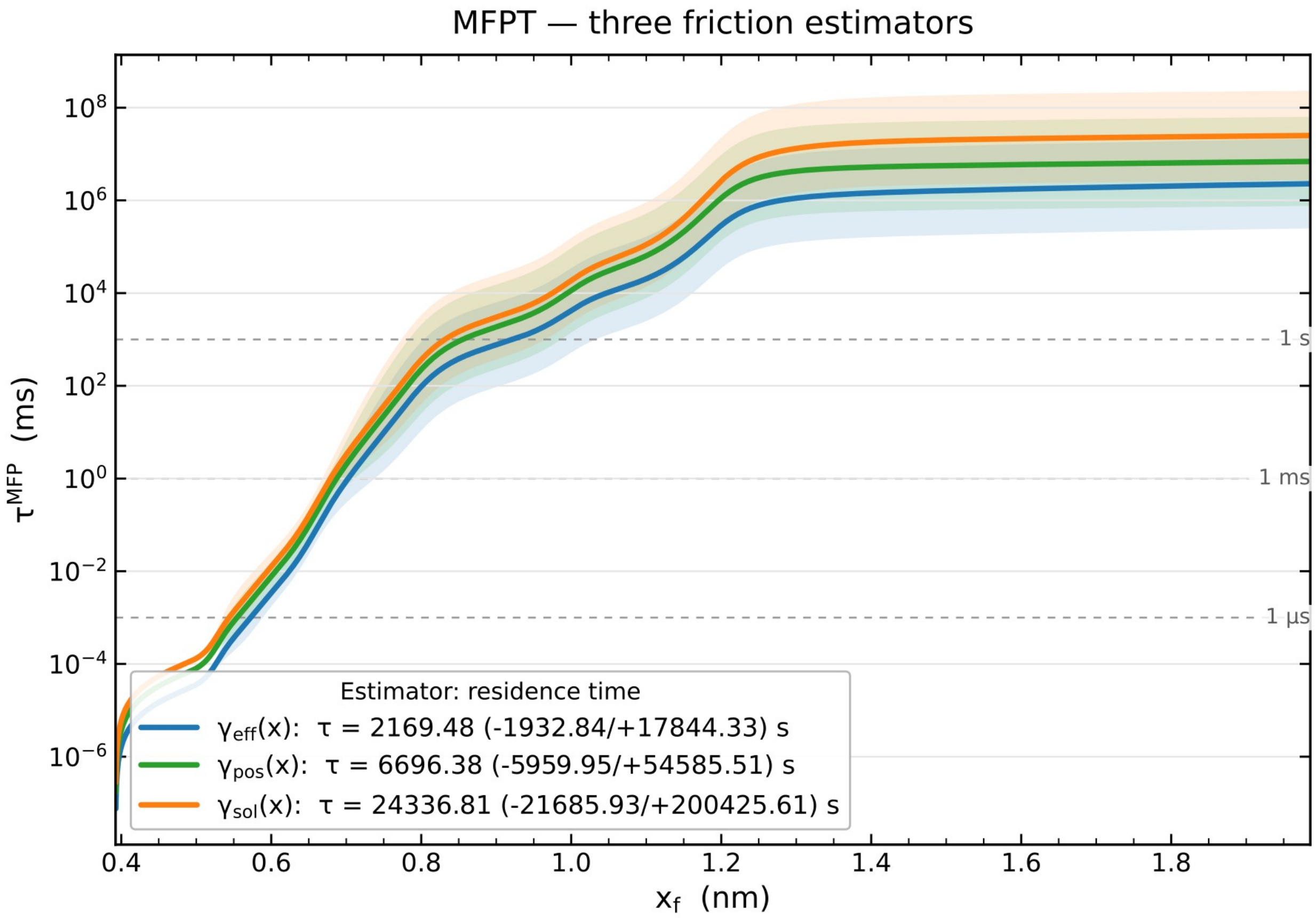

Supplementary Fig. 16 continued.

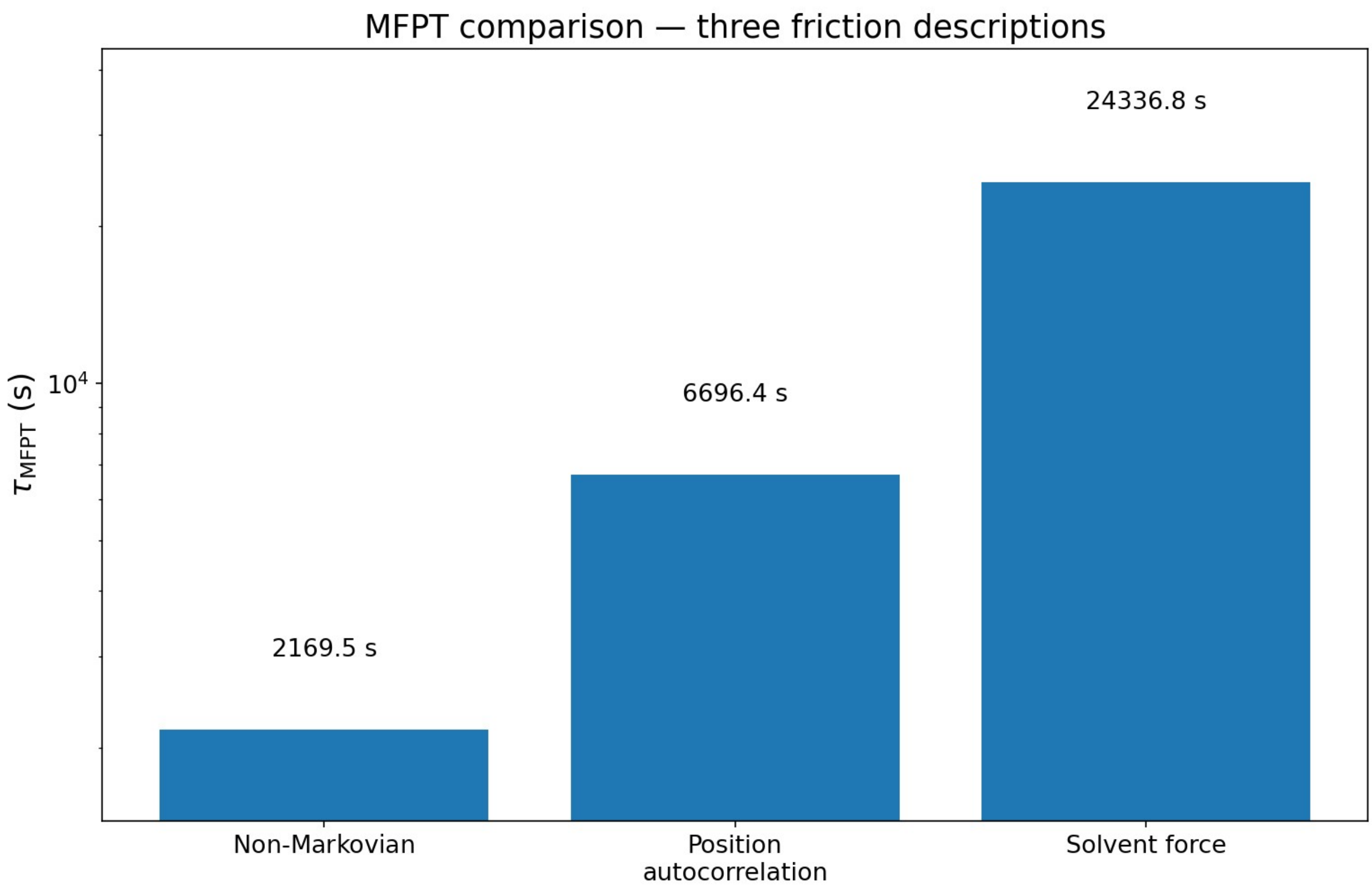

**Supplementary Fig. 17 | N-HSP90–compound 18 autocorrelation model-order and robustness diagnostics.** Panels show AIC/BIC and fit-quality metrics for two to six exponential components, effective-friction/MFPT sensitivity to model order, integral-quality metrics and an overall friction/MFPT robustness summary.

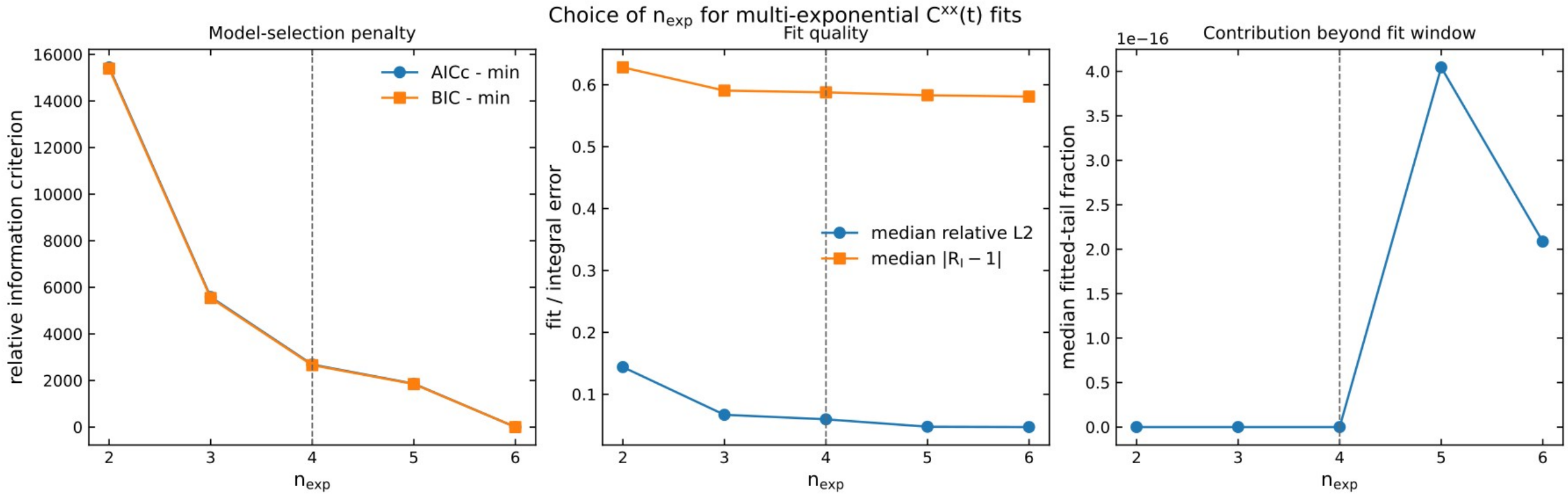

Supplementary Fig. 17 continued.

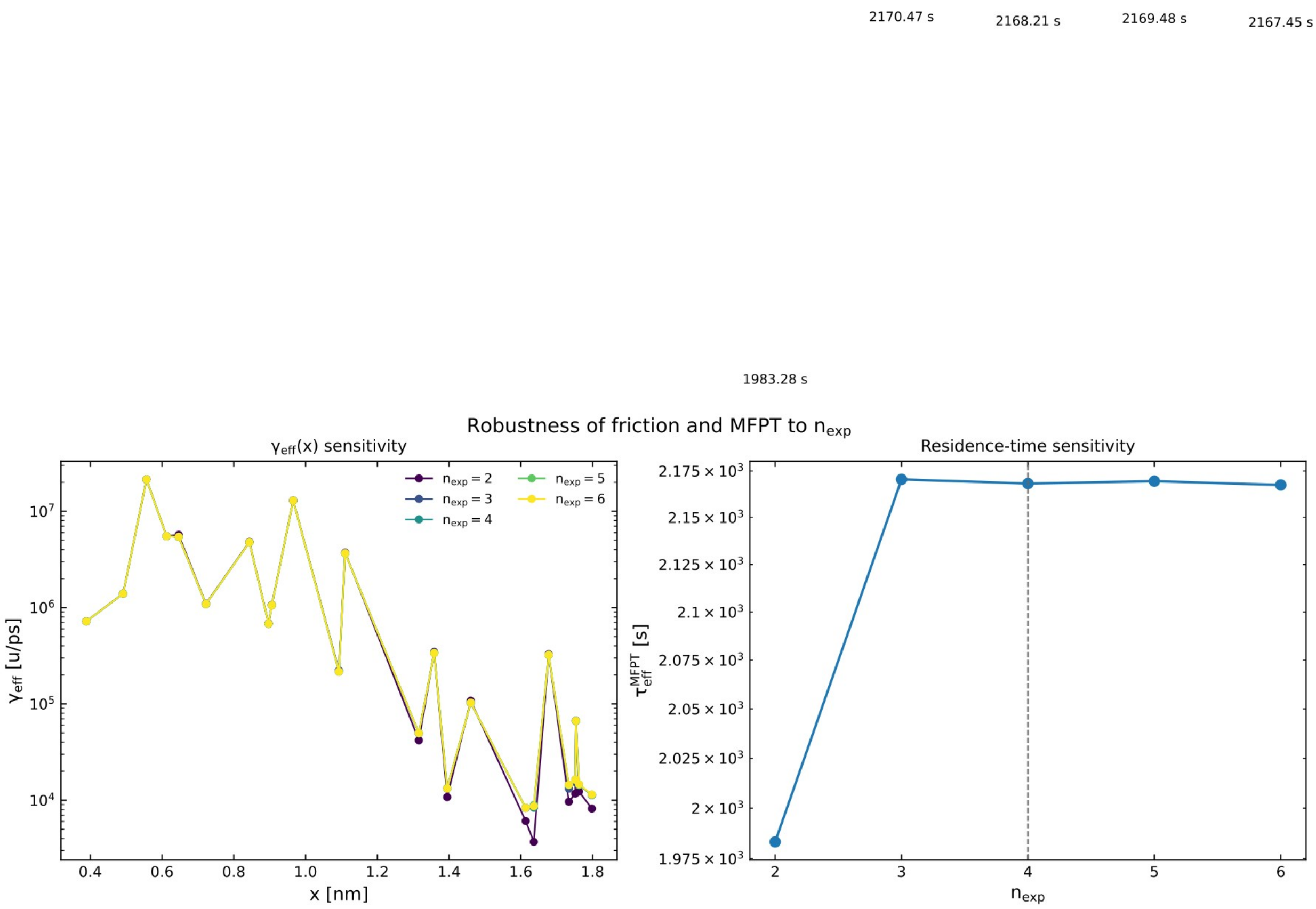

Supplementary Fig. 17 continued.

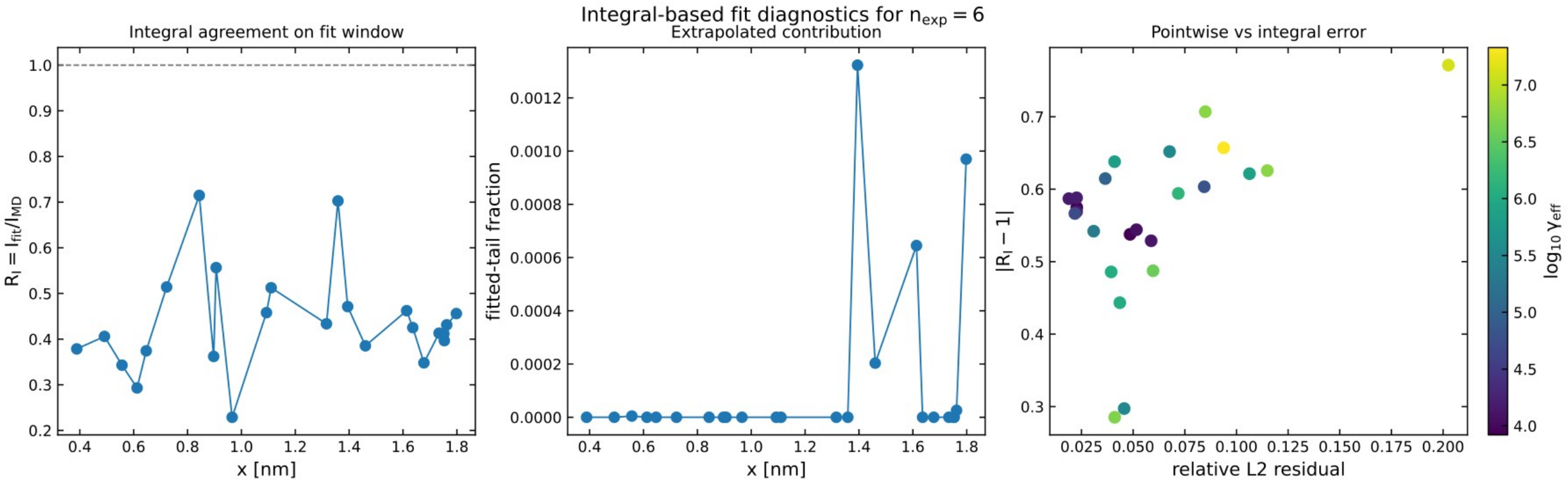

Supplementary Fig. 17 continued.

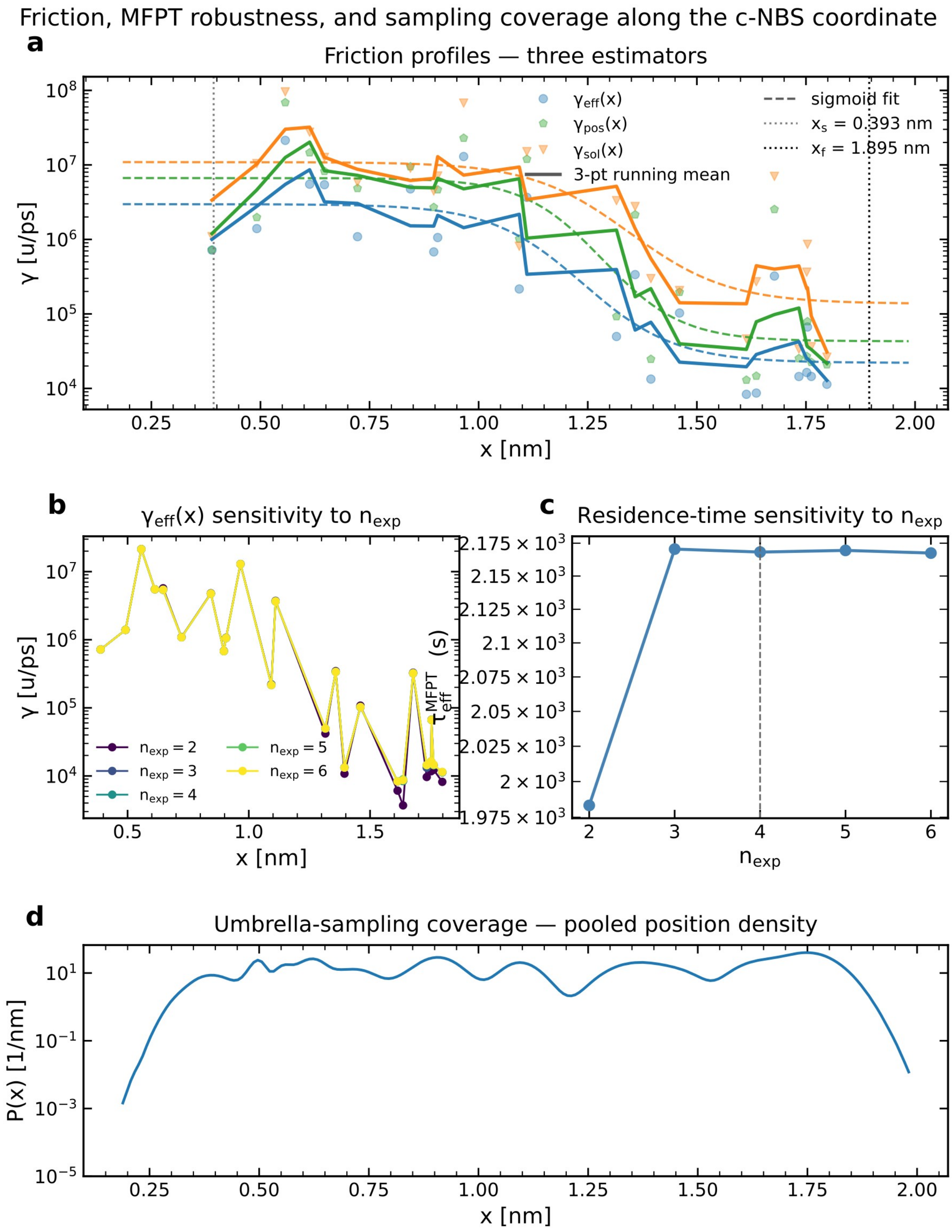

**Supplementary Fig. 18 | ** RNase S–S15 umbrella sampling and PMF diagnostics. Panels show umbrella-window coordinate distributions with harmonic reference distributions, the WHAM PMF with pointwise bootstrap mean ± one standard deviation, and cumulative coordinate coverage.

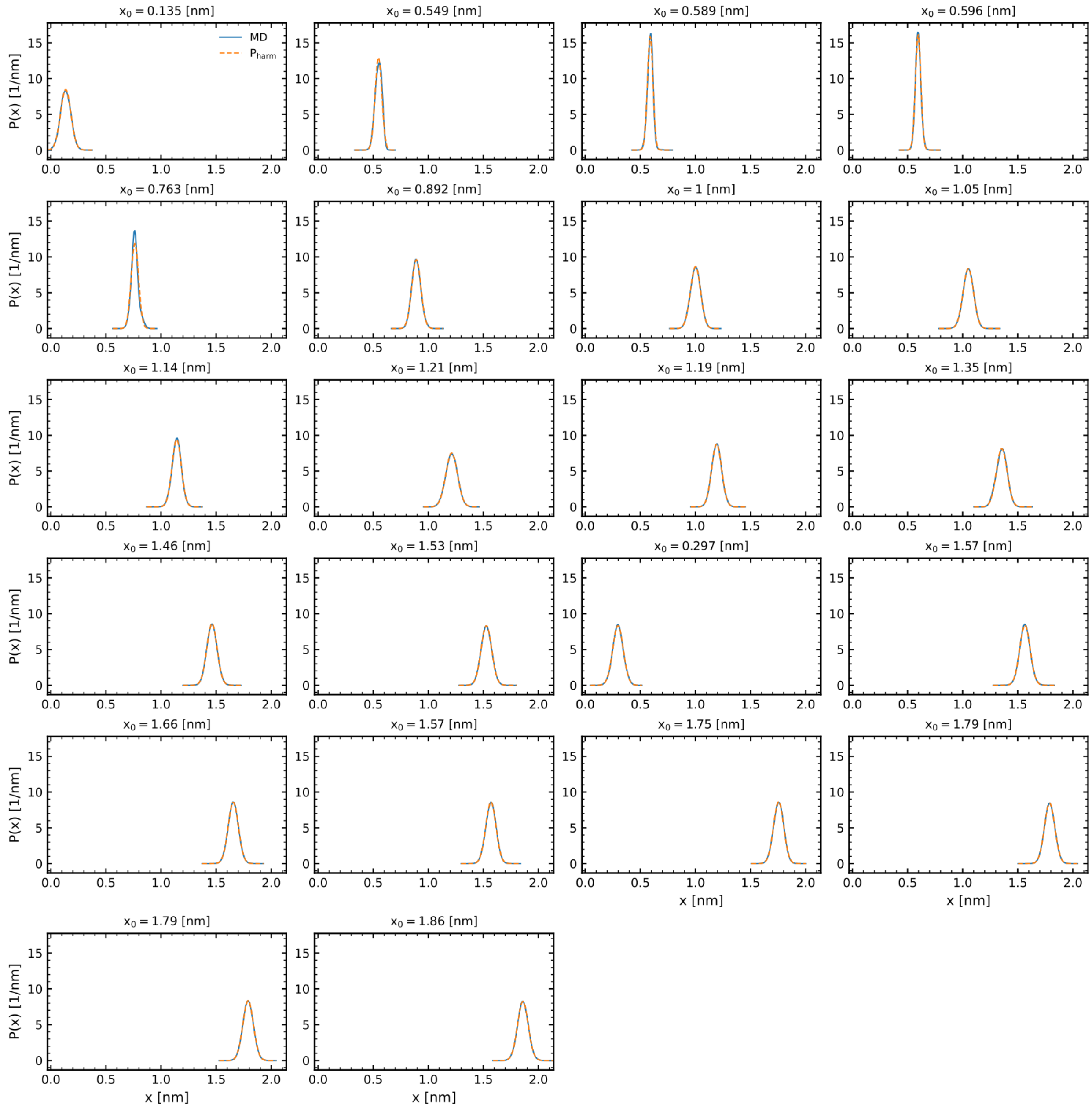

Supplementary Fig. 18 continued.

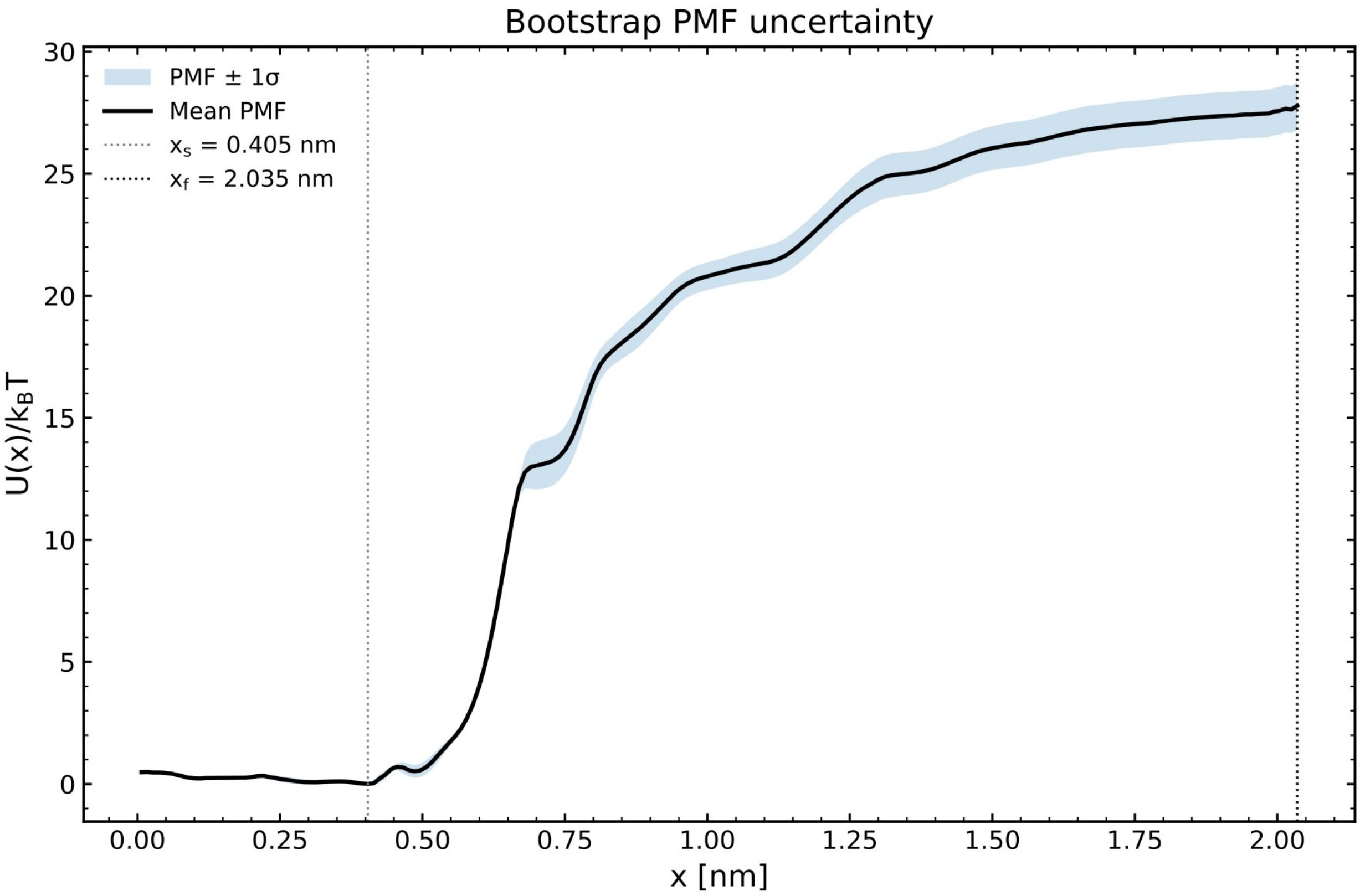

Supplementary Fig. 18 continued.

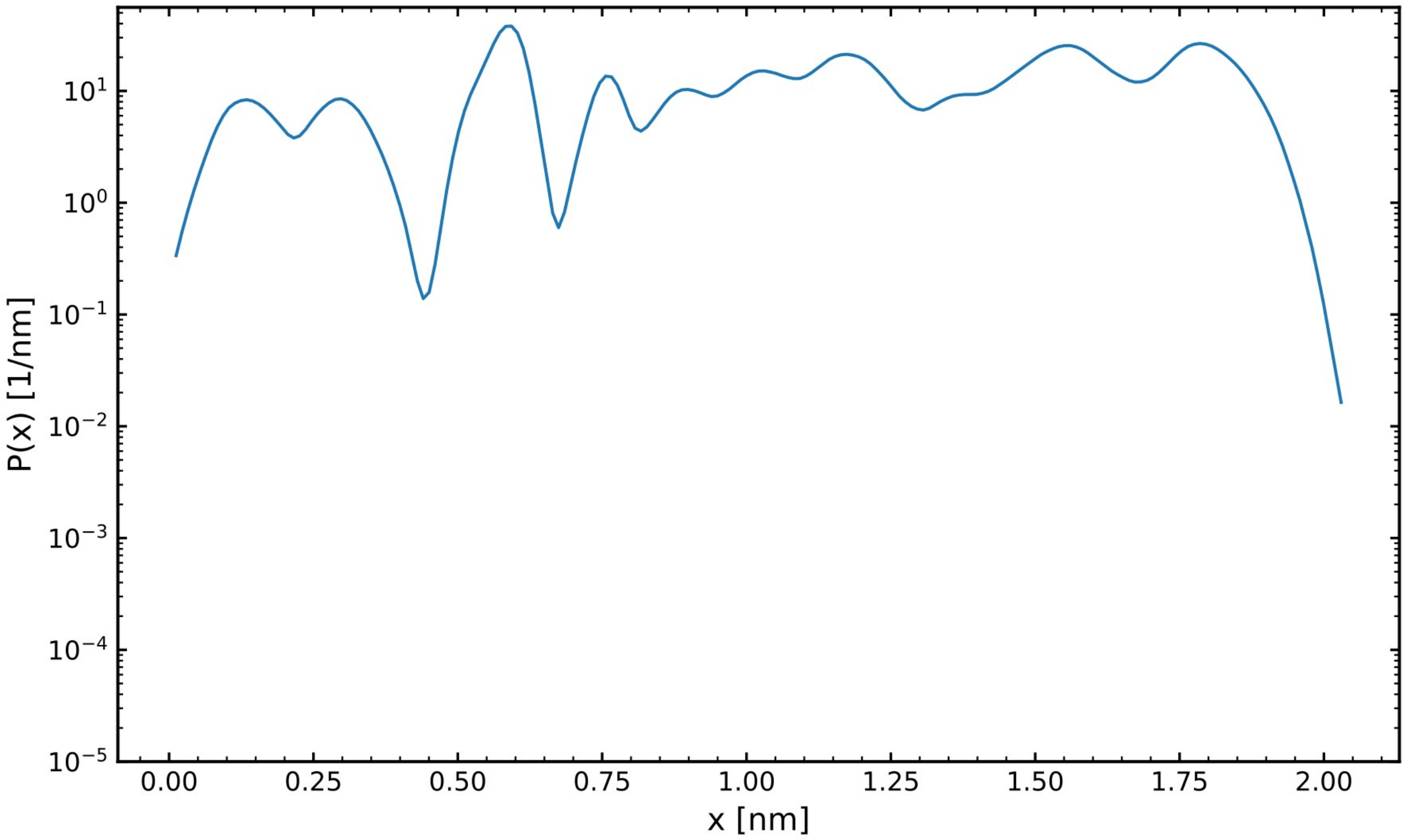

**Supplementary Fig. 19 |** RNase S–S15 position-autocorrelation diagnostics. Panels show per-window position-autocorrelation fits, running integrals and squared-autocorrelation running integrals used for the non-Markovian effective-friction calculation.

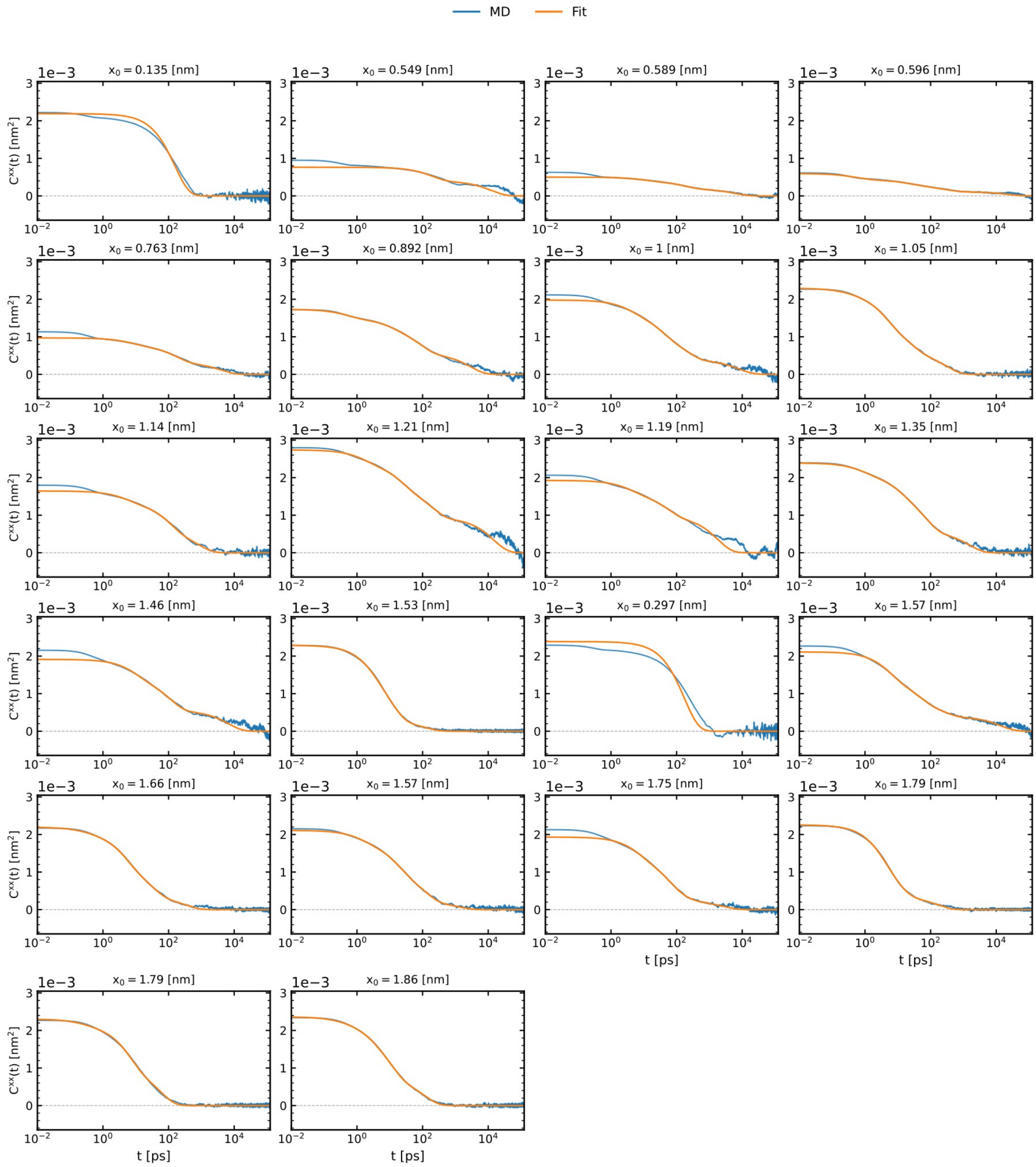

Supplementary Fig. 19 continued.

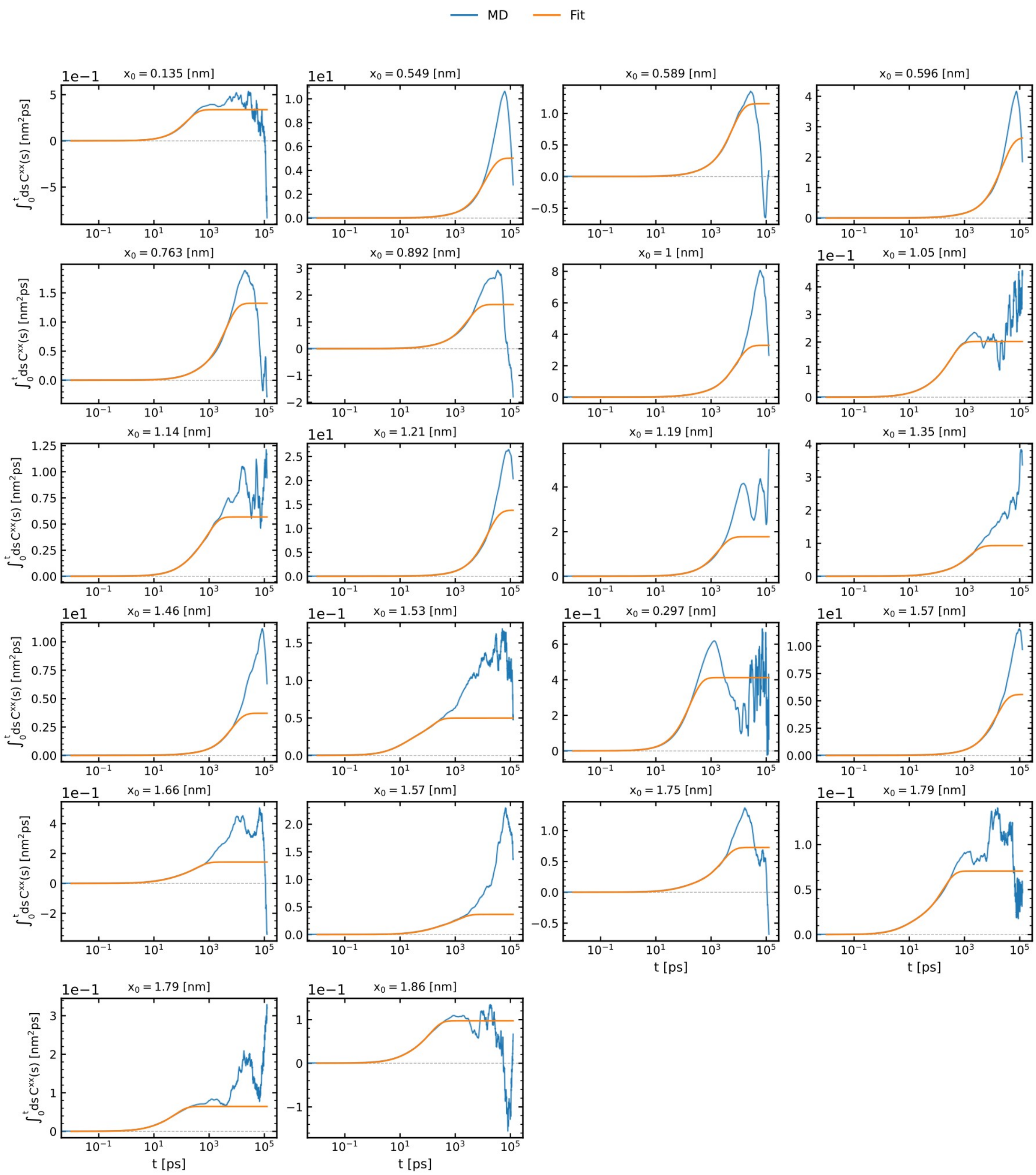

Supplementary Fig. 19 continued.

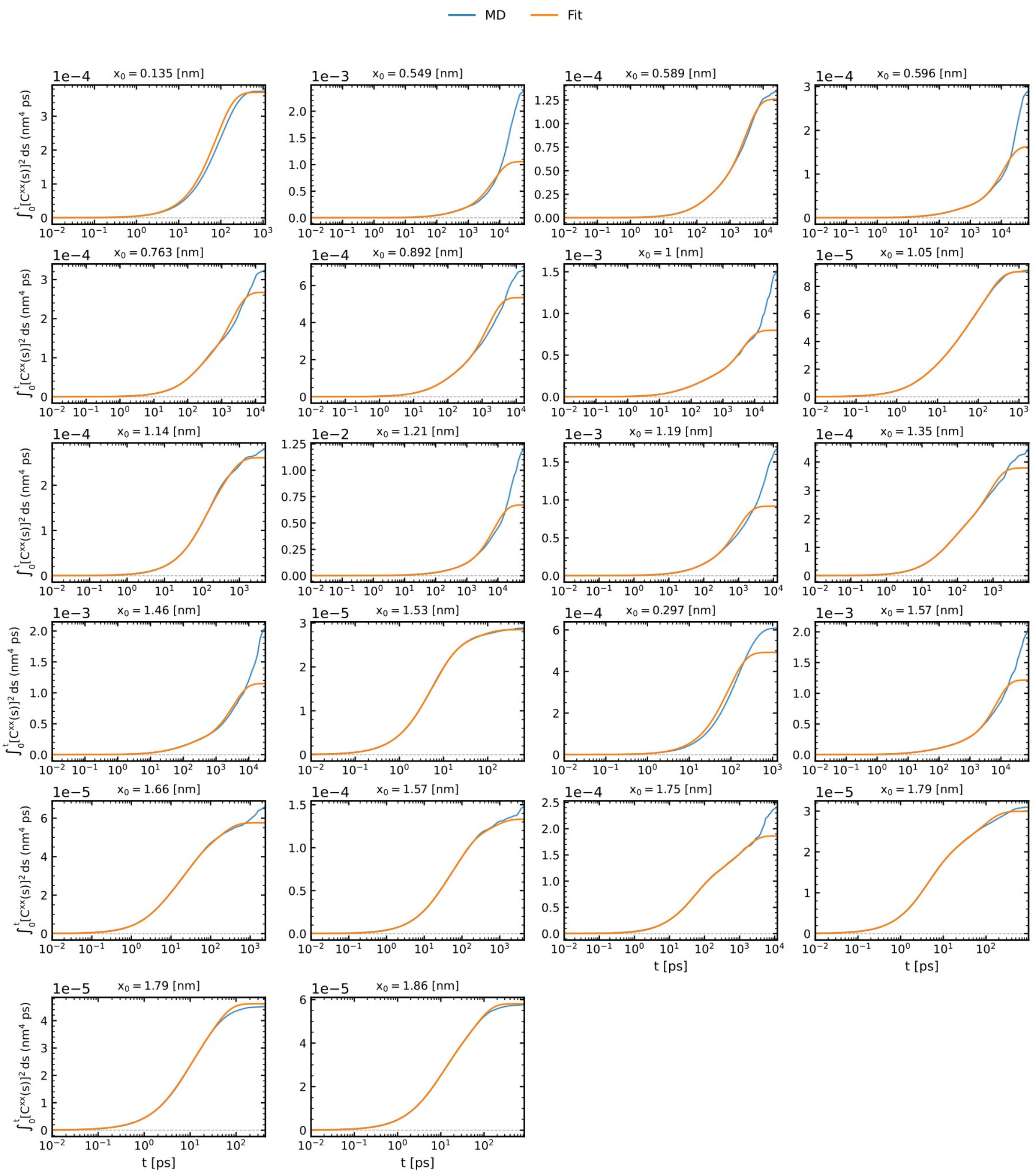

**Supplementary Fig. 20 | ** RNase S–S15 solvent-force and effective-mass diagnostics. Panels show solvent-force autocorrelation fits, solvent-force running integrals and the effective mass estimated for each umbrella window.

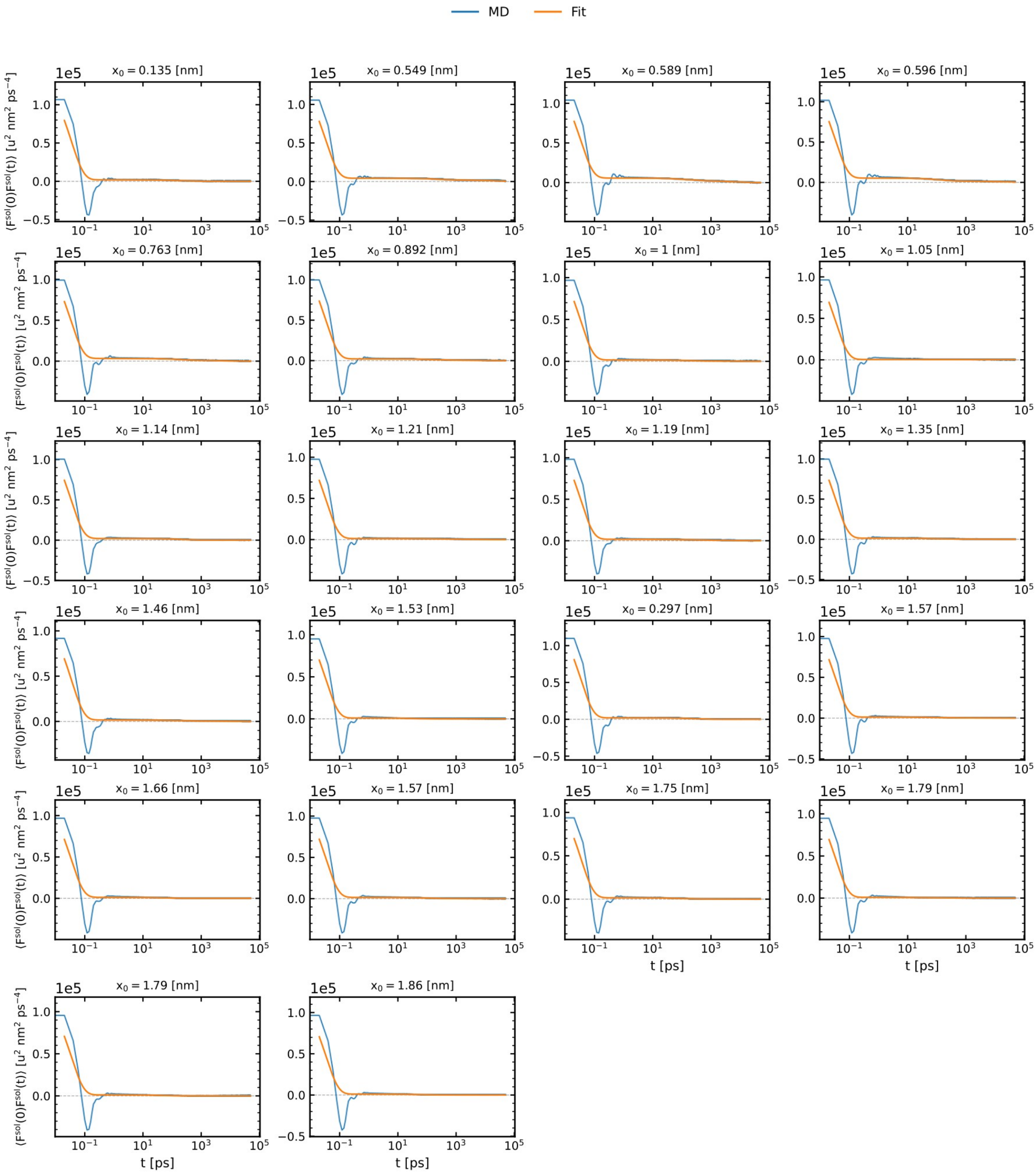

Supplementary Fig. 20 continued.

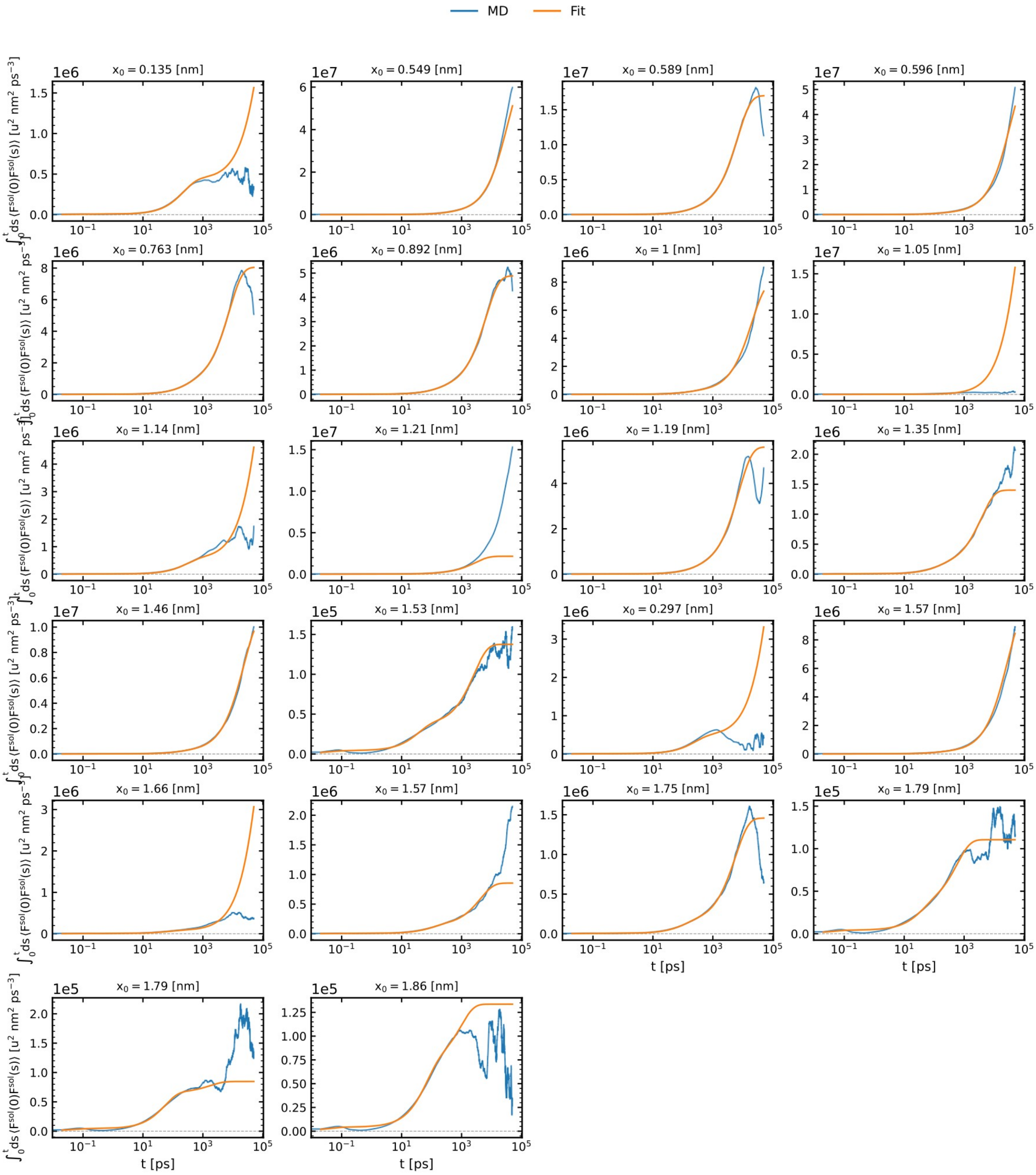

Supplementary Fig. 20 continued.

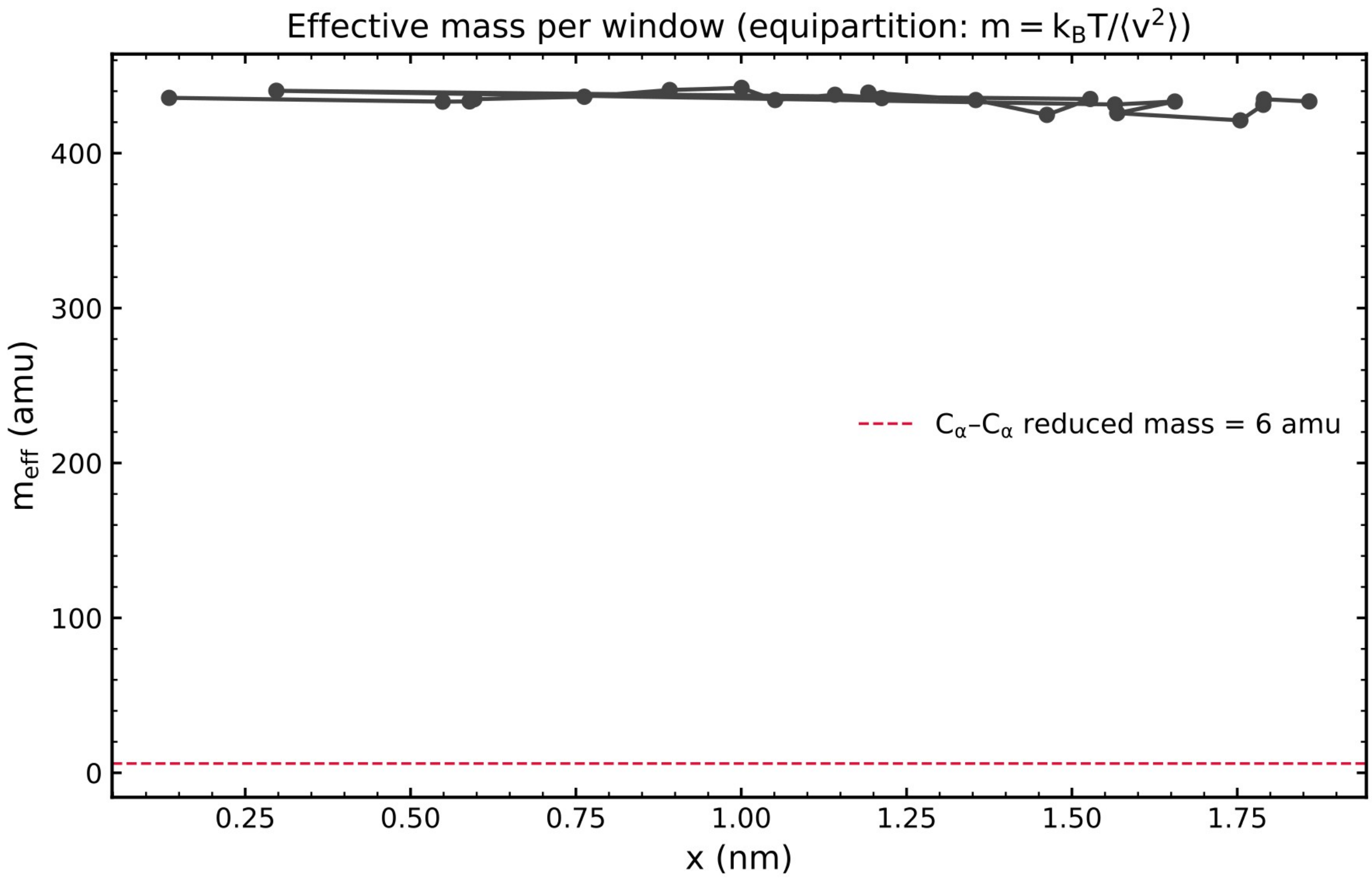

**Supplementary Fig. 21 |** RNase S–S15 friction and MFPT diagnostics. Panels show effective, position-autocorrelation and solvent-force friction profiles, cumulative MFPT accumulation and the final MFPT comparison among the three friction descriptions.

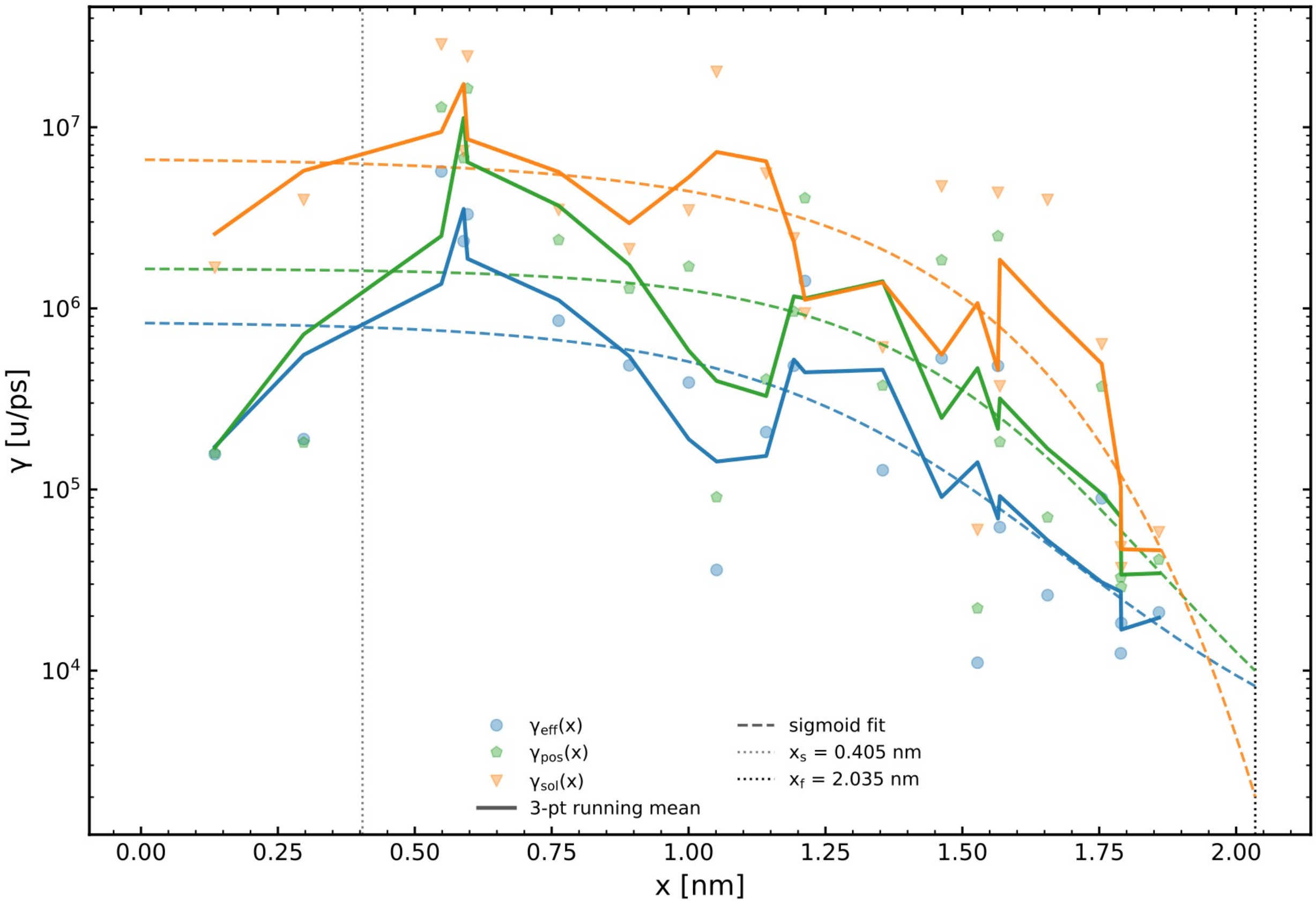

Supplementary Fig. 21 continued.

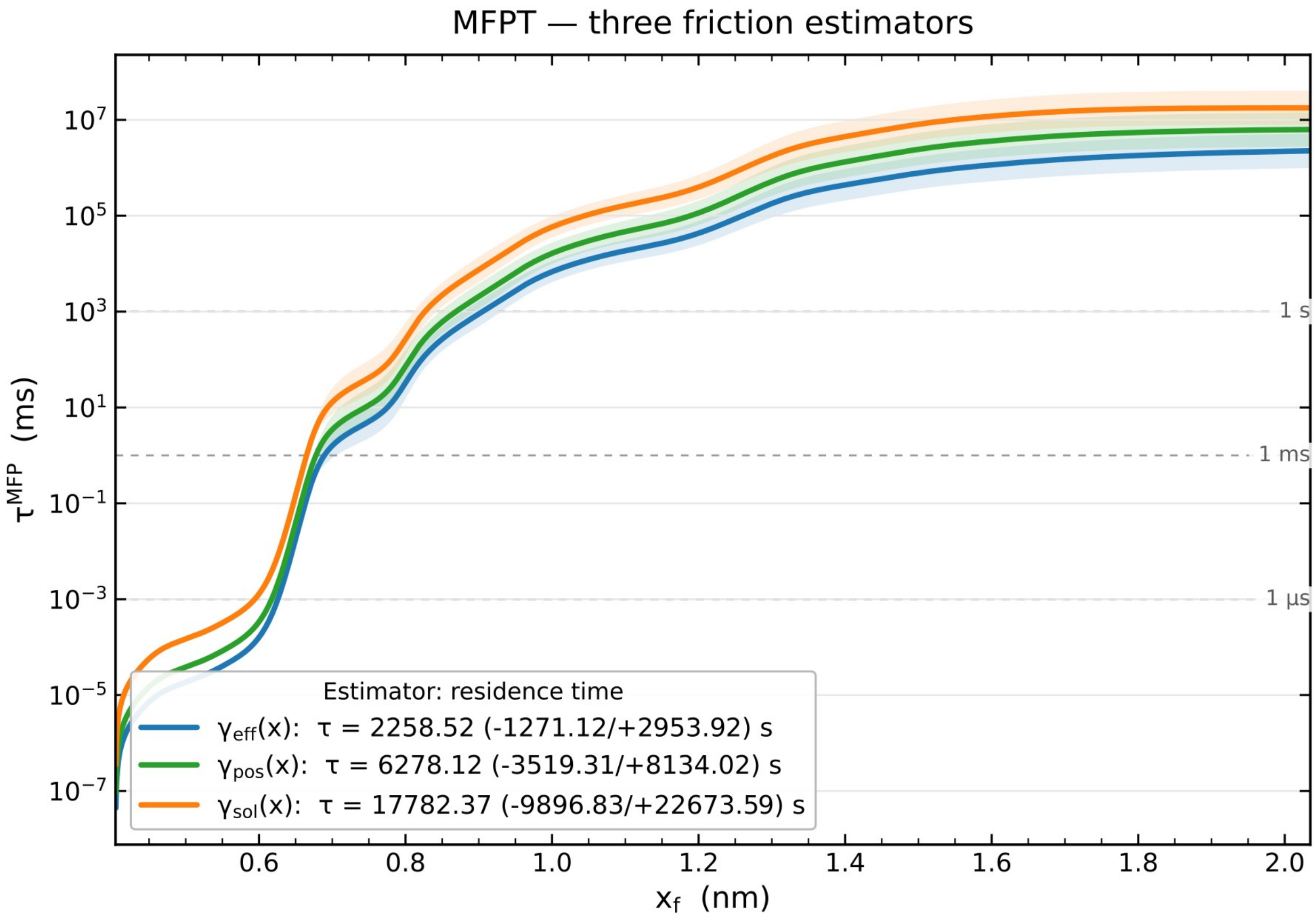

Supplementary Fig. 21 continued.

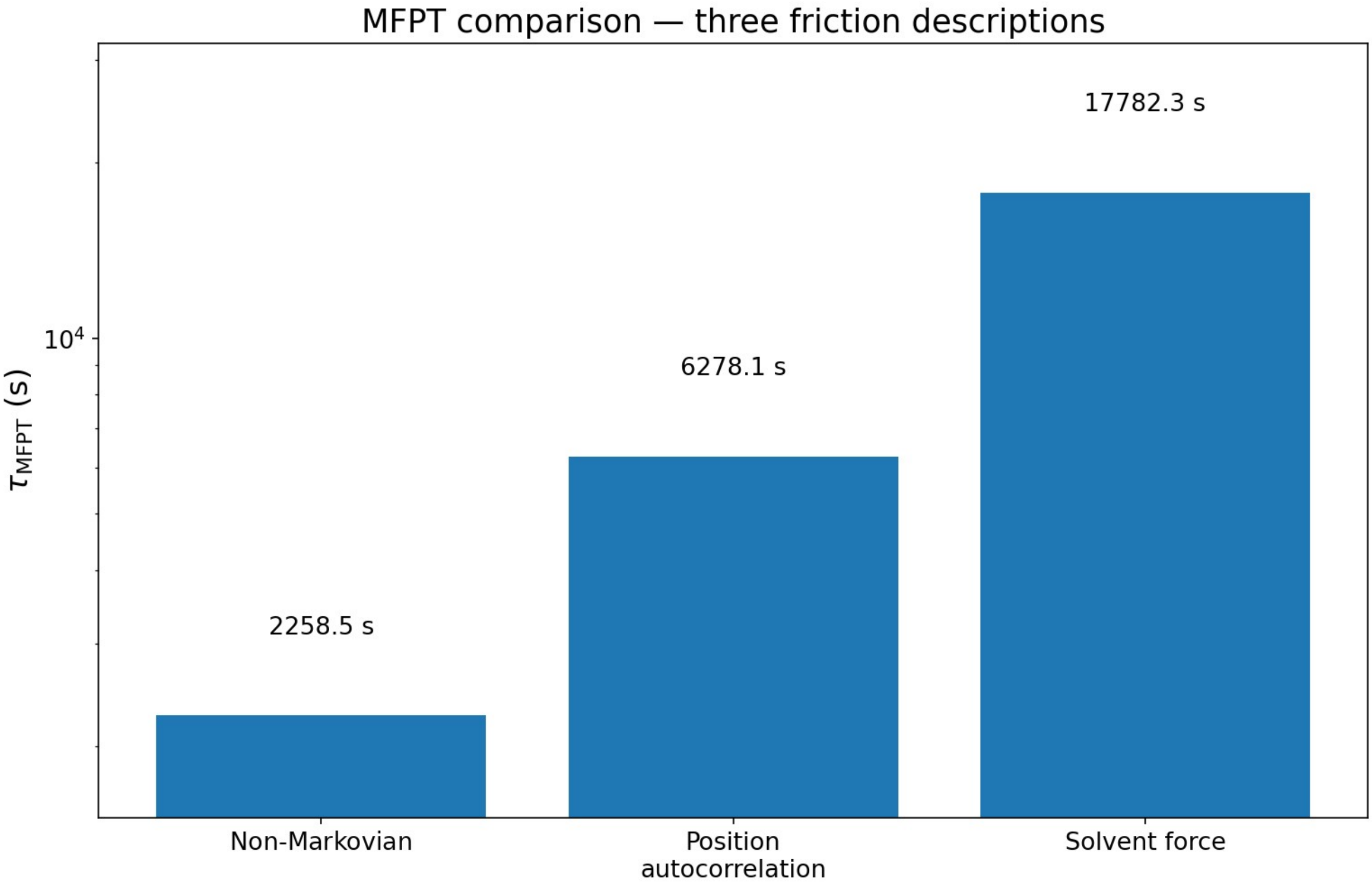

**Supplementary Fig. 22 | RNase S–S15 autocorrelation model-order and robustness diagnostics.** Panels show AIC/BIC and fit-quality metrics for two to six exponential components, effective-friction/MFPT sensitivity to model order, integral-quality metrics and an overall friction/MFPT robustness summary.

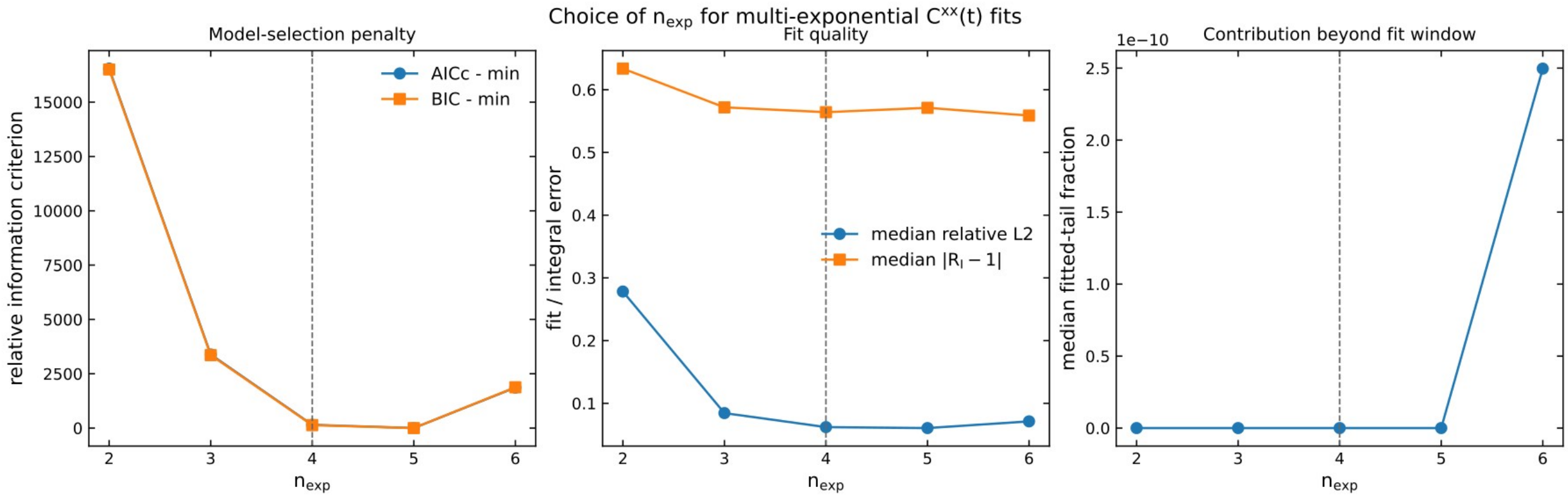

Supplementary Fig. 22 continued.

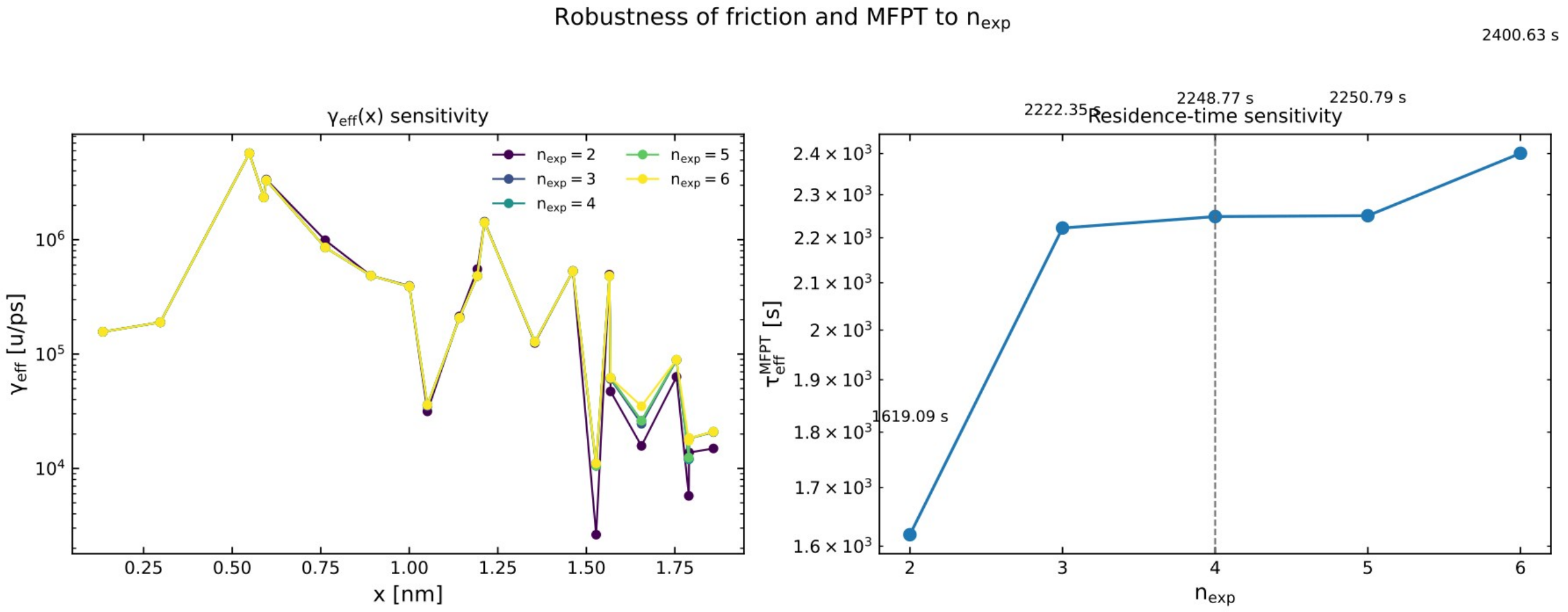

Supplementary Fig. 22 continued.

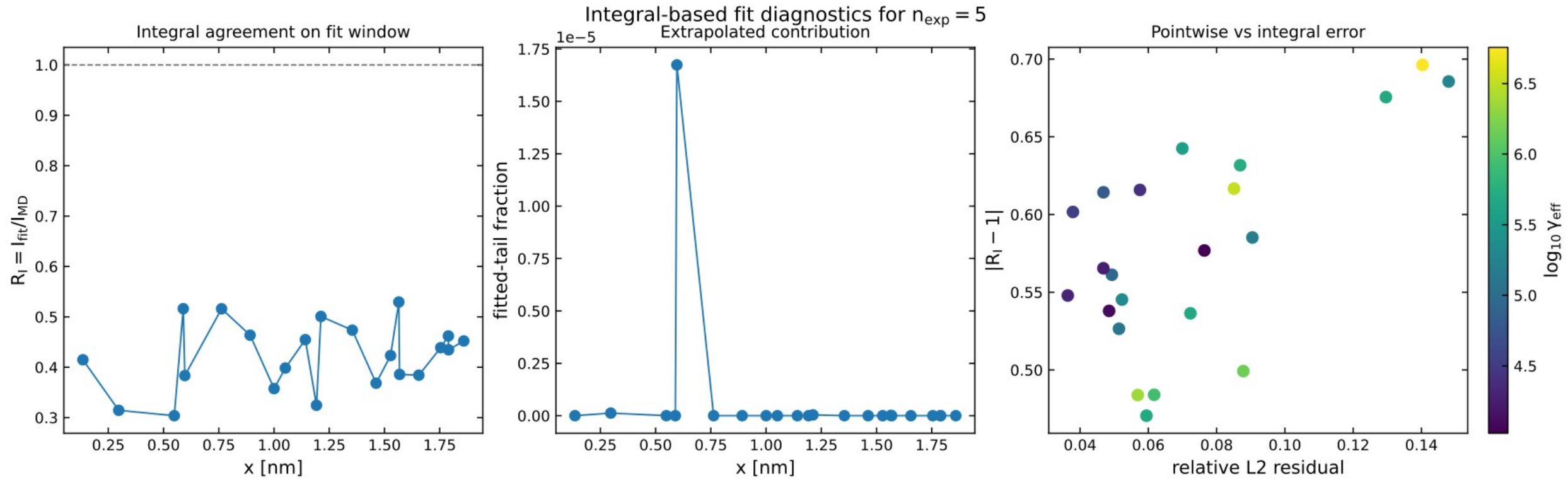

Supplementary Fig. 22 continued.

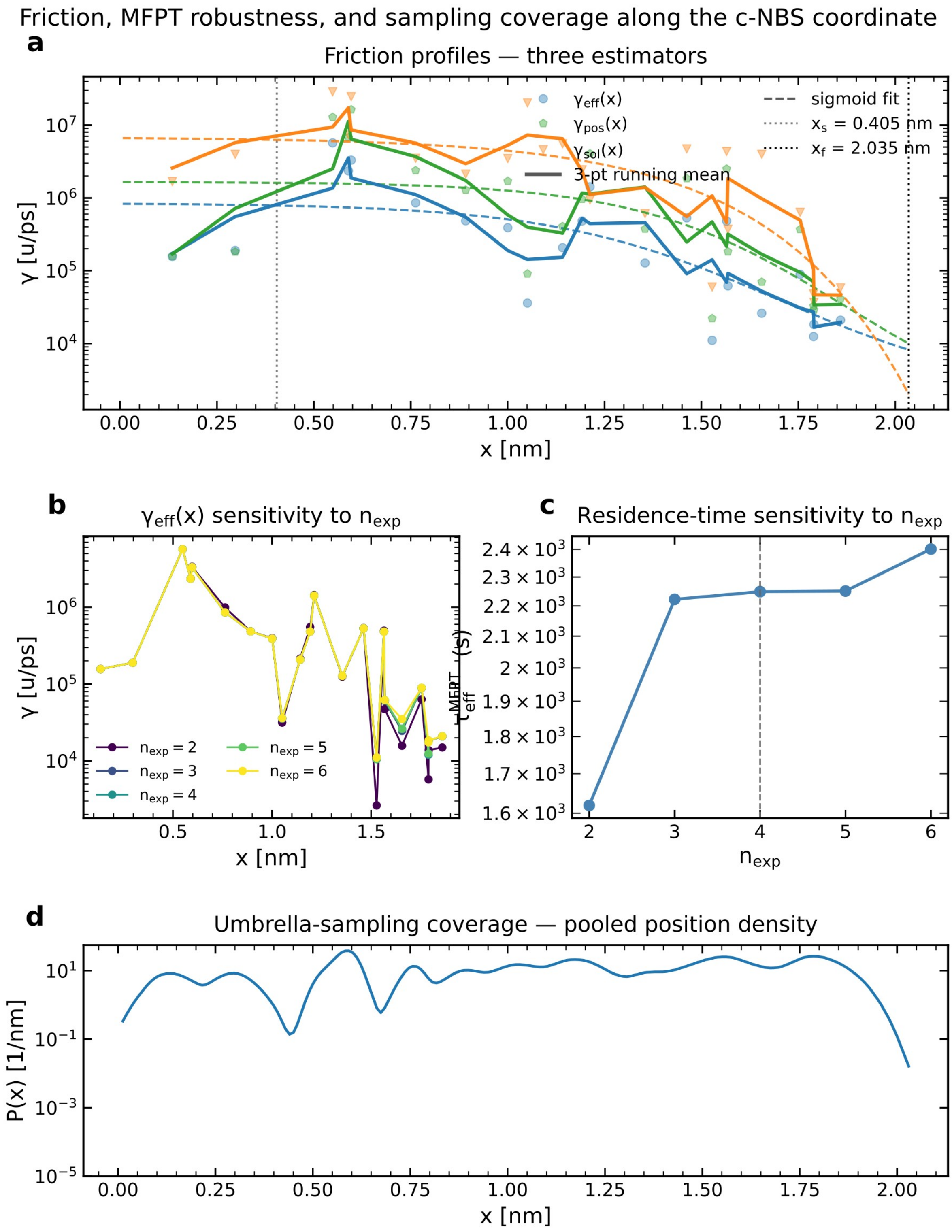

**Supplementary Fig. 23 | Cross-system comparison of predicted and experimental residence times.** Predicted residence time divided by the corresponding experimental residence time for $A_2A$ WT, N-HSP90–compound 18 and RNase S–S15. Separate curves show the non-Markovian friction $\gamma_{eff}$ and the position-autocorrelation and solvent-force friction profiles $\gamma_{pos}$ and $\gamma_{sol}$. The horizontal dashed line denotes equality between prediction and experiment.

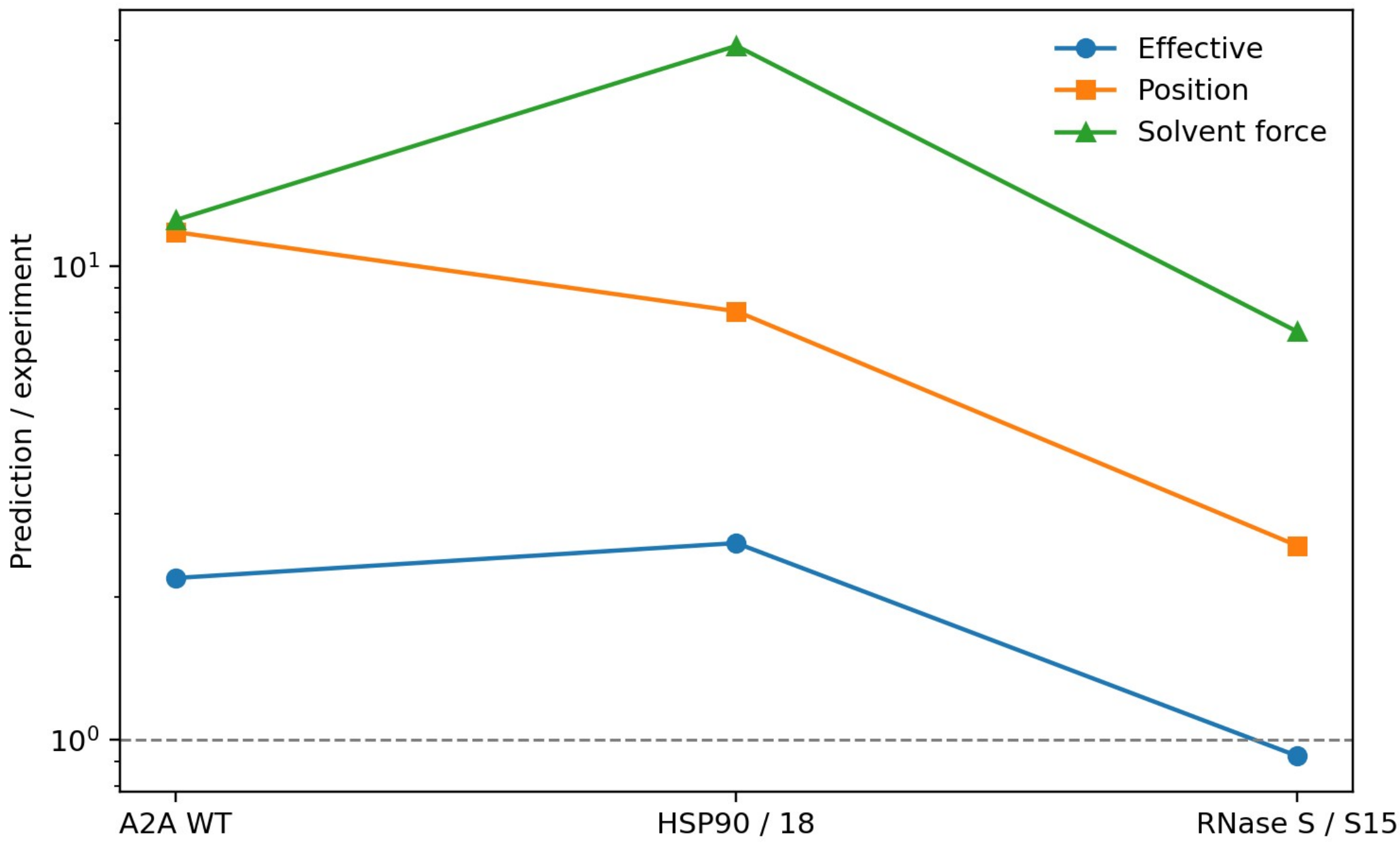

## Supplementary References